\documentclass[10pt]{article}

\usepackage{newtxtext}       %

\usepackage[paperwidth=210mm, paperheight=297mm]{geometry}

\usepackage{type1cm}        
\usepackage{makeidx}         
\usepackage{graphicx}        
\usepackage{multicol}        
\usepackage[bottom]{footmisc}

\usepackage{amsmath,amssymb,amsfonts,amsthm}

\usepackage{array,stmaryrd,booktabs,mathtools,paralist,siunitx,tablefootnote}

\usepackage{tikz,pgfplots,epstopdf}
\definecolor{myblue}{RGB}{17,64,111}
\definecolor{myred}{RGB}{150,30,15}

\newtheorem{example}{Example}[section]
\newtheorem{theorem}{Theorem}[section]
\newtheorem{proposition}{Proposition}[section]
\newtheorem{remark}{Remark}[section]
\newtheorem{algorithm}{Algorithm}[section]

\numberwithin{equation}{section}

\usepackage{comment}

\usepackage{titlesec}
\usepackage{hyperref}

\newcommand{\R}{\mathbb{R}}

\def\vects#1{\mbox{\scriptsize \boldmath$ #1$}}

\newcommand{\beqa}{\begin{eqnarray}}
\newcommand{\eeqa}{\end{eqnarray}}

\newcommand{\beq}{\begin{equation}}
\newcommand{\eeq}{\end{equation}}

\newcommand{\eref}[1]{(\ref{#1})}

\def\vect#1{\mbox{\boldmath$ #1$}}                   
\newcommand{\bA}{{\vect A}}
\newcommand{\ba}{{\vect a}}
\newcommand{\bB}{{\vect B}}
\newcommand{\bbb}{{\vect b}}
\newcommand{\bc}{{\vect c}}
\newcommand{\bC}{{\vect C}}
\newcommand{\bD}{{\vect D}}
\newcommand{\bd}{{\vect d}}
\newcommand{\be}{{\vect e}}
\newcommand{\bE}{{\vect E}}
\newcommand{\bff}{{\vect f}}
\newcommand{\bF}{{\vect F}}
\newcommand{\bG}{{\vect G}}

\newcommand{\bI}{{\vect I}}
\newcommand{\bL}{{\vect L}}
\newcommand{\bl}{{\vect l}}
\newcommand{\bn}{{\vect n}}
\newcommand{\bP}{{\vect P}}
\newcommand{\bQ}{{\vect Q}}
\newcommand{\bq}{{\vect q}}
\newcommand{\bR}{{\vect R}}
\newcommand{\br}{{\vect r}}
\newcommand{\bS}{{\vect S}}
\newcommand{\bs}{{\vect s}}
\newcommand{\bt}{{\vect t}}
\newcommand{\bT}{{\vect T}}
\newcommand{\bu}{{\vect u}}
\newcommand{\bU}{{\vect U}}
\newcommand{\bv}{{\vect v}}
\newcommand{\bV}{{\vect V}}
\newcommand{\bw}{{\vect w}}
\newcommand{\bx}{{\vect x}}
\newcommand{\by}{{\vect y}}
\newcommand{\bone}{{\vect 1}}

\newcommand{\balpha}{{\vect{\alpha} }}
\newcommand{\bbeta}{{\vect{\beta} }}
\newcommand{\bgamma}{{\vect{\gamma} }}
\newcommand{\bLambda}{{\vect{\Lambda} }}

\newcommand{\bzero}{{\vect 0 }} 

\newcommand{\oalpha}{{\overline{\alpha} }}
\newcommand{\orho}{{\overline{\rho}}}

\newcommand{\obl}{{\overline{\bl}}}
\newcommand{\obr}{{\overline{\br}}}
\newcommand{\obv}{{\overline{\bv}}}
\newcommand{\obw}{{\overline{\bw}}}
\newcommand{\oby}{{\overline{\by}}}

\newcommand{\oE}{{\overline{E}}}
\newcommand{\oee}{{\overline{e}}}
\newcommand{\oh}{{\overline{h}}}

\newcommand{\ot}{{\overline{t}}}
\newcommand{\oz}{{\overline{z}}}

\newcommand{\ogamma}{\overline{\gamma}}

\newcommand{\te}{\tilde{e}}   

\newcommand{\Tref}{T_{ref}}

\newcommand{\cbar}{\overline{c}}    
\newcommand{\cisoT}{\tilde{c}}    

\newcommand{\cp}{{\check{p}}}
\newcommand{\cT}{{\check{T}}}

\newcommand{\calG}{{\mathcal{G}}}

\newcommand{\oPi}{{\overline{\Pi}}}

\newcommand{\hbv}{\hat \bv}

\newcommand{\pdiff}[2]{\partial_{#2} #1}
\newcommand{\odiff}[2]{\frac{d\,#1}{d #2}}

\newcommand{\odifft}[1]{{\odiff{#1}{t}}}
\newcommand{\pdifft}[1]{{\partial_t #1}}
\newcommand{\pdiffx}[2]{\partial_{#2} #1}

\newcommand{\pdiv}[1]{{\nabla\cdot #1}}

\newcommand{\pgrad}[1]{{\nabla #1}}

\newcommand{\veloI}{\bV_I}

\newcommand{\diag}{{\mbox{diag}}}

\newcommand{\Source}[2]{\bS_{#1,#2}}

\newcommand{\source}[2]{S_{#1,#2}}
\newcommand{\osource}[2]{\overline{S}_{#1,#2}}
\newcommand{\sourcemy}[2]{S_{#1,#2}}

\newcommand{\dyade}[1]{{#1}\,#1^T}

\newcommand{\identity}{\bI}
\newcommand{\vprod}[2]{{#1}\cdot #2}

\newcommand{\relaxp}{\theta_p}
\newcommand{\relaxv}{\theta_v}

\newcommand{\relaxG}{\theta_{\mu}}
\newcommand{\relaxall}{\theta_{\xi}}

\newcommand{\velo}[1]{\bv_{#1}}

\def\calDD{D}

\newcommand{\proofendmy}{\hfill$\square$ }

\title{A survey on isothermal and isentropic Baer-Nunziato-type models}
\author{Maren Hantke\footnote{Institute for Mathematics, Martin-Luther-Universit\"at Halle-Wittenberg, Halle (Saale), D-06099, Germany, \textit{email: maren.hantke@mathematik.uni-halle.de} }, \
Siegfried M\"uller\footnote{Institut f\"ur Geometrie und Praktische Mathematik, RWTH Aachen, Templergraben 55, Aachen, D-52056, Germany, \textit{email: mueller@igpm.rwth-aachen.de} }, \
Aleksey Sikstel\footnote{Division of Mathematics, Universit\"at K\"oln, Weyertal 86 -- 90, K\"oln, D-50931, Germany, \textit{email: a.sikstel@uni-koeln.de}}, \
Ferdinand Thein\footnote{
Johannes Gutenberg-Universit\"at Mainz, Staudingerweg 9, 55122 Mainz, Germany, \textit{email: fthein@uni-mainz.de} }}

\usepackage{titlesec}

\begin{document}

\date{}

\maketitle

\bigskip

\abstract{
Multi-component Baer-Nunziato-type  models for isothermal and isentropic fluids
are investigated. These are given by balance equations for volume fractions, density and momentum for each component accounting for the relaxation to equilibrium by means of  relaxation terms.
Mathematical properties of the models are derived such as hyperbolicity and symmetrization. The fields are characterized and corresponding Riemann invariants are determined. Appropriate entropy-entropy flux pairs are derived taking into account the phasic energy equations including the heat flux. Physically meaningful constraints are presented that ensure  the entropy inequality to hold.
Instantaneous relaxation to equilibrium is investigated and appropriate algorithms are presented.
Numerical results for the isothermal Baer-Nunziato model are compared to an isothermal Euler model
and to an isothermal phase-field model.
}

\tableofcontents

\section{Introduction }
\label{sec:intro}


Compressible multi-component flows where the fluid is a mixture of several components
all of which may be present in different aggregate states
have a wide range of applications,
for instance a mixture of reacting gases or a mixture of a liquid and a gas. They have been successfully modeled by Baer-Nunziato-type models. Originally, these models have been introduced by Baer and Nunziato \cite{Baer-Nunziato:1986} proposing a two-fluid model for detonation waves in granular explosives. This model falls into the class of ensemble averaged models introduced by Drew and Passman \cite{Drew:1983,Drew-Passman:1999} to provide some gross features of the flow.  In contrast to volume averaging, time averaging or statistical averaging, the ensemble averaging allows the interpretation of flow phenomena in terms of repeatability of multi-component flows.

Baer-Nunziato-type  models  are full non-equilibrium models where each component has its own pressure, velocity and temperature and is governed by its own set of fluid equations. In contrast to other two-fluid models separating the fluids, here the fluids are assumed to coexist in each location due to an averaging procedure. The presence of more than one component in the mixture is modeled by the concept of volume fractions $\alpha_k$ describing the ratio of volume of component $k$ to the total volume in a ball
when the volume of the ball tends to zero, see \cite{Drew-Passman:1999}.

To model the relaxation process between the components relaxation terms have been introduced.
For instance, Saurel and Abgrall \cite{Saurel-Abgrall:1999}
included relaxation terms for the pressure and the velocities of the components.
Further relaxation procedures can be used to drive the temperatures and the Gibbs free energies
into equilibrium,
see Abgrall et al.~\cite{Saurel-Petitpas-Abgrall:2008,Rodio-Abgrall:2015} or
Zein et al.~\cite{Zein-Hantke-Warnecke:2010}.
By instantaneous relaxation equilibrium values for the pressure, the velocity, the temperature and the 
chemical potentials
can be found.

Since mechanical relaxation, thermal relaxation and relaxation of the chemical potentials proceed on different time scales, cf.~\cite{Saurel-Le,Petitpas-Saurel-Franquet-Chinnayya:2009,Saurel-Petitpas-Abgrall:2008}, reduced models have been derived assuming zero relaxation times for some of the non-equilibrium quantities relaxing much faster than  the remaining quantities, see \cite{Kapila-Menikoff-Bdzil-Son-Stewart:2001,Hantke-Mueller-Grabowsky:2020}. Thus, the stiffness inherent in the non-equilibrium model is avoided  that allows for  a much faster numerical simulation of the reduced model.
Usually reduced models suffer from some short-comings. For instance, conservation of energy might be violated, the system loses its hyperbolicity or the reduced model is not the asymptotic correct limit of the nonlinear model.
For more details we refer the interested reader to \cite{Zein:2010}
and the references cited therein.

In \cite{Flatten-Lund:2011,Linga-Flatten:2019} a  hierarchy of
two-fluid models  derived from the Baer-Nunziato model \cite{Baer-Nunziato:1986} is investigated  where mechanical, thermal and chemical relaxation is assumed to proceed in different order. This hierarchy splits into two branches of models distinguished by the assumption of dynamical velocity equilibrium and local velocity equilibrium, where either  both  fluids have the \emph{same} velocity or the fluids have \emph{different} velocities that coincide for a particular state.

Since Baer-Nunziato-type models are averaged models, the resulting balance laws are underdetermined. To compensate for the loss in information \emph{closure conditions} have to be imposed additionally.
There exists a rich literature on Baer-Nunziato-type models, see for
instance \cite{Gallouet-Herard-Seguin:04, Herard:07, Saleh:2012, Coquel_Herard_Saleh-Seguin:2014,
Mueller-Hantke-Richter:16, Hantke-Mueller:19}
investigating their properties, in particular, \emph{hyperbolicity} and \emph{thermodynamical consistency}. These properties are essential from an analytical, numerical and physical point of view. For instance, hyperbolicity is needed in the analysis of the Riemann problem and the construction of (approximate) Riemann solvers whereas a non-negative entropy production ensures consistency with the 2nd law of thermodynamics  and also might ensure well-posedness, although not yet mathematically rigorously verified. The analysis helps to identify physically admissible closure conditions for the interfacial pressure and the interfacial velocity that occur as model parameters and cannot be closed due to the averaging procedure.
However,  a unique closure 
does not exist. Typically, the interfacial velocity is modeled as a convex combination of the phasic velocities.
Then the interfacial pressures can be uniquely determined
assuming that the induced entropy production term due to interfacial velocity and interfacial pressures vanishes,
cf.~\cite{Herard:07,Mueller-Hantke-Richter:16}.

In recent years, Baer-Nunziato-type models have been applied to different types of fluids. Originally, the components of the fluids are assumed to be immiscible. In  \cite{Herard-Hurisse:2021,Herard-Mathise:2021}
hybrid mixtures consisting of both immiscible and miscible components are investigated. A mixture of weakly compressible fluids is considered in \cite{ReAbgrall:2020}. Our main interest in the
present work is on 
isothermal and isentropic fluids. 
These fall in the larger class of barotropic fluids.
For a one-dimensional model
H{\`e}rard \cite{herard:2016} investigates some properties of a $K$-component Baer-Nunziato-type model for barotropic fluids. Choosing the interfacial velocity as a convex combination of the phasic velocities 
and assuming that the entropy production due to the interfacial pressures and 
the interfacial velocity vanishes
there
uniquely exist interfacial pressures for $K=2,3$ and 4. In case of 
barotropic fluids a mathematical
entropy inequality 
for multi-phase fluids is verified. For  a special class of pressure-velocity relaxation the induced entropy production is consistent with the entropy inequality. This also holds when taking into account mass transfer.
Also for a one-dimensional model Saleh and Seguin \cite{saleh-seguin:2020} have proven weak hyperbolicity by determining eigenvalues and eigenvectors, the existence of a convex mathematical entropy in case of 
barotropic
fluids and the existence of a symmetric form. The analysis is confined to a particular choice of the interfacial velocity and interfacial pressures.
In contrast to \cite{herard:2016,saleh-seguin:2020} a multi-dimensional three-component model is considered in \cite{boukili-herard:2018}.
A mathematical
entropy inequality is proven. Moreover, the corresponding quasi-1D model is verified to be weakly hyperbolic and Riemann invariants corresponding to the linearly degenerated field associated with the interfacial velocity are  given. Again, the results are subject to a specific choice for the interfacial velocity and interfacial pressures.
In a recent work a conservative compressible two phase mixture model has been discussed analytically in \cite{Thein2022} and numerically in \cite{RioMartin2023}. This model is the barotropic version of a model studied in \cite{Romenski2007,Romenski2009} and
belongs to
the class of Symmetric Hyperbolic and Thermodynamically Compatible (SHTC) systems.
An overview on the theory of SHTC systems
can be found in \cite{Godunov:1995a,Godunov1996,Godrom2003,Peshkov2018,Romenski1998,Romenski2001} and we
refer to the aforementioned literature for further references. In \cite{Thein2022} the instantaneous relaxation to the Kapila model limit is studied. Further it is shown in \cite{Thein2022} that for smooth solutions the conservative
isothermal/isentropic
 two fluid SHTC model is equivalent to the
isothermal/isentropic
 Baer-Nunziato system for a particular choice of interface quantities and numerical studies are provided. Therefore these results necessitate a further thorough study of the
 isothermal/isentropic
 Baer-Nunziato system to provide a sound basis for further comparative studies, also including isothermal sharp interface models studied in \cite{Hantke2013,Hantke2019a}. For further literature on symmetric hyperbolic systems for mixtures we also refer to \cite{Ruggeri2021}.

The main objective of the current work is to fill some gaps in the case of
barotropic
fluids and to extend the analysis
in the particular case of
isothermal fluids as well. In detail, the issues we are addressing in our work are as follows:
(i) eigenvalues and eigenvectors for the quasi-1D model are determined, hyperbolicity away from resonance states is verified,
the corresponding characteristic fields are investigated and a complete set of linearly independent Riemann invariants is given for all fields; note that the analysis
is not restricted to isothermal and isentropic fluids 
but applies to general barotropic fluids;
(ii) a mathematical entropy inequality according to Lax's entropy-entropy flux pair is verified independently for isentropic and isothermal fluids;
(iii) unique interfacial pressures are determined ensuring that the entropy production induced by the interfacial states  is thermodynamically consistent; the interfacial pressures depend on the  interfacial velocity chosen as a convex combination of the phasic velocities;
(iv) the entropy production due mechanical relaxation and relaxation of chemical potentials ($K=2$ (liquid-vapor), $K=3$
(liquid-vapor-gas), isothermal fluids) is verified to be thermodynamically consistent;
(v) the quasi-1D model is symmetrized;
(vi) procedures for instantaneous relaxation of velocity, pressure and chemical potentials (isothermal fluids) are provided.
We emphasize that all investigations are performed for mixtures with an arbitrary number of components $K$ except for the relaxation of the chemical potentials.
Furthermore we want to remark that the system is Galilean invariant and the proof is in complete analogy to the one for the full system discussed in \cite{Mueller-Hantke-Richter:16}.

Thus, the paper is organized as follows:
In Sect.~\ref{sec:baer-nunziato} we present the model. Mathematical properties are discussed in Sect.~\ref{sec:bn-math-prop}. An entropy-entropy-flux pair is derived in Sect.~\ref{subsec:bn-neq-thermo}. In particular, we present a thermodynamically consistent closure for the interfacial pressures and the interfacial velocity. Then in Sect.~\ref{subsec:relax} we investigate instantaneous relaxation to equilibrium and provide appropriate algorithms. Numerical simulations are performed and discussed in Sect.~\ref{sec:bn-neq-results}. Finally, in Sect.~\ref{sec:conclusion} we summarize our findings and give an outlook on future work.

\section{Baer-Nunziato-type models with barotropic equation of state}
\label{sec:baer-nunziato}

First of all, we summarize the balance laws for the multi-component Baer-Nunziato-type model where the number of components is arbitrary. 
The phasic fluids are assumed to be barotropic, i.e., each of the single fluids is governed by a pressure only depending on the  density of the fluid, see Section \ref{subsec:bn-evolution}.
To close the model we have to provide appropriate equations of state (EoS) for the phasic pressures. Starting from fundamental relations of thermodynamics  we provide the stiffened gas EoS for both isothermal fluids and isentropic fluids, see Section \ref{thermodynamics-properties-entropy}.
Finally, in Section \ref{subsec:bn-neq-relax} we discuss appropriate relaxation models.

\subsection{Evolution equations}
\label{subsec:bn-evolution}


The multi-component flow is described by a  non-equilibrium model where all components are present in each point of the  space-time continuum. Each component $k=1,\ldots,K$ has its own density $\rho_k$ and velocity $\velo{k}$. The amount of each component is determined by its volume fraction $\alpha_k$. The volume fractions are related by the \textit{saturation constraint}
\begin{equation}
\label{eq:saturation}
\sum_{k=1}^{K} \alpha_k=1,\quad \alpha_k\in (0,1) .
\end{equation}
The evolution of the volume fractions is characterized by the non-conservative equations
\begin{equation}
\label{eq:sa-model-alphak}
 \pdifft{\alpha_k} + \vprod{\veloI}{\pgrad{\alpha_k}} = \source{\alpha}{k} ,\
  k = 1,\ldots, K .
\end{equation}
Due to the saturation condition \eref{eq:saturation} we only need $K-1$ equations.
Without loss of generality we express $\alpha_K$ by the other volume fractions, i.e.,
\begin{equation}
\label{eq:convention}
  \alpha_K = 1 - \sum_{k=1}^{K-1}\alpha_k,\
  \nabla \alpha_K =  - \sum_{k=1}^{K-1}\nabla\alpha_k,\
  \source{\alpha}{K} =  - \sum_{k=1}^{K-1}\source{\alpha}{k} .
\end{equation}

In analogy to the two-phase model of Saurel and Abgrall \cite{Saurel-Abgrall:1999}
the fluid equations for each component can be written as
\begin{subequations}
\label{eq:sa-model}
\vspace*{-5mm}
\begin{align}
\label{eq:sa-model-rhok}
&\pdifft{(\alpha_k\,\rho_k)} + \pdiv{(\alpha_k\,\rho_k\,\velo{k})} = \source{\alpha\rho}{k} ,\\
\label{eq:sa-model-mk}
&\pdifft{(\alpha_k\,\rho_k\,\velo{k})} + \pdiv{(\alpha_k\,\rho_k\,\dyade{\velo{k}} + \alpha_k\,p_k\,\identity)} +
\sum_{l=1,\ne k}^{K} P_{k,l}\,\pgrad{\alpha_l} =
 \Source{\alpha\rho\vects{\velo{}}}{k} ,
\end{align}
\end{subequations}
where we neglect effects due to viscosity, surface tension and gravity.
The fluid equations are supplemented by a barotropic equation of state
\begin{equation}
\label{eq:eos-k}
 p_k = p_k(\rho_k)\quad\mbox{resp.}\quad \rho_k = \rho_k(p_k)
\end{equation}
for each of the components.
In the present work we will consider both \textit{isentropic} fluids with constant entropies, i.e.,
\beq
\label{eq:isentropicfluid}
 s_k= const.\quad\forall k=1,\ldots, K
\eeq
and
\textit{isothermal} fluids with constant temperatures, i.e.,
\beq
\label{eq:isothermalfluid}
 T_k = \Tref =const.\quad\forall k=1,\ldots, K .
\eeq
The above model is a simplification of the full Baer-Nunziato system,
see e.g.~\cite{Mueller-Hantke-Richter:16},
where the system consists of a mass balance, an energy balance and a balance of momentum for each component and additionally $K-1$ equations for the volume fractions.
In particular, the temperature and entropy may also vary and, hence, the pressure does not depend on the density alone as in the barotropic case.
Neglecting viscosity but accounting for heat exchange
the energy equation of the full Baer-Nunziato system
reads
\begin{align}
\label{eq:sa-model-full-Ek-local}
\pdifft{(\alpha_k\,\rho_k\,E_k)}
+ \pdiv{(\alpha_k\,\rho_k\,\velo{k}\,(E_k+p_k/\rho_k))}
+\sum_{l=1,\ne k}^{K} P_{k,l}\, \vprod{\veloI} \pgrad{\alpha_l} =
- \pdiv{(\alpha_k\, \bq_k)}
+ \sourcemy{\alpha\rho E}{k}
\end{align}
with the heat flux $\bq_k$ and total energy $E_k = e_k + \velo{k}^2/2$ determined by
the internal energy  $e_k$ and kinetic energy $u_k=\velo{k}^2/2$. We emphasize that in case of an isothermal fluid the energy equation becomes a conditional  equation for the heat flux.
This is employed in the investigation of the entropy production, see Section \ref{subsec:bn-neq-thermo-isothermal}.
For isentropic single-component flows the energy equation becomes redundant. Note that this is not the case for multi-component or multi-phase flows. Nevertheless, energy balance equations typically are not taken into account when considering isentropic systems.

The gradient terms on the left-hand side in
\eref{eq:sa-model-alphak}, \eref{eq:sa-model-mk}  and \eref{eq:sa-model-full-Ek-local}
account for the interaction between different components.
In particular, the term $P_{k,l}$ accounts for different pressures at the phase interface between phase $k$ and $l$.
The interfacial  velocity is denoted by $\veloI$.

The source terms
$\Source{\alpha\rho\vects{\velo{}}}{k}$
$\source{\alpha}{k}$, $\source{\alpha\rho}{k}$, $\Source{\alpha\rho\vects{\velo{}}}{k}$  and $\source{\alpha\rho E}{k}$
on the right-hand sides of
\eref{eq:sa-model-alphak}, \eref{eq:sa-model} and \eref{eq:sa-model-full-Ek-local}
describe the interaction of the components corresponding to mass, momentum and energy transfer
via different relaxation types, e.g., due to velocity, pressure and chemical potentials
$\xi\in\{v,p,\mu\}$
\[
\source{\alpha}{k}:=\sum_\xi\source{\alpha}{k}^\xi, \,
\source{\alpha\rho}{k}:=\sum_\xi\source{\alpha\rho}{k}^\xi, \,
\Source{\alpha\rho\vects{\velo{}}}{k} := \sum_\xi\Source{\alpha\rho\vects{\velo{}}}{k}^\xi, \,
\source{\alpha\rho E}{k}:=\sum_\xi\source{\alpha\rho E}{k}^\xi.
\]
These depend on the specific components at hand that will be discussed in Section
\ref{subsec:bn-neq-relax}.

Obviously, the evolution equations are not in conservative form. To respect the physical principles of conservation of mass, momentum and energy we impose the following \textit{conservation constraints} on the source terms
\begin{equation}
\label{eq:constraints}
\sum_{k=1}^{K} \source{\alpha}{k}^\xi=0,\quad
\sum_{k=1}^{K} \source{\alpha\rho}{k}^\xi=0,\quad
\sum_{k=1}^{K} \Source{\alpha\rho\vects{\velo{}}}{k}^\xi=\bzero,\quad
\sum_{k=1}^{K} \source{\alpha\rho E}{k}^\xi=0
\end{equation}
for each relaxation type as well as the interfacial pressures
\begin{align}
\label{eq:constraints-interfacial-pressures-gen}
\sum_{k=1}^K \sum_{l=1,\ne k}^K P_{k,l} \pgrad{\alpha_l} =
\sum_{l=1}^{K-1} \left( \sum_{k=1,\ne l}^K  P_{k,l} - \sum_{k=1}^{K-1}  P_{k,K}  \right) \pgrad{\alpha_l} =
0 .
\end{align}
Since by the saturation constraint \eref{eq:saturation} the gradients $\pgrad{\alpha_l}$, $l=1,\ldots,K-1$, are linearly independent, the conservation constraint \eref{eq:constraints-interfacial-pressures-gen} is equivalent to
\beq
\label{eq:constraints-interfacial-pressures}
  \sum_{k=1,\ne l}^{K} P_{k,l} = \sum_{k=1}^{K-1} P_{k,K} =: P_I   \quad\forall l=1,\ldots,K-1.
\eeq
Then the evolution equations for the mixture quantities
\begin{equation}
\label{eq:mixture-cons}
             \rho := \sum_{k=1}^{K} \alpha_k\,\rho_k,\
             \rho\velo{} := \sum_{k=1}^{K} \alpha_k\,\rho_k\,\velo{k}
\end{equation}
obtained by  summation of the single-component fluid equations \eref{eq:sa-model-rhok}, \eref{eq:sa-model-mk}
and employing the saturation constraint \eref{eq:saturation} as well as the conservation constraints
\eref{eq:constraints} and \eref{eq:constraints-interfacial-pressures}
are in conservative form. See \cite{Mueller-Hantke-Richter:16} for details.

So far, the model is not yet closed. For this purpose, we have to find closing conditions for the interfacial pressures $P_{k,l}$,
the interfacial velocity $\veloI$ and the relaxation terms $\sourcemy{\alpha}{k}$, $\sourcemy{\alpha\rho}{k}$,
$\Source{\alpha\rho\vects{\velo{}}}{k}$, $\sourcemy{\alpha\rho E}{k}$.
In the following sections we will derive appropriate constraints. However, these  will not specify a unique model but some options are still remaining for the choice of the interfacial velocity and the relaxation terms.

\begin{remark}
\label{rem:energy-barotropic-fluid}
For
a given barotropic pressure law \eqref{eq:eos-k} we always can introduce an ``internal energy'' function $\oee_k=\oee_k(\rho_k)$ as solution of the ODE $\oee'_k(\rho_k) = p_k/\rho_k^2$. Defining the `` total energy'' $\oE_k=\oee_k +\bv_k^2/2$ an ``energy''  equation can be derived from \eqref{eq:sa-model}
\begin{align}
\label{eq:sa-model-barotropic-Ek-local}
\pdifft{(\alpha_k\,\rho_k\,\oE_k)}
+ \pdiv{(\alpha_k\,\rho_k\,\velo{k}\,(\oE_k+p_k/\rho_k))}
+\sum_{l=1,\ne k}^{K} P_{k,l}\, \vprod{\bv_k} \pgrad{\alpha_l} =
p_k \vprod{(\veloI - \bv_k)}\pgrad{\alpha_k}
+\osource{\alpha \rho \oE}{k}.
\end{align}
This balance law is different to \eqref{eq:sa-model-full-Ek-local}. In particular,  $\osource{\alpha \rho \oE}{k}$ differs from
$\sourcemy{\alpha\rho E}{k}$. In case of an isentropic fluid the ``internal and total energies'' $\oee_k$ and $\oE_k$  coincide with the energies $e_k$ and $E_k$, respectively. In Section \ref{subsec:bn-neq-thermo-isentrop} for isentropic fluids we use $ \sum_{k=1}^K \alpha_k \rho_k \oE_k$ for a mathematical entropy in Lax' concept of entropy-entropy flux pairs.
\end{remark}

\subsection{Some remarks on thermodynamics}
\label{thermodynamics-properties-entropy}

In order to investigate thermodynamical properties of the model \eref{eq:sa-model-alphak}, \eref{eq:sa-model},
we assume that the equation of state \eref{eq:eos-k} is in agreement with thermodynamical principles. For this purpose, we recall from equilibrium thermodynamics that for each component $k$ the internal energy $e_k$ is a function of the entropy and the volume $\tau_k$ satisfying
\beq
  \label{thermo-energy}
   d e_k = T_k d s_k - p_k d\tau_k.
\eeq
Thus, the pressure and the temperature are the partial derivatives of $e_k(\tau_k,s_k)$ that are assumed to be positive, i.e.,
\begin{align}
\label{thermo-pres-temp}
  p_k(\tau_k,s_k) := - \frac{\partial e_k}{\partial \tau_k}(\tau_k,s_k) \ge 0,\quad
  T_k(\tau_k,s_k) :=   \frac{\partial e_k}{\partial    s_k}(\tau_k,s_k) \ge 0.
\end{align}
Furthermore, to ensure thermodynamical stability we assume that the Hessian of $e_k$ is a convex function with respect to $\tau_k$ and $s_k$, i.e.,
\begin{align}
   \label{thermo-energy-convex}
 \frac{\partial^2 e_k}{\partial^2 \tau_k}(\tau_k,s_k) \ge 0,\
  \frac{\partial^2 e_k}{\partial^2    s_k}(\tau_k,s_k) \ge 0,\
  \frac{\partial^2 e_k}{\partial^2 \tau_k}(\tau_k,s_k) \frac{\partial^2 e_k}{\partial^2    s_k}(\tau_k,s_k) \ge
  \left( \frac{\partial^2 e_k}{\partial \tau_k \partial s_k}(\tau_k,s_k) \right)^2  .
\end{align}
Finally, in agreement with the third law of thermodynamics we assume
\[
  \tau_k \ge 0,\quad s_k\ge 0.
\]

Alternatively, assuming that $p_k$ and $T_k$ are strictly positive, then $e_k$ becomes a monotone function in $\tau_k$ and $s_k$ and we may change variables, i.e.,
$s_k=s_k(\tau_k,e_k)$ satisfying
\beq
  \label{thermo-entropy}
  T_k d s_k = d e_k  + p_k d\tau_k 
\eeq
with partial derivatives
\beq
  \label{thermo-entropy-1}
  \frac{\partial s_k}{\partial \tau_k}(\tau_k,e_k) = \frac{\cp_k(\tau_k,e_k)}{\cT_k(\tau_k,e_k)} > 0,\quad
  \frac{\partial s_k}{\partial    e_k}(\tau_k,e_k) = \frac{1}{\cT_k(\tau_k,e_k)} > 0.
\eeq
and
\[
  \cp_k(\tau_k,e_k) := p_k(\tau_k,s_k(\tau_k,e_k)),\quad
  \cT_k(\tau_k,e_k) := T_k(\tau_k,s_k(\tau_k,e_k)).
\]
It is well-known that $s_k=s_k(\tau_k,e_k)$ is a concave function, i.e., the Hessian is negative-definite
\begin{align}
  \label{thermo-entropy-1b}
 \frac{\partial^2 s_k}{\partial^2 \tau_k}(\tau_k,e_k) \le 0,\
  \frac{\partial^2 s_k}{\partial^2    e_k}(\tau_k,e_k) \le 0,\
 \frac{\partial^2 s_k}{\partial^2 \tau_k}(\tau_k,e_k) \frac{\partial^2 s_k}{\partial^2    e_k}(\tau_k,e_k) \ge
  \left( \frac{\partial^2 s_k}{\partial \tau_k \partial e_k}(\tau_k,e_k) \right)^2  ,
\end{align}
if and only if $e_k(\tau_k,s_k)$ is a convex function, i.e., thermodynamic stability holds.

Finally, following \cite{Thein:2018}, p.~21 and 23, we introduce the speed of sound $c_k$ and the fundamental derivative of thermodynamics $\calG_k$ defined by means of \eref{thermo-pres-temp} as
\begin{align}
\label{eq:sound-fdt-s-tau}
c_k^2(\tau_k,s_k) := -\tau_k^2 \frac{\partial p_k}{\partial \tau_k}(\tau_k,s_k),\quad
\calG_k(\tau_k,s_k) := \frac{1}{2} \tau_k^3 \frac{\frac{\partial^2 p_k}{\partial \tau_k^2}(\tau_k,s_k)}{c_k^2(\tau_k,s_k)}
\end{align}
or, equivalently,
\begin{align}
\label{eq:sound-fdt-s-rho}
c_k^2(\rho_k,s_k) = \frac{\partial p_k}{\partial \rho_k}(\rho_k,s_k),\quad
\calG_k(\rho_k,s_k) =  \frac{1}{c_k(\rho_k,s_k)} \frac{\partial (\rho_k c_k)}{\partial \rho_k}(\rho_k,s_k) ,
\end{align}
or, equivalently,
\begin{align}
\label{eq:sound-fdt-T-rho-general}
c_k^2(\rho_k,T_k) = \frac{\partial p_k}{\partial \rho_k}(\rho_k,T_k) +
                    \frac{T_k}{\rho_k^2 c_{v,k}} \left(\frac{\partial p_k}{\partial T_k}(\rho_k,T_k)\right)^2,\quad
\calG_k(\rho_k,T_k) = 1 + \frac{\rho_k}{2\, c_k(\rho_k,T_k)} \frac{\partial c_k}{\partial \rho_k}(\rho_k,T_k).
\end{align}

\begin{example}[Stiffened gas EoS]
\label{ex:stiff-gas-Eos}
Let the constant material parameters  $c_{v,k}$, $\gamma_k$, $\pi_k$ and $q_k$ denoting the specific
heat capacity at constant volume, the adiabatic exponent, the minimal pressure and the heat of formation of component $k$, respectively, chosen such that
$c_{v,k}>0$, $\pi_k\ge0$, $q_k$ and $\gamma_k>1$. Let further $\rho_{0,k}$ be some reference density at reference temperature $T_{0,k}$.
By means of these coefficients the internal energy, the pressure and the entropy
are determined by
\begin{subequations}
\label{stiff}
\begin{align}
\label{eq:eos-stiff-e}
& e_k = c_{v,k} T_k + \pi_k/\rho_k + q_k\,,\\
\label{eq:eos-stiff-p}
& p_k = \rho_k(\gamma_k-1)c_{v,k}T_k-\pi_k\,,\\
\label{eq:eos-stiff-s}
& s_k = c_{v,k} +  c_{v,k}\ln \frac{T_k}{T_{0,k}} +
  q_k \frac{1}{T_{0,k}} -
  (\gamma_k-1)c_{v,k}  \ln \frac{\rho_k}{\rho_{0,k}} -
   \frac{\pi_k}{\rho_{0,k}} \frac{1}{T_{0,k}}\,.
\end{align}
\end{subequations}
Then the speed of sound and the fundamental derivative of gas dynamics  are determined  by \eref{eq:sound-fdt-T-rho-general} as
\begin{align}
\label{eq:sound-fdt-T-rho-stiff}
c_k = \sqrt{\gamma_k T_k \, c_{v,k}\, (\gamma_k-1)},
\quad
\calG_k = \frac{\gamma_k+1}{2}.
\end{align}
For a derivation we refer to \cite{Hantke-Mueller:18}.
\end{example}

In the present work we are interested in
isothermal and isentropic fluids where 
the pressure is only a function of the density, i.e.,
\begin{align}
p_k(\rho_k)=p_k(\rho_k,s_k=const) \quad \text{or} \quad p_k(\rho_k)=p_k(\rho_k,T_k=const)
\end{align}
in \eqref{eq:sound-fdt-s-rho} and \eqref{eq:sound-fdt-T-rho}.
Thus, these fluids fall in the class of barotropic fluids.
In the following we assume for barotropic fluids that
\begin{subequations}
  \label{eq:eos-barotropic-ass}
\begin{align}
  \label{eq:eos-barotropic-ass-1}
  & p_k'(\rho_k) > 0, \\[2mm]
\label{eq:eos-fdt}
  &\calG_k(\rho_k) =1+\frac{\rho_k}{2 c_k^2(\rho_k)} p_k''(\rho_k) > 0.
\end{align}
\end{subequations}
This ensures positivity of the barotropic speed of sound
\begin{align}
\label{eq:eos-c}
  \cbar_k(\rho_k) = \sqrt{p_k'(\rho_k)} > 0
\end{align}
and convexity of 
the isentropes or isotherms, in the special case of an isentropic or isothermal fluid, respectively.\\
In case of an isothermal fluid the isothermal speed of sound and the fundamental derivative of thermodynamics are defined as
\begin{align}
\label{eq:sound-fdt-T-tau}
\cisoT_k^2(\tau_k) := -\tau_k^2 \frac{\partial p_k}{\partial \tau_k}(\tau_k,\Tref),\quad
\overline{\calG}(\tau_k := \frac{1}{2} \tau_k^3 \frac{\frac{\partial^2 p_k}{\partial \tau_k^2}(\tau_k,\Tref)}{\cisoT_k^2(\tau_k,\Tref)}
\end{align}
or, equivalently,
\begin{align}
\label{eq:sound-fdt-T-rho}
\cisoT_k^2(\rho_k) = \frac{\partial p_k}{\partial \rho_k}(\rho_k,\Tref),\quad
\overline{\calG}_k(\tau_k) = 1 + \frac{\rho_k}{\cisoT_k(\rho_k,Tref)} \frac{\partial \cisoT_k}{\partial \rho_k}(\rho_k,\Tref).
\end{align}
Note that  the barotropic speed of sound $\cbar_k$, see \eqref{eq:eos-c},
is identical to the thermodynamical speed of sound $c_k(\rho_k,s_k=const)$,
see \eqref{eq:sound-fdt-s-rho}, for an isentropic fluid whereas it is identical to the isothermal speed of sound $\cisoT_k(\rho_k)$,
see \eqref{eq:sound-fdt-T-rho},  for an  isothermal fluid.

For our numerical investigations in Section \ref{sec:bn-neq-results} we use the isentropic stiffened gas
EoS and the isothermal stiffened gas EoS which have been verified to be thermodynamically consistent,
see \cite{Hantke-Mueller:18}. These are summarized in the following.
\begin{example}[Isothermal stiffened gas EoS]
\label{ex:isotherm-stiff-gas-Eos}

Starting from the fixed temperature $\Tref$, i.e., $T_k=\Tref$ for all $k$ and the corresponding saturation pressures $p_{ref,k}$, we choose
the ratio of specific heats $\gamma_k>1$, the minimal pressure $\pi_k\ge 0$, the specific heat at constant volume $c_{v,k}$, and the heat of formation $q_k$ for every component at the reference state.
By means of these coefficients the pressure and the entropy are determined by
\begin{align}
	\label{eq:eos-iso-p}
  p_k(\rho_k) &= \cisoT_k^2 \rho_k - \pi_k,\\
  \label{eq:eos-iso-s}
  s_k(\rho_k)  &= -(\gamma_k-1)c_{v,k}\,\ln\left( \frac{p_k+\pi_k}{p_{ref,k}+\pi_k}\right) + c_{v,k}+\frac{q_k}{\Tref}.
\end{align}
In this particular case, the isothermal speed of sound and the isothermal fundamental derivative of gas dynamics become constants that are determined  by \eref{eq:sound-fdt-T-rho} as
\begin{align}
\label{eq:sound-fdt-T-rho-isotherm}
\cisoT_k = \sqrt{\Tref \, c_{v,k}\, (\gamma_k-1)},
\quad
\overline{\calG}_k = 1.
\end{align}
We emphasize that the speed of sound $c_k$ and the isothermal speed of sound $\cisoT_k$ defined by \eqref{eq:sound-fdt-T-rho-general} and \eqref{eq:sound-fdt-T-rho}, respectively, are related by
\begin{align}
\label{eq:diff-sound-speed-stiff}
c_k^2 = \gamma_k \cisoT_k^2 .
\end{align}
The fundamental derivatives $\calG_k$ and $\overline{\calG}_k$ have the same sign.
For later use we also give the internal energies and the Gibbs free energies 
\begin{align}
e_k(\rho_k) &= c_{v,k}\Tref+\frac{\pi_k}{\rho_k}+e_{0,k}, \quad e_{0,k}:=q_k-(\gamma_k-1)c_{v,k}\Tref,
\\
g_k(\rho_k) &= (\gamma_k-1)c_{v,k}\Tref\ln\left(\frac{p_k(\rho_k)+\pi_k}{p_{ref,k}+\pi_k}\right).
\end{align}

\end{example}

\begin{example}[Isentropic stiffened gas EoS]
\label{ex:isentrop-stiff-gas-Eos}
Again, starting from the uniform reference temperature $\Tref$ and the corresponding saturation pressures
$p_{ref,k}$, we choose
the ratio of specific heats $\gamma_k>1$, the minimal pressure $\pi_k\ge 0$, the specific heat at constant volume $c_{v,k}$, and the heat of formation $q_k$ for every component at the reference state.
By means of these coefficients the pressure and the temperature are determined by
\begin{align}
  p_k(\rho_k) &= A_k \rho_k^{\gamma_k}-\pi_k,\\
  T_k(\rho_k) &= B_k \rho_k^{\gamma_k-1}
\end{align}
with
\begin{subequations}
\label{eq:isentr-stiffgas-coeff}
\begin{align}
\label{eq:isentr-stiffgas-coeff-Ak}
  A_k   &= \frac{T_{0,k}^{\gamma_k}}{(p_{0,k}+\pi_k)^{\gamma_k-1}} \left( c_{v,k} (\gamma_k-1) \right)^{\gamma_k} ,\\
\label{eq:isentr-stiffgas-coeff-Bk}
  B_k   &= \frac{T_{0,k}^{\gamma_k}}{(p_{0,k}+\pi_k)^{\gamma_k-1}} \left( c_{v,k} (\gamma_k-1) \right)^{\gamma_k-1} .
\end{align}
\end{subequations}
For sake of completeness, we give the constant entropy at the reference state as well as the internal energy
\begin{align}
  s_{0,k} &=
c_{v,k} + \frac{q_k}{\Tref}
\\
e_k(\rho_k) &= c_{v,k}T_k(\rho_k)+\frac{\pi_k}{\rho_k}+e_{0,k}, \quad e_{0,k}:=q_k-(\gamma_k-1)c_{v,k}\Tref.
\end{align}
The speed of sound and the fundamental derivative of gas dynamics are determined  by \eref{eq:sound-fdt-s-rho} as
\begin{align}
c_k(\rho_k) =
\sqrt{\gamma_k\, A_k\, \rho_k^{\gamma_k-1}}
,\quad
\calG_k = \frac{\gamma_k+1}{2} .
\end{align}

\end{example}

So far we have discussed the individual components. However, since we are considering a mixture of $K$ components the specific internal energy $e$ of the mixture is given by the \emph{Gibbs-Duhem-relation}, cf.~\cite{Mueller:2009},
\beq
\label{eq:GD}
	de = Tds - pd\frac{1}{\rho} + \sum_{k=1}^K\mu_k dN_k
\eeq
where $e$, $s$, $p$ denote the mixture quantities of the energy, the entropy and the pressure defined analogously   to \eqref{eq:mixture-cons} as
\begin{equation}
\label{eq:mixture-pres}
  \rho e = \sum_{i=1}^K \alpha_k \rho_k e_k,\
  \rho s = \sum_{i=1}^K \alpha_k \rho_k s_k,\
  p = \sum_{i=1}^K \alpha_k p_k .
\end{equation}
We do not comment on the mixture temperature $T$ here in detail, since we are considering an isothermal process.
However, in the case of different temperatures for each component we refer to \cite[Def.~28.1]{Ruggeri2021} for a definition of the mixture temperature. There are also cases when a single temperature for all components is considered. In this
case the temperature is obtained as usual as the derivative of the mixture energy with respect to the mixture entropy.
The quantity $\mu_k$ denotes the chemical potential of component $k$ and $N_k$ is the mole number of the respective component. The chemical potential $\mu_k$ quantifies the change of the $k$th component due to phase change.
For later use we want to specify the chemical potential for the mixture of ideal gases.

\begin{example}[Chemical potential for a mixture of two isothermal ideal gases]
Assume, components $k$ and $j$ form a mixture of two isothermal ideal gases, then from \eref{eq:GD} we obtain
\beq
\label{eq:mu}
\mu_k ( \alpha_k, \rho_k , \alpha_j, \rho_j ) = g_k ( \rho_k ) + 
(\gamma_k-1)c_{v,k} \Tref \ln\left( \frac{\alpha_k p_k(\rho_k)}{\alpha_k p_k(\rho_k)+\alpha_j p_j(\rho_j)} \right).
\eeq
In particular, if component $j$ vanishes, then we have the pure substance $k$ and $\mu_k ( \rho_k ) = g_k ( \rho_k )$.
\end{example}
This example becomes relevant in Sect.~\ref{sec:chem-relax} where we discuss the relaxation of chemical potentials in the context of a 3-component mixture consisting of two isothermal ideal gases (water vapor ($k=1$) and some inert gas ($k=3$)) and liquid water ($k=2$) modeled by an isothermal stiffened gas.

\subsection{Relaxation Model}
\label{subsec:bn-neq-relax}

The non-equilibrium model without relaxation terms allows for different values for velocities,  pressures, temperatures as well as chemical potentials at the same point. The relaxation mechanism described by the source terms drives all these quantities into equilibrium. Typically it is distinguished between  mechanical and thermal
relaxation processes that relax either pressures and velocities or temperatures and chemical potentials to equilibrium.
In the following we give the relaxation terms for the full Baer-Nunziato model containing the energy equations. The energy relaxation will be needed in case of isothermal fluids, see Section \ref{subsec:bn-neq-thermo-isothermal}, whereas it does not enter the relaxation process of the isentropic model.

\textit{Mechanical relaxation.}
The \emph{pressure relaxation} implies volume variations that induce energy variations due to the interfacial pressure work. Following previous work in \cite{Saurel-Abgrall:1999,Zein:2010,Mueller-Hantke-Richter:16}
we model the pressure relaxation in case of barotropic fluids by
\begin{equation}
\label{pres-relax-comp}
\source{\alpha}{k}^p = \relaxp\,\alpha_k\,(p_k- p),\
\source{\alpha\rho}{k}^p = 0,\
\Source{\alpha\rho\vects{\velo{}}}{k}^p = \bzero,\
\source{\alpha\rho E}{k}^p = \relaxp\,\alpha_k\,p(p-p_k)\,.
\end{equation}
Here $\relaxp$ is the pressure relaxation parameter and
$p$ the  mixture pressure defined in Eqn.~\eqref{eq:mixture-pres}
Similarly we apply for the \emph{velocity relaxation}
\begin{equation}
\label{velo-relax-comp}
\source{\alpha}{k}^v = 0,\
\source{\alpha\rho}{k}^v = 0,\
\Source{\alpha\rho\vects{\velo{}}}{k}^v = \relaxv\, \alpha_k\,\rho_k\,(\velo{}-\velo{k})\,,
\source{\alpha\rho E}{k}^v = \relaxv\, \alpha_k\,\rho_k\,\vprod{\velo{}}(\velo{}-\velo{k})
\end{equation}
with the velocity relaxation parameter $\relaxv$ and the mixture velocity $\velo{}$ determined by the conserved
mixture quantities \eref{eq:mixture-cons}. Performing velocity and pressure relaxation the fluid mixture is in
mechanical equilibrium, i.e., $p_k=p^\infty$ and $\bv_k = \velo{}^\infty$, $k=1,\ldots,K$.
Obviously, the mechanical relaxation terms satisfy the conservation constraints
\eref{eq:constraints} due to the definition of the mixture quantities \eref{eq:mixture-cons} and \eref{eq:mixture-pres}.

We like to mention that in \cite{herard:2016} for
barotropic fluids
a general class of models for mechanical relaxation is considered:
\begin{subequations}
\label{mech-relax-comp-herard}
\vspace*{-5mm}
\begin{align}
\vspace*{-5mm}
\label{pres-relax-comp-herard}
&\source{\alpha}{k}^p = \relaxp\,\sum_{l=1}^K d_{k,l} (p_k-p_l),\
\source{\alpha\rho}{k}^p = 0,\
\Source{\alpha\rho\vects{\velo{}}}{k}^p = \bzero\, \\
\label{velo-relax-comp-herard}
&\source{\alpha}{k}^v =
\source{\alpha\rho}{k}^v = 0,\
\Source{\alpha\rho\vects{\velo{}}}{k}^v = \relaxv\, \sum_{l=1}^K e_{k,l} (\velo{l} - \velo{k})\,,
\end{align}
\end{subequations}
with state-dependent coefficients $d_{k,l}>0$ and $e_{k,l} \ge 0$.
In particular, choosing $d_{k,l}=\alpha_k \alpha_l$ and
$e_{k,l} = \alpha_k \rho_k \alpha_l \rho_l/\sum_{j=1}^K \alpha_j \rho_j$
we note that the mechanical relaxation model
\eref{pres-relax-comp} and \eref{velo-relax-comp} falls into the class of models \eref{mech-relax-comp-herard}.

\textit{Thermal relaxation.}
According to \eref{eq:isothermalfluid} for isothermal fluids the temperatures are all in equilibrium. For isentropic fluids the temperatures cannot be in equilibrium because all entropies are constant resulting in an overdetermined system. Thus, there exists no thermal relaxation process.

\textit{Chemical potential relaxation.}
In the following we consider only isothermal fluids because there is already no thermal relaxation for isentropic fluids.
Mass transfer between different phases of the same substance occurs, whenever these phases are not in chemical equilibrium, i.e., when their chemical potentials are not the same. This physical fact is the key idea to model the mass transfer by relaxation of the chemical potentials.
It is obvious, that from now on it is necessary to identify the components.
In particular, we are interested in the
two-phase model, i.e., $K=2$,  with water vapor ($k=1$), liquid water ($k=2$), and in the
three-component model, i.e., $K=3$,  with water vapor ($k=1$), liquid water ($k=2$) and inert gas ($k=3$).

\textit{Two-component mixtures.}
For a two-component mixture ($K=2$) the relaxation of chemical potentials is
modeled according to \cite{Zein:2010}
by
\begin{subequations}
\label{mass-relax-two-comp}
\begin{align}
&
\sourcemy{\alpha}{1}^\mu = \relaxG\, \frac{\dot m}{\varrho},\
\sourcemy{\alpha\rho}{1}^\mu = \relaxG\, \dot m,\
\Source{\alpha\rho\vects{\velo{}}}{1}^\mu = \relaxG\, \dot m\, \hbv,\
\sourcemy{\alpha\rho E}{1}^\mu = \relaxG\, \dot m\,\left(\epsilon+\frac{\hbv^2}{2}\right),    \\
&
\sourcemy{\alpha}{2}^\mu = -\sourcemy{\alpha}{1}^\mu,\
\sourcemy{\alpha\rho}{2}^\mu = -\sourcemy{\alpha\rho}{1}^\mu,\
\Source{\alpha\rho\vects{\velo{}}}{2}^\mu = -\Source{\alpha\rho\vects{\velo{}}}{1}^\mu,
\sourcemy{\alpha\rho E}{2}^\mu = -\sourcemy{\alpha\rho E}{1}^\mu
\end{align}
\end{subequations}
with the relaxation parameter $\relaxG$.
For the velocity $\hbv$ we choose a convex combination
\beq
  \label{chem-relax-velo}
  \hbv = \sum_{k=1}^{K} \beta_k^v \velo{k},\quad
  \beta_k^v\in [0,1],\quad \sum_{k=1}^{K} \beta_k^v = 1.
\eeq
Formulas for the parameters  $\varrho$ and $\epsilon$ can be found in \cite{Zein:2010} where they
are determined such that pressure and thermal equilibrium is maintained during the relaxation.
Obviously, the conservation constraints \eref{eq:constraints} are satisfied.
Note that due to these constraints we are not allowed to introduce $\epsilon_k$ and $\varrho_k$
differently for each component $k=1,2$.
In \cite{Han-Hantke-Mueller:2017} another choice for these parameters has been introduced
that ensures non-negativity of the entropy production due to chemical relaxation.

Since for a two-component mixture the Gibbs free energy coincides with the chemical potential, chemical equilibrium is
achieved, if the  Gibbs free energies of the two components  coincide, i.e., $g_1=g_2$.

\textit{Three-component mixtures.}
Here we  consider three components, i.e., $K=3$, with
water vapor ($k=1$), liquid water ($k=2$) and inert gas ($k=3$),
where the gas phase is assumed to be a mixture of water vapor and some inert gas.
The relaxation terms of chemical potentials for isothermal fluids are given by
\begin{subequations}
\label{mass-relax-three-comp}
\begin{align}
\!\!\!&
\source{\alpha}{1}^\mu = \relaxG\, \frac{\dot m}{\varrho_1},\
\source{\alpha\rho}{1}^\mu = \relaxG\, \dot m,\
\Source{\alpha\rho\vects{\velo{}}}{1}^\mu = \relaxG\, \dot m\, \hbv,\
\source{\rho E}{1}^\mu = \relaxG\, \dot m\,\left(\epsilon_1+\frac{\hbv^2}{2}\right),   \\
\!\!\!&
\source{\alpha}{2}^\mu = \relaxG\, \frac{\dot m}{\varrho_2},\
\source{\alpha\rho}{2}^\mu = -\relaxG\, \dot m,\
\Source{\alpha\rho\vects{\velo{}}}{2}^\mu = -\relaxG\, \dot m\, \hbv,\
\source{\alpha\rho E}{2}^\mu = -\relaxG\, \dot m\,\left(\epsilon_2+\frac{\hbv^2}{2}\right),  \\
\!\!\!&
\source{\alpha}{3}^\mu = -\relaxG\, \dot m\, \left(\frac{1}{\varrho_1} + \frac{1}{\varrho_2}\right),\
\source{\alpha\rho}{3}^\mu = 0,\
\Source{\alpha\rho\vects{\velo{}}}{3}^\mu = \bzero\, ,
\source{\alpha\rho E}{3}^\mu = \relaxG\, \dot m\,\left(\epsilon_2-\epsilon_1\right)
\end{align}
\end{subequations}
with the relaxation parameter $\relaxG$. Formulas for the parameters $\varrho_1,\varrho_2$
can be found in \cite{Zein:2010,Zein-Hantke-Warnecke:2013}.
For details on the physics see the book of M\"uller and M\"uller \cite{MuellerMueller}.

Chemical equilibrium is
achieved, if the chemical potential of the water vapor phase equals the Gibbs free energy of the liquid
water phase. In the limit case of no inert gas the expression of the
chemical potential of the water vapor phase reduces to the Gibbs free energy.
The mass flux $\dot m$ between the liquid and the vapor phase is driven by the difference of their chemical
potentials, i.e., $\dot m=\dot m(\mu_1-\mu_2)$.
In particular, the mass flux vanishes if 
the difference of the chemical potentials is zero, i.e., $\mu_1=\mu_2$ or if one of the phases is completely extinct.

Note that the relaxation terms \eref{mass-relax-two-comp} resp.\ \eref{mass-relax-three-comp} differ from the terms given in
\cite{Han-Hantke-Mueller:2017} due to the different form of the chemical potentials in the isothermal case.

The relaxation terms \eref{pres-relax-comp}, \eref{velo-relax-comp}, \eref{mass-relax-two-comp},
\eref{mass-relax-three-comp} are of major importance when dealing with interface problems,
see for instance Saurel and Abgrall
\cite{Saurel-Abgrall:1999} or Lallemand and Saurel \cite{Lallemand-Saurel:2000}
for mechanical relaxation terms. Typically, it is assumed that
pressure and velocity relax instantaneously, see \cite{Saurel-Abgrall:1999}, whereas the thermal
relaxation and the relaxation of chemical potentials are much slower, see Zein \cite{Zein:2010}.
Here we are interested only in the equilibrium state that is characterized by vanishing relaxation terms
rather than the transient relaxation process itself.
Since the equilibrium state does not depend on the order of relaxation, the  relaxation parameters
$\theta_{\xi}$, $\xi\in\{p, v,  \mu \}$,
drop out and have not to be known explicitly. The same is true for the parameters $\varrho_k$, $k=1,2$.

Furthermore we point out, that our modeling of the mass transfer is physically correct
because we take into account chemical potentials instead of Gibbs free energies.
In \cite{Zein:2010,Zein-Hantke-Warnecke:2013}
the Gibbs free energy is relaxed, which neglects the effect of mixture
entropy in cases of impure substances.

We would
like to mention that in \cite{herard:2016} for
barotropic fluids
a general class of models for chemical relaxation (mass transfer) is considered:
\begin{align}
\label{mass-relax-comp-herard}
&\source{\alpha}{k}^\mu = 0,\
\source{\alpha\rho}{k}^\mu = \relaxG\,\sum_{l=1,\ne k}^K \Gamma_{k,l} ,\
\Source{\alpha\rho\vects{\velo{}}}{k}^\mu = \relaxG\,\sum_{l=1,\ne k}^K \bV_{k,l} \Gamma_{k,l}
\end{align}
with interfacial mass transfer $\Gamma_{k,l}$ and interfacial velocities $\bV_{k,l}=\beta_{k,l}\bv_k+(1-\beta_{k,l}) \bv_l$   with state-dependent coefficients $\beta_{k,l}\in [0,1]$. To ensure the conservation constraint \eref{eq:constraints} it is assumed that $\Gamma_{k,l}=-\Gamma_{l,k}$ and
$\bV_{k,l}=\bV_{l,k}$.
In particular, it is suggested to use $\beta_{k,l}=1/2$ and $ \Gamma_{k,l} = f_{k,l}\left(  (p_l/\rho_l+e_l(\rho_l)) - (p_k/\rho_k+e_k(\rho_k)) \right)$ for an entropy-consistent closure.
Since in the relaxation models \eref{mass-relax-two-comp} and \eref{mass-relax-three-comp}
the change of volume  is accounted for in the evolution equations of the volume fractions, i.e., $\source{\alpha}{k}^\mu \ne 0$, these models do not fit into the
class \eref{mass-relax-comp-herard}.
From a physical point of view, the volume will change when the mass of a component changes due to evaporation or condensation. This has to be accounted for in the relaxation terms $\source{\alpha}{k}^\mu$. 
Furthermore, in \cite{herard:2016} the existence of an equilibrium to \eref{mass-relax-comp-herard} is not investigated.

Finally, we would like to mention that a physically meaningful relaxation model
has to be Galilean invariant. For the models \eqref{pres-relax-comp}, \eqref{velo-relax-comp}, \eqref{mass-relax-two-comp} and \eqref{mass-relax-three-comp} this was investigated in \cite{Mueller-Hantke-Richter:16}.
In case of
barotropic fluids
H{\'e}rard's relaxation model determined by \eqref{mech-relax-comp-herard} and  \eqref{mass-relax-comp-herard}  can also be verified to be Galilean invariant.

\begin{remark}
\label{rem:relaxation-energy-barotropic-fluid}
For the closure of the barotropic model \eqref{eq:sa-model-alphak}, \eqref{eq:convention} and\eqref{eq:sa-model} with barotropic pressure law \eqref{eq:eos-k} there is no need to give the relaxation terms 
$\sourcemy{\alpha\rho E}{k}^\xi$, $\xi\in\{p,v,\mu\}$. These relaxation terms are only needed in the entropy investigation in the isothermal case in Sect.~\ref{subsec:bn-neq-thermo-isothermal}.
Thus, H{\'e}rard's relaxation models \eqref{mech-relax-comp-herard} and \eqref{mass-relax-comp-herard} cannot be considered in this context.
\end{remark}

\section{Mathematical Properties}
\label{sec:bn-math-prop}

In this section we investigate some mathematical properties of the barotropic Baer-Nunziato model. First of all, we derive in Sec.~\ref{subsec:bn-neq-hyp} the eigenvalues and eigenvectors for the transport part of the model and verify that it is hyperbolic if the non-resonance condition holds.
Then we consider the corresponding characteristic fields, see Sec.~\ref{subsec:bn-neq-char-field}, and derive for each field Riemann invariants, see Sec.~\ref{subsec:bn-neq-Riemann-Inv}. Finally, we derive a symmetrizer in Sec.~\ref{subsec:bn-neq-Symmetrization}.

\subsection{Hyperbolicity: eigenvectors and eigenvalues}
\label{subsec:bn-neq-hyp}

Neglecting the relaxation processes in
the evolution equations  \eref{eq:sa-model-alphak}, \eref{eq:sa-model} the model reduces
to a homogeneous first order system describing transport effects only. Therefore this system
should be hyperbolic to ensure that all wave speeds are finite and the system may be locally decoupled. From a
mathematical point of view, this property is helpful in the construction of Riemann solvers.
Therefore we determine here the eigenvalues and eigenvectors. Since the corresponding characteristic fields are essential in the construction of a Riemann solver we investigate these fields in more detail.
This will also provide a physically reasonable closure for the interfacial velocity $\veloI$.
%
Note that the following analysis holds 
for a general barotropic equation of state \eqref{eq:eos-k}  satisfying the constraints
\eqref{eq:eos-barotropic-ass} and \eqref{eq:eos-c}. In particular, it is applicable to both isentropic and isothermal fluids.

To investigate hyperbolicity we may consider the evolution equations for the primitive variables
$\alpha_k$,  $\bv_k$, $p_k$ instead of the variables $\alpha_k$, $\alpha_k\rho_k$, $\alpha_k\rho_k\bv_k$. This is justified because the change of variables is a bijective mapping under which the eigenvalues are invariant and the eigenvectors are related by the Jacobian of the transformation.
Using the equation of state \eref{eq:eos-k} we derive from \eref{eq:sa-model-alphak}, \eref{eq:sa-model} the system in primitive variables.
\begin{subequations}
\label{eq:sa-model-transport}
\begin{align}
\label{eq:sa-model-transport-alphak}
&\pdifft{\alpha_k} + \sum_{i=1}^{d}
V_{I,i}\,\pdiff{\alpha_k}{x_i}  = \source{\alpha}{k}\\
\label{eq:sa-model-transport-vk}
&\pdifft{\bv_k} + \sum_{i=1}^{d}
\left( v_{k,i}\,\pdiff{\bv_k}{x_i} +
\sum_{l=1,l\ne k}^{K}
\frac{1}{\alpha_k\,\rho_k} (P_{k,l}-p_k)\,\be_{d,i}\,\pdiff{\alpha_l}{x_i} +
       \frac{1}{\rho_k}\,\be_{d,i}\,\pdiff{p_k}{x_i} \right) = \Source{\vects{\velo{}}}{k}, \\
\label{eq:sa-model-transport-pk}
&\pdifft{p_k} + \sum_{i=1}^{d}
\left( v_{k,i}\,\pdiff{p_k}{x_i}
- \sum_{l=1,l\ne k}^{K}\frac{\rho_k}{\alpha_k}\, \cbar^2_k\, (v_{k,i}-V_{I,i})\,\pdiff{\alpha_l}{x_i} +
       \rho_k\,\cbar_k^2\,\pdiff{v_{k,i}}{x_i}
\right) = \source{p}{k},
\end{align}
\end{subequations}
where $\be_{d,i}\in\R^d$ denotes the unit vector in the $i$th coordinate direction.
The relaxation terms are determined by
\begin{align}
\source{p}{k} := \frac{c_k^2}{\alpha_k} (\source{\alpha\rho}{k} - \rho_k \source{\alpha}{k}) ,\quad
 \Source{\vects{\velo{}}}{k} :=
   \frac{1}{\alpha_k \rho_k} \left( \Source{\alpha\rho\vects{\velo{}}}{k} - \bv_k \source{\alpha\rho}{k}\right) .
\end{align}
Neglecting the relaxation terms, we now consider the projection of the homogeneous transport system
onto normal direction $\xi:=\vprod{\bx}{\bn}$ for arbitrary unit direction $\bn\in\R^d$.
\begin{remark}
\label{rem:speed-sound-barotropic}
For a general barotropic fluid the speed of sound $\cbar_k$ is determined by
\eqref{eq:eos-c}.
We  emphasize that  it may differ in general from the thermodynamical speed of sound  \eqref{eq:sound-fdt-s-tau}.
In case of the isothermal stiffened gas it is 
identical to
 the isothermal speed of sound \eqref{eq:sound-fdt-T-tau},
see \eqref{eq:eos-iso-p}, \eqref{eq:sound-fdt-T-rho-isotherm} and \eqref{eq:diff-sound-speed-stiff}.
\end{remark}

Introducing the vector of primitive variables
\beq
  \label{eq:sa-model-primitive-coupled-variables}
  \bw = (\alpha_1,\ldots, \alpha_{K-1},\bw_1^t,\ldots, \bw_K^t)^t,\
  \bw_k=(\bv_k^t,p_k)^t
\eeq
the projected system can be written in quasi-conservative form as
\beq
  \label{eq:sa-model-primitive-coupled-normal}
  \pdifft{\bw} + \bB_n(\bw) \, \pdiff{\bw}{\xi} = \bzero .
\eeq
The matrix $\bB_n$ is determined by the block matrix
\beq
  \label{eq:sa-model-primitive-Bn}
  \bB_n := \sum_{i=1}^{d} \bB_i\,n_i =
  \left(
  \begin{matrix}
  V_{I,n}\,\bI_{K-1} &           &        &           \\
  \bA_{1,n}          & \bB_{1,n} &        &           \\
  \vdots             &           & \ddots &           \\
  \bA_{K,n}          &           &        & \bB_{K,n} \\
  \end{matrix}
  \right)
\eeq
with the blocks defined as
\begin{align}
  \label{eq:akn}
  \bA_{k,n} :=
  \left(
  \begin{matrix}
  \bn \bbeta_k^t\\[2mm]
   \frac{\rho_k}{\alpha_k}\, (v_{k,n}-V_{I,n})\,\bgamma_k^t\\
  \end{matrix}
  \right) ,\quad
  \bB_{k,n} :=
  \left(
  \begin{matrix}
  v_{k,n}\,\bI_{d}         & \frac{1}{\rho_k}\,\bn \\[2mm]
  \rho_k\,\cbar_k^2\bn^t       & v_{k,n}
  \end{matrix}
  \right) .
\end{align}
{Here
$\bI_{d}$ and  $\bI_{K-1}$ are the unit matrices in $\R^{d\times d}$ and $\R^{(K-1)\times (K-1)}$,
respectively, and $\bzero_{d}$ and  $\bone_{K-1}$ are vectors in $\R^d$ and  $\R^{K-1}$
with value 0 or 1, respectively. The vectors $\bbeta_k$ and $\bgamma_k$
are defined by their components $l=1,\ldots,K-1$ as
\begin{align}
\label{eq:beta_kl}
  &\beta_{k,l} :=
   \frac{1}{\alpha_k \rho_k}
   \left(
    (P_{k,l}-p_k) (1-\delta_{k,l}) - (P_{k,K}-p_k)(1-\delta_{k,K})
  \right),\\
\label{eq:gamma_kl}
  &\gamma_{k,l}:=
    \cbar_k^2 (\delta_{k,l}-\delta_{k,K})
\end{align}
with $\delta_{k,l}$ the Kronecker symbol. The normal components of the velocities and the interfacial velocity
are defined as
\beq
  v_{k,n} := \bv_k\cdot \bn,\quad V_{I,n} := \bV_I\cdot \bn .
\eeq
Obviously, the eigenvalues of the matrix \eref{eq:sa-model-primitive-Bn} can now be explicitly determined.
\begin{theorem}(Eigenvalues)
\label{theroem:eigenvalues}
Let be $\bn\in\R^d$ with $|\bn|$=1. Then the eigenvalues of the matrix $\bB_n$ read
\begin{subequations}
\label{eq:eigenvalues}
\begin{align}
  \label{eq:model-eigenvalue-I}
  \lambda_{I,k} &= V_{I,n},\ k=1,\ldots, K-1,\\
  \label{eq:model-eigenvalue-i}
  \lambda_{k,i} &= v_{k,n},\ k=1,\ldots, K,\ i=1,\ldots, d-1,\\
  \label{eq:model-eigenvalue-pm}
  \lambda_{k,\pm} &= v_{k,n} \pm \cbar_k,\ k=1,\ldots, K .
\end{align}
\end{subequations}
\end{theorem}
\proof
Employing the block structure the characteristic polynomial reads
\beq
  \det(\bB_n - \lambda\,\bI) =
  \det(V_{I,n}\,\bI_{K-1} - \lambda\,\bI_{K-1})\,\prod_{k=1}^{K} \det(\bB_{k,n}-\lambda\,\bI_{d+1}) = 0 .
\eeq
The eigenvalues of the matrices $\bB_{k,n}$ are determined by
\beq
  \det(\bB_{k,n}-\lambda\,\bI_{d+1}) = (v_{k,n}-\lambda)^{d-1}\,( (v_{k,n}-\lambda)^2-\cbar_k^2) .
\eeq

\proofendmy

Next we need to verify that the left and right eigenvectors corresponding to these eigenvalues are linearly
independent. The derivation of the eigenvectors is in complete analogy to the Baer-Nunziato model for general fluids,
see \cite{Mueller-Hantke-Richter:16}. For this purpose, we introduce the notation
\begin{align}
\label{eq:drift}
  \delta_k^n:=v_{k,n}-V_{I,n},\ \sigma_k:=(\delta_k^n)^2 - \cbar_k^2,\
  \kappa_0 := \prod_{l=1}^{K} \alpha_l \sigma_l,\
  \kappa_k := \prod_{l=1,l\ne k}^{K} \alpha_l \sigma_l
\end{align}
for $k=1,\ldots, K$.
Note that $\delta_k^n$ is referred to as drift velocity in the literature. Then we can prove the following

\begin{theorem}(Hyperbolicity)
\label{theroem:hyperbolicity}
Let be $\bn\in\R^d$ with $|\bn|=1$.
Let the interfacial velocity $\veloI$ be chosen such that
the normal interfacial velocity $V_{I,n}$ does not coincide with phasic normal velocities $v_{k,n}$
and the phasic acoustic wave speeds $v_{k,n}\pm \cbar_k$, i.e.,
the non-resonance condition
\begin{align}
\label{eq:non-resonance}
   V_{I,n} \ne v_{k,n} \pm \cbar_k,\qquad k=1,\ldots,K,
\end{align}
is satisfied. Then the homogeneous first order system \eref{eq:sa-model-transport} is hyperbolic,
i.e., (i) the eigenvalues of the matrix $\bB_n$ are all real but not necessarily distinct and
(ii) there exist bases of  corresponding  left and right eigenvectors that are orthonormal.
\end{theorem}

\proof
Motivated by the block structure of the matrix $\bB_{n}$ we make the following ansatz for
computing the corresponding left and right eigenvectors
\beq
  \bR_n:=
  \left(
  \begin{matrix}
     \bR^0_{I,n} &           &       & \\[1mm]
     \bR^1_{I,n} & \bR_{1,n} &       & \\
     \vdots      &           & \ddots & \\[1mm]
     \bR^K_{I,n} &           &       & \bR_{K,n}
  \end{matrix}
  \right),\quad
  \bL_n:=
  \left(
  \begin{matrix}
     \bL^0_{I,n} &           &       & \\[1mm]
     \bL^1_{I,n} & \bL_{1,n} &       & \\
     \vdots      &           & \ddots & \\[1mm]
     \bL^K_{I,n} &           &       & \bL_{K,n}
  \end{matrix}
  \right)
\eeq
with  blocks
\begin{eqnarray*}
 && \bR^0_{I,n}:= \kappa_0 \bI_{K-1},\ \kappa_0\in\R,\ \bL^0_{I,n}\in\R^{(K-1)\times(K-1)},\\
&&  \bR^k_{I,n}, \bL^k_{I,n}\in\R^{(d+1)\times(K-1)},\quad
  \bR_{k,n}, \bL_{k,n}\in\R^{(d+1)\times(d+1)},\ k=1,\ldots,K.
\end{eqnarray*}
Here the matrices $\bR_n$ and $\bL_n$ are composed of the right and left eigenvectors in their columns and rows,
respectively. To determine the blocks $\bR_{k,n}$ we note that $\lambda_{k,i}$ and
$\lambda_{k,\pm}$ are also eigenvalues of the matrix $\bB_{k,n}$ and
the eigenvalue problem for $\bB_n$ decouples into eigenvalue problems for
the matrices $\bB_{k,n}$ corresponding to a single component.
In a first step, we therefore compute the eigenvectors to these sub-problems where we first determine
an orthonormal basis $\{\bn, \bt_1,\ldots,\bt_{d-1}\}$ of $\R^d$ such that
$\vprod{\bt_i}{\bt_j} = \delta_{i,j}$ and $\vprod{\bt_i}{\bn} = 0$.
Then the right and left eigenvectors to the eigenvalues \eref{eq:model-eigenvalue-i}
and \eref{eq:model-eigenvalue-pm} are
\beqa
  \hspace*{-10mm} &&
  \br_{k,i} = (\bt_i^t,0)^t,\
  \br_{k,\pm} = (\pm \cbar_k/\rho_k \bn^t,\cbar_k^2)^t, \\[2mm]
  \hspace*{-10mm}&&
  \bl_{k,i} = (\bt_i^t,0)^t,\
  \bl_{k,\pm} = 0.5 \cbar_k^{-2} (\pm \cbar_k \rho_k \bn^t,1)^T
\eeqa
for $ i=1,\ldots,d-1$.
Thus, there exists an eigenvalue decomposition of the matrix $\bB_{k,n}$, i.e.,
\beq
  \bL_{k,n} \bB_{k,n} \bR_{k,n} = \bLambda_{k,n} ,
\eeq
where $\bL_{k,n}$ and $\bR_{k,n}$ are defined by the left and right eigenvectors
and $\bLambda_{k,n}$ is a diagonal matrix with eigenvalues on the diagonal
\beq
  \label{eq:rkn}
  \bR_{k,n}:=
  \left(
  \begin{matrix}
    -\cbar_k/\rho_k \bn & \bt_1 & \ldots & \bt_{d-1} & \cbar_k/\rho_k \bn \\
    \cbar_k^2           & 0     & \ldots & 0         & \cbar_k^2 \\
  \end{matrix}
  \right),
\eeq
\beq
  \label{eq:lkn}
  \bL_{k,n}:= \frac{1}{2 \cbar_k^2}
  \left(
  \begin{matrix}
    -\cbar_k \rho_k \bn & 2 \cbar_k^2\bt_1 & \ldots & 2 \cbar_k^2 \bt_{d-1} & \cbar_k \rho_k \bn \\
    1                 & 0              & \ldots & 0                   & 1 \\
  \end{matrix}
  \right)^t,
\eeq
\beq
   \label{eq:lambdakn}
 \bLambda_{k,n}:=
  \left(
  \begin{matrix}
    v_{k,n}-\cbar_k     & \bzero_{d-1}^t     & 0         \\
    \bzero_{d-1}    & v_{k,n} \bI_{d-1}  & \bzero_{d-1} \\
    0               & \bzero_{d-1}^t     & v_{k,n}+\cbar_k \\
  \end{matrix}
  \right) .
\eeq

To calculate the eigenvectors to the multiple eigenvalue $\lambda_{I,i}$
we employ the knowledge of the matrices $\bR_{k,n}$. According to the block structure
of the matrix $\bR_n$ the matrix of corresponding right eigenvectors needs to satisfy
\[
(\bB_n - \lambda_{I,i} \bI)\bR_n=\bzero
\Leftrightarrow
(\bB_{k,n} - \lambda_{I,i} \bI_{d+1})\bR^k_{I,n}= - \bA_{k,n} \bR^0_{I,n} = -\kappa_0 \bA_{k,n},\ k=1,\ldots,K .
\]
Assuming that the eigenvalue $\lambda_{I,i}$ does not coincide with one of the eigenvalues
$\lambda_{k,i}$ and $\lambda_{k,\pm}$, then
$\bB_{k,n} - \lambda_{I,i} \bI_{d+1}$
is regular and there exists a unique solution for $\bR^k_{I,n}$.
With \eref{eq:drift} these matrices are given by
\beqa
  \bR^0_{I,n} &:=& \kappa_0\, \bI_{K-1},\\
  \bR^k_{I,n} &:=& -\kappa_0 \left(\bB_{k,n}- V_{I,n} \bI_{d+1} \right)^{-1} \bA_{k,n} 
  =
  \kappa_0\,
  \left(
  \begin{array}{c}
     \bn ( \bgamma_k^t - \alpha_k \bbeta_k^T) \delta_k^n/(\alpha_k \sigma_k) \\[1mm]
    (\alpha_k \cbar_k^2 \bbeta_k^t - (\delta_k^n)^2 \bgamma_k^t )   \rho_k/(\alpha_k \sigma_k)
\end{array}
  \right) .
\eeqa
Since $\alpha_k\in (0,1)$ according to \eref{eq:saturation}, these matrices are regular, i.e., the columns are
linearly independent, if and only if $\sigma_k \ne 0$  for $ k=1,\ldots,K$, i.e.,
the non-resonance condition \eref{eq:non-resonance} holds.
In particular, the inverse
\begin{align*}
\left(\bB_{k,n}- V_{I,n} \bI_{d+1} \right)^{-1} =
  \left(
  \begin{matrix}
    (\delta_k^n)^{-1} \bI_d + \cbar_k^2 (\delta_k^n \sigma_k)^{-1} \bn\otimes \bn  & &
    - (\rho_k \sigma_k)^{-1} \bn \\[2mm]
    -\rho_k \cbar_k^2 \sigma_k^{-1} \bn^T            &  & \delta^n_k \sigma_k^{-1}
  \end{matrix}
  \right)
\end{align*}
exists.

Thus the corresponding left eigenvectors are determined by the rows of the inverse of $\bR_n$.
Since $\bL_n \bR_n =\bI$, the blocks turn out to be
\beqa
  \bL_{k,n} &=& \bR^{-1}_{k,n},\ k=1,\ldots, K,\\
  \bL^0_{I,n} &=& (\bR^0_{I,n})^{-1} = \kappa_0^{-1} \, \bI_{K-1},\\
   \bL^{k}_{I,n} &=& -\bL_{k,n} \bR^{k}_{I,n}(\bR^0_{I,n})^{-1} =
   -\kappa_0^{-1} \,\bL_{k,n} \bR^{k}_{I,n} ,\ k=1,\ldots, K,
\eeqa
and we obtain for  the right eigenvectors
\beq
\label{eq:model-right-ev}
\br_{I,l} := \kappa_0 \left( (\obr_{I,l}^0)^t, (\obr_{I,l}^1)^t, \ldots, (\obr_{I,l}^K)^t\right)^t,\
 l=1,\ldots,K-1.
\eeq
where $\obr_{I,l}^0=\be_{l,K-1}$ and $\obr_{I,l}^k=(\oby_{k,l}^t, \oz_{k,l})^t$
is determined by the components
\begin{align}
  \label{eq:model-right-ev-y}
  \oby_{k,l}&= \bn \left(\gamma_{k,l} -  \alpha_k \beta_{k,l} \right) \delta^n_k/(\alpha_k \sigma_k), \\
  \label{eq:model-right-ev-z}
 \oz_{k,l} &=
\left(  \alpha_k \cbar_k^2 \beta_{k,l}- (\delta_k^n)^2 \gamma_{k,l}\right) \rho_k/ (\alpha_k \sigma_k)
\end{align}
Similar to \eref{eq:model-right-ev} the columns of $\bL_n$ are then given by
\beq
\label{eq:model-left-ev}
\bl_{I,l} :=
\left(
(\bl_{I,l}^{0})^t,
(\bl_{I,l}^1)^t, \ldots, (\bl_{I,l}^K)^t\right)^t,\ l=1,\ldots,K-1
\eeq
with
\beq
  \bl^0_{I,l} = \kappa_0^{-1}\,
\be_{l,K-1}, \quad
  \bl^k_{I,l} =
  \frac{\rho_k}{2 \cbar_k^2 \alpha_k \sigma_k}
 ( a_{k,l}^{-},\bzero_{d-1}^t,a_{k,l}^{+})^t \nonumber
\eeq
and
\begin{align}
\label{eq:coeff-aklpm}
  a_{k,l}^{\pm} :=
\left( \alpha_k \cbar_k \beta_{k,l} \mp \delta_k^n \gamma_{k,l}\right)
 \left( \cbar_k \pm \delta_k^n \right) .
\end{align}

After having determined the eigenvalues and the corresponding linearly independent right and left eigenvectors
we finally end up with the eigenvalue decomposition of the matrix $\bB_n$
\beq
  \label{eq:model-transport-decomposition}
  \bL_{n} \bB_{n} \bR_{n} = \bLambda_{n}
\eeq
with the block-diagonal matrix $\bLambda_{n} = \diag(\bLambda_{0,n},\bLambda_{1,n},\ldots,\bLambda_{K,n})$
and $\bLambda_{0,n}:=V_{I,n} \bI_{K-1}$. To verify this decomposition we make use of the identity
$\bB_{k,n} \bR^k_{I,n} = \bR^k_{I,n}\bLambda_{0,n} - \bA_{k,n} \bR^0_{I,n}$.

\proofendmy

For convenience we summarize the representation of the eigenvectors because we want to investigate the type of the characteristic fields.

\begin{proposition}(Eigenvectors)
\label{prop:eigenvectors}
Let be $\bn\in\R^d$ with $|\bn|=1$ and let be $\bt_l\in\R^d$, $l=1,\ldots,d-1$, such that  $\{\bn, \bt_1,\ldots,\bt_{d-1}\}$ is an orthonormal basis of $\R^d$, i.e.,
$\vprod{\bt_i}{\bt_j} = \delta_{i,j}$ and $\vprod{\bt_i}{\bn} = 0$.
\begin{enumerate}

\item (\textit{Transport waves}) The  right  and left eigenvalues to the eigenvalues $\lambda_{k,i}$, $k=1,\ldots, K$, $i=1,\ldots, d-1$, are
\begin{align}
\label{eq:RI-final-transport}
  \br_{k,i} =\bl_{k,i}=
 ( \bzero_{K-1}^t, \underbrace{\bzero_{d+1}^t,\ldots, \bzero_{d+1}^t}_{(k-1)-times},(\bt_i^t,0),\underbrace{\bzero_{d+1}^T,\ldots, \bzero_{d+1}^T}_{(K-k)-times} )^t .
\end{align}

\item (\textit{Acoustic waves}) The  right  and left eigenvalues to the eigenvalues
$\lambda_{k,\pm}$, $k=1,\ldots, K$, are
\begin{align}
\label{eq:RI-final-acoustic}
  \br_{k,\pm} &=
 ( \bzero_{K-1}^t, \underbrace{\bzero_{d+1}^t,\ldots, \bzero_{d+1}^t}_{(k-1)-times},\obr_{k,\pm}^t,\underbrace{\bzero_{d+1}^t,\ldots, \bzero_{d+1}^t}_{(K-k)-times} )^t \\
  \bl_{k,\pm} &=
 ( \obl_{I,k,\pm}^t, \underbrace{\bzero_{d+1}^t,\ldots, \bzero_{d+1}^t}_{(k-1)-times},\obl_{k,\pm}^t,\underbrace{\bzero_{d+1}^t,\ldots, \bzero_{d+1}^t}_{(K-k)-times} )^t
\end{align}
 with
\begin{align}
 &\obr_{k,\pm} = (\pm \cbar_k\rho_k^{-1} \bn^t, \cbar_k^2)^t ,\\
 &\obl_{k,\pm} = 0.5 \cbar_k^{-2} (\pm \cbar_k\rho_k\bn^t,1)^t,\
  \obl_{I,k,\pm} =0.5 \rho_k \cbar_k^{-2} (\alpha_k \sigma_k)^{-1} \ba_k^{\pm}
 \end{align}
and the components of the vector $\ba_k^{\pm}\in\R^{K-1}$ determined by \eref{eq:coeff-aklpm}.

\item (\textit{Material waves}) If the non-resonance condition \eref{eq:non-resonance} holds, then the right and left
eigenvalues to the eigenvalues $\lambda_{I,k}$, $k=1,\ldots, K-1$, are
\begin{align}
\label{eq:RI-final-material}
  \br_{I,k}^T &=
 \kappa_0 \left( \be_{K-1,k}^t, (\obr_{I,k}^1)^t, \ldots, (\obr_{I,k}^K)^t\right)^{t},\\
  \bl_{I,k} &=
  \kappa_0^{-1} (\be_{K-1,k}^t, \underbrace{\bzero_{d+1}^t,\ldots, \bzero_{d+1}^t}_{K-times} )^t ,
\end{align}
where $\obr_{I,l}^k=(\oby_{k,l}^t, \oz_{k,l})^t$
is determined by the components \eref{eq:model-right-ev-y} and \eref{eq:model-right-ev-z}.
Here $\be_{K-1,k}$ is the $k$th unit vector in $\R^{K-1}$.
\end{enumerate}

In particular, the left and right eigenvectors are orthonormal.

\end{proposition}

\subsection{Characteristic fields}
\label{subsec:bn-neq-char-field}

By means of the eigenvalues and the right eigenvectors we can now characterize the characteristic fields.

\begin{proposition}(Characteristic fields)
\label{prop:charfield}
\begin{enumerate}
  \item The fields corresponding to the eigenvalue $\lambda_{k,i}$ are linearly degenerated.
  \item If the assumptions \eref{eq:eos-barotropic-ass}
  hold, then  the acoustic waves are genuinely nonlinear.
   \item If the interfacial velocity is a convex combination of the velocities $\bv_k$, i.e.,
\beq
  \label{thermodynamics-properties-interfacial-2}
  \veloI = \sum_{k=1}^{K} \beta_k \velo{k},\quad
  \beta_k\in [0,1],\quad \sum_{k=1}^{K} \beta_k = 1
\eeq
with coefficients
\beq
\label{thermodynamics-properties-interfacial-2a}
  \beta_k:=\frac{\ogamma_k \alpha_k \rho_k}{{\hat\rho}},\quad {\hat\rho}:= \sum_{k=1}^{K} \ogamma_k \alpha_k \rho_k ,
\eeq
for some arbitrary but fixed convex combination $\sum_{k=1}^{K} \ogamma_k = 1$, 
$\ogamma_k\in[0,1]$,
then the material waves are linearly degenerated.
\end{enumerate}
\end{proposition}

\proof
The characteristic fields corresponding to the different waves  are characterized by the directional derivative $\nabla _\bw \lambda(\bw) \cdot \br(\bw)$.

Obviously, the derivatives of the eigenvalues $\lambda_{k,i}=v_{k,n}$, $k=1,\ldots, K$, $i=1,\ldots, d$, are
\[
  \pdiff{\lambda_{k,i}}{\alpha_l}=0,\
  \pdiff{\lambda_{k,i}}{v_{l,j}}=\delta_{k,l}\, n_j,\
  \pdiff{\lambda_{k,i}}{p_l}=0,\
\]
 and, thus, we conclude from the corresponding right eigenvectors \eref{eq:RI-final-transport}
\[
 \nabla_\bw \lambda_{k,i}(\bw) \cdot \br_{k,i}(\bw) = 0,
\]
i.e., these fields are linearly degenerated.

Next we consider the acoustic waves corresponding to the eigenvalues $\lambda_{k,\pm}=v_{k,n}\pm \cbar_k$. Obviously, it holds
\[
  \pdiff{v_{k,n}}{\alpha_l}=0,\
   \pdiff{\lambda_{k,i}}{v_{l,j}}=\delta_{k,l}\, n_j,\
 \pdiff{v_{k,n}}{p_l}=0.
\]
To determine the derivatives of the 
barotropic speed of sound
$\cbar_k$ we perform a change of variables $\rho_k=\rho_k(p_k)$.
Note that $p_k=p_k(\rho_k)$ is a monotone function according to the assumption \eref{eq:eos-barotropic-ass-1}.
Since it holds
\[
  \pdiff{\cbar_k^2(\rho_k)}{*} = 2 \cbar_k(\rho_k) \pdiff{\cbar_k(\rho_k)}{*} = p_k''(\rho_k) \pdiff{\rho_k}{*},
\]
the derivatives of $c_k$ with respect to the primitive variables $\bw$ read
\[
  \pdiff{\cbar_k}{\alpha_l}=0,\
  \pdiff{\cbar_k}{v_{l,j}}= 0,\
  \pdiff{\cbar_k}{p_l}=\delta_{k,l}\frac{1}{2\, \cbar_k^3(\rho_k)}p_k''(\rho_k)
\]
and we obtain by the right eigenvectors \eref{eq:RI-final-acoustic}
\[
 \nabla_\bw \lambda_{k,\pm}(\bw) \cdot \br_{k,\pm}(\bw) =
\pm \frac{\cbar_k}{\rho_k} \calG_k
\]
with $\calG_k$ the fundamental derivative of gas dynamics for a pure barotropic fluid corresponding to component $k$.
According to the assumptions
\eref{eq:eos-barotropic-ass}
both the barotropic speed of sound
\eref{eq:eos-c} and the fundamental derivative \eref{eq:eos-fdt}
do not vanish, i.e., the acoustic waves are genuinely nonlinear.

Finally, we consider the characteristic fields corresponding to the eigenvalues $\lambda_{I,k}=V_{I,n}$. From the corresponding right eigenvectors \eref{eq:RI-final-material} we conclude
\beq
\label{eq:linear-fields}
  \br_{I,i}\cdot \nabla_{\bw} \lambda_{I,i} =
   \kappa_0 \left( \pdiff{V_{I,n}}{\alpha_i}+\sum_{k=1}^{K} \left(
\oby_{k,i}\cdot\nabla_{\bv_k}V_{I,n}  + \oz_{k,i} \pdiff{V_{I,n}}{p_k}\right) \right) .
\eeq
For the gradient of the interfacial velocity it holds
\beq
  \pdiff{V_{I,n}}{\alpha_i} = \frac{\beta_i}{\alpha_i} \delta_i^n - \frac{\beta_K}{\alpha_K}\delta^n_K,\
  \nabla_{\bv_k}V_{I,n} = \bn,\
  \pdiff{V_{I,n}}{p_k} = \frac{\beta_k}{\rho_k} \frac{1}{\cbar^2_k} \delta_k^n.
\eeq
By a straightforward calculation incorporating the coefficients \eref{eq:model-right-ev-y}, \eref{eq:model-right-ev-z} and using the convex combination \eref{thermodynamics-properties-interfacial-2} with coefficients \eref{thermodynamics-properties-interfacial-2a} we can verify that the right-hand side in \eref{eq:linear-fields} vanishes, i.e., the material waves are linearly degenerated.

\proofendmy

\begin{remark}
If the material field is genuinely nonlinear field, then there is no way to cope with the  non-conservative products in the phasic momentum, see Eqn.~\eref{eq:sa-model-mk}. However, if fields associated to the non-conservative products only occur in linearly degenerated fields, then Riemann invariants of the associated field can be enforced and, thus, the exact solution of the Riemann problem exists as has been verified for Baer-Nunziato-type models in \cite{Goatin-LeFloch:2004}. Therefore, it is suggested in \cite{Gallouet-Herard-Seguin:04,Herard:07} to determine the interfacial velocity such that the  fields associated with the eigenvalues $\lambda_{I,i}$, $i=1,\ldots,K-1$, are linearly degenerated.
Motivated by Gallou\"et et al.~\cite{Gallouet-Herard-Seguin:04} and
H{\'e}rard \cite{Herard:07} for a two-phase and a three-phase model, respectively, the interfacial velocity is chosen as a convex combination of the  velocities of the components, see Eqn.~\eref{thermodynamics-properties-interfacial-2}.
For the coefficients of the convex combination the ansatz of Saleh \cite{Saleh:2012}, Eqn.~(4.3.40), in case of a two-phase mixture can be generalized providing \eref{thermodynamics-properties-interfacial-2a}.
\end{remark}

\subsection{Riemann invariants}
\label{subsec:bn-neq-Riemann-Inv}

In the following proposition we give for all characteristic fields a set of Riemann invariants whose gradients are linearly independent.

\begin{proposition}(Riemann invariants)
\label{prop:RiemannInvariants}
The resonance condition \eref{eq:non-resonance} is assumed to hold. Introduce the functions 
\begin{subequations}
\label{eq:RiemannInvariants}
\begin{align}
  & \Psi_{j,l}:=  \velo{j} \cdot \bt_l,\ j=1,\ldots, K,\ l=1,\ldots, d-1,\\
  & \Psi_{I,j} := \alpha_j,\ j=1,\ldots, K-1,\\
  & \Psi_{j,\pm} := \velo{j} \cdot \bn \pm \int (\cbar_j \rho_j)^{-1} d p_j ,\ j=1,\ldots, K.
\end{align}
\end{subequations}
\begin{enumerate}
\item 
For the transport waves corresponding to the right eigenvectors $\br_{k,i}$, $k=1,\ldots, K$, $i=1,\ldots, d-1$, the functions 
\begin{subequations}
\label{eq:RiemannInvariants-transport}
\begin{align}
  &\Psi_{j,l},\ j=1,\ldots, K,\ j\ne k,\ l=1,\ldots, d-1,\\
  &\Psi_{k,l},\  l=1,\ldots, d-1,\ l\ne i,\\
  &\Psi_{I,j},\ j=1,\ldots, K-1,\\
  &\Psi_{j,\pm},\ j=1,\ldots, K
\end{align}
\end{subequations}
form a system of $K(d+2)-2$ Riemann invariants whose gradients are linearly independent.
Other Riemann invariants are  
\begin{subequations}
\label{eq:RiemannInvariants-transport-other}
\begin{align}
  &\Psi = \bv_k \cdot \bn = v_{k,n} = \lambda_{k,i},\\
  &\Psi = v_{l,j},\ l=1,\ldots, K,\ l\ne k,\  j=1,\ldots,d,\\
  &\Psi = p_l,\ l=1,\ldots, K .
\end{align}
\end{subequations}
\item 
For the acoustic waves corresponding to the right eigenvectors $\br_{k,\star}$, $k=1,\ldots, K$, $\star\in\{+,-\}$, the functions 
\begin{subequations}
\label{eq:RiemannInvariants-acoustic}
\begin{align}
  &\Psi_{j,l},\ j=1,\ldots, K,\ l=1,\ldots, d-1,\\
  &\Psi_{I,j},\ j=1,\ldots, K-1,\\
  &\Psi_{j,\pm},\ j=1,\ldots, K,\ (j,\pm)\ne (k,\star)
\end{align}
\end{subequations}
form a system of $K(d+2)-2$ Riemann invariants whose gradients are linearly independent.
Other Riemann invariants are  
\begin{subequations}
\label{eq:RiemannInvariants-acoustic-other}
\begin{align}
  &\Psi = v_{l,j},\ l=1,\ldots, K,\ l\ne k,\  j=1,\ldots,d,\\
  &\Psi = p_l,\ l=1,\ldots, K,\ l\ne k .
\end{align}
\end{subequations}
\item 
For the material waves corresponding to the right eigenvectors $\br_{I,k}$, $k=1,\ldots, K-1$, the functions \\[-5mm]
\begin{subequations}
\label{eq:RiemannInvariants-material}
\begin{align}
  &\Psi_{j,l},\ j=1,\ldots, K,\ l=1,\ldots, d-1,\\
  &\Psi_{I,j},\ j=1,\ldots, K-1,\ j\ne k,\\
  &\Psi_{j,\pm},\ j=1,\ldots, K
\end{align}
\end{subequations}
form a system of $K(d+2)-2$ Riemann invariants whose gradients are linearly independent.\\
Additionally assuming that  the interfacial velocity is a convex combination of the velocities $\bv_k$ according to \eref{thermodynamics-properties-interfacial-2} with coefficients \eref{thermodynamics-properties-interfacial-2a} other Riemann invariants are
\begin{subequations}
\label{eq:RiemannInvariants-material-other}
\begin{align}
\label{eq:RiemannInvariants-material-other-a}
  &\Psi = V_{I,n} = \lambda_{I,k},\\
\label{eq:RiemannInvariants-material-other-b}
  & \Psi=\alpha_j\rho_j(\bv_{j,n}-V_{I,n} ),\ j=1,\ldots, K-1,\\
\label{eq:RiemannInvariants-material-other-c}
  & \Psi = \sum_{j=1}^{K} \left( \alpha_j p_j + \alpha_j \rho_j (\bv_{j,n}- V_{I,n} )^2 \right) .
\end{align}
Additionally assuming that the interfacial velocity and the interfacial pressures are
$\bV_I= \bv_K$ and $P_{k,l} = \beta_k p_l + p_k (1 -\beta_k)$, $k,l=1,\ldots,K$, $k\ne l$, then for
a general barotropic fluid \eqref{eq:eos-k}
another family of Riemann invariants is given by
\begin{align}
\label{eq:RiemannInvariants-material-other-d}
  \Psi = \oee_j + \frac{p_j}{\rho_j} + \frac{1}{2}  (\bv_{j,n}- V_{I,n} )^2 ,\ j=1,\ldots, K-1 ,
\end{align}
where $\oee_j$ is the ``internal energy'' according to Remark \ref{rem:energy-barotropic-fluid}.
\end{subequations}
\end{enumerate}
\end{proposition}

\proof
First of all, we determine the gradients of the functions defined in \eqref{eq:RiemannInvariants}
\begin{align*}
  \nabla_{\bw} \Psi_{j,i} =  \bl_{j,i},\
  \nabla_{\bw} \Psi_{I,j} =  \kappa_0 \bl_{I,j},\
  \nabla_{\bw} \Psi_{j,\pm} =
    \pm  \left( 2 \frac{\cbar_j}{\rho_j} \bl_{j,\pm} -  \frac{\kappa_0}{\alpha_j \sigma_j \cbar_j}  \sum_{m=1}^{K} a_{j,m}^{\pm} \bl_{I,m} \right),
\end{align*}
where $\kappa_0$, $\sigma_j$ and $a_{j,m}^{\pm}$ are defined by \eqref{eq:drift} and \eqref{eq:coeff-aklpm}.
Since these gradients are linear combinations of the left eigenvectors it follows by the construction of the left and right eigenvectors, see Proposition \ref{prop:eigenvectors},
\begin{align*}
  \br_{k,i} \cdot \nabla_{\bw} \Psi_{j,i} =0,\
  \br_{k,\star} \cdot \nabla_{\bw} \Psi_{I,j}=0,\
  \br_{I,k} \cdot \nabla_{\bw} \Psi_{j,\pm} =0
\end{align*}
according to the systems
\eqref{eq:RiemannInvariants-transport}, \eqref{eq:RiemannInvariants-acoustic}  and \eqref{eq:RiemannInvariants-material}, respectively. Therefore these functions are Riemann invariants. Linear independence of the corresponding gradients follows from the hyperbolicity,  see Proposition \ref{prop:eigenvectors},  ensuring that the left eigenvectors and, thus, the gradients, are linearly independent.\\
Next we verify that \eqref{eq:RiemannInvariants-transport-other} are also Riemann invariants
for
the material wave corresponding to the right eigenvector $\br_{k,i}$, i.e.,
the following needs to hold
\begin{align*}
  \br_{k,i}\cdot \nabla_{\bw} \Psi =
  \pdiff{\Psi}{\bv_k} \cdot \bt_i  = 0.
\end{align*}
For the functions $\Psi$ given in \eqref{eq:RiemannInvariants-transport-other} this condition can be easily verified using that the vectors $\{\bn,\bt_1,\ldots,\bt_{d-1}\}$ are orthonormal to each other.\\
To verify that \eqref{eq:RiemannInvariants-acoustic-other} are also Riemann invariants
for
the acoustic wave corresponding to the right eigenvector $\br_{k,\star}$ we need
\begin{align*}
  \br_{k,\pm}\cdot \nabla_{\bw} \Psi =
  \pm \cbar_k \rho_k^{-1}\pdiff{\Psi}{\bv_k} \cdot \bn  + \cbar_k^2 \pdiff{\Psi}{p_k} = 0.
\end{align*}
For the functions $\Psi$ given in
\eqref{eq:RiemannInvariants-acoustic-other}
this condition can be easily verified using
similar arguments
as before.\\
Finally, we verify   that \eqref{eq:RiemannInvariants-material-other} are also Riemann invariants to the material wave corresponding to the right eigenvector $\br_{I,k}$, i.e.,
\begin{align*}
  \br_{I,k}\cdot \nabla_{\bw} \Psi =
     \kappa_0 \left( \pdiff{\Psi}{\alpha_i}+\sum_{l=1}^{K} \left(
\oby_{l,i}\cdot\nabla_{\bv_l}\Psi  + \oz_{l,i} \pdiff{\Psi}{p_l}\right) \right)   = 0.
\end{align*}
For \eqref{eq:RiemannInvariants-material-other-a} this holds true by construction of the coefficients \eref{thermodynamics-properties-interfacial-2a}.\\
For \eqref{eq:RiemannInvariants-material-other-b}  
the gradient is determined by
\begin{align*}
  &\pdiff{\Psi}{\alpha_l} =
  \delta_{l,j} \rho_j\left(v_{j,n}-V_{I,n}\right) - \alpha_j \rho_j \pdiff{V_{I,n}}{\alpha_l} ,\\
  &\nabla_{\bv_l} \Psi = \alpha_j \rho_j \delta_{l,j} \bn - \alpha_j \rho_j  \nabla_{\bv_l} V_{I,n},\\
  &\pdiff{\Psi}{p_l} = \alpha_j \rho_j'(p_j) \delta_{l,j} \left(v_{j,n}-V_{I,n}\right) - \alpha_j\rho_j \pdiff{V_{I,n}}{p_l}.
\end{align*}
Then we obtain
\begin{align*}
  \br_{I,k}\cdot \nabla_{\bw} \Psi =
  \kappa_0 \left( \delta_{j,k} \rho_j \left(v_{j,n}-V_{I,n}\right) +
                  \alpha_j\rho_j \oby_{j,k}\cdot \bn +
                  \alpha_j \rho_j'(p_j) \oz_{j,k} \left(v_{j,n}-V_{I,n}\right)
           \right)
     - \alpha_j\rho_j \br_{I,k}\cdot \nabla_{\bw} V_{I,n} . 
\end{align*}
The last term on the right-hand side vanishes because $\Psi=V_{I,n}$ is a Riemann invariant.
The remaining terms vanish when plugging in \eref{eq:model-right-ev-y}, \eref{eq:model-right-ev-z}, \eref{eq:drift} and using $\rho_j'(p_j) = 1/p_j'(\rho_j) = 1/\cbar_j^2$.\\
For \eqref{eq:RiemannInvariants-material-other-c} the gradient is determined by
\begin{align*}
  &\pdiff{\Psi}{\alpha_l} =
   p_l-p_K + \rho_l \left(v_{l,n}-V_{I,n}\right)^2 -
   2 \sum_{j=1}^{K} \alpha_j \rho_j \left(v_{j,n}-V_{I,n}\right) \pdiff{V_{I,n}}{\alpha_l},\\
  &\nabla_{\bv_l} \Psi = 2 \alpha_l\rho_l \left(v_{l,n}-V_{I,n}\right) \bn -
   2 \sum_{j=1}^{K} \alpha_j \rho_j \left(v_{j,n}-V_{I,n}\right)\nabla_{\bv_l}  V_{I,n},\\
  &\pdiff{\Psi}{p_l} = \alpha_l \left( 1 + \rho_l'(p_l) \left(v_{l,n}-V_{I,n}\right)^2 \right) -
   2 \sum_{j=1}^{K} \alpha_j \rho_j \left(v_{j,n}-V_{I,n}\right) \pdiff{V_{I,n}}{p_l}.
\end{align*}
This yields
\begin{align*}
  &\br_{I,k}\cdot \nabla_{\bw} \Psi = \\
  &\kappa_0 \left( p_k-p_K + \rho_i \left(v_{k,n}-V_{I,n}\right)^2
   + \sum_{j=1}^K \alpha_j \left( 2 \rho_j \left(v_{j,n}-V_{I,n}\right) \oby_{j,k}\cdot \bn +
                   \left( 1 + \rho_j'(p_j) \left(v_{j,n}-V_{I,n}\right)^2 \right)  \oz_{j,k}\right)
           \right) \\
  &   - 2\sum_{j=1}^K \alpha_j\rho_j  \left(v_{j,n}-V_{I,n}\right)   \br_{I,k}\cdot \nabla_{\bw} V_{I,n} .
\end{align*}
The last term on the right-hand side vanishes because $\Psi=V_{I,n}$ is a Riemann invariant.
Again, using $\rho_j'(p_j) =  1/\cbar_j^2$ and employing \eref{eq:drift}, the remaining terms simplify to
\begin{align*}
  \br_{I,k}\cdot \nabla_{\bw} \Psi =
  \kappa_0 \left( p_k-p_K + \rho_k \left(\delta_k^n\right)^2  -
                  \sum_{j=1}^K \rho_l \left(  \gamma_{j,k} \left(\frac{\delta_j^n}{\cbar_j} \right)^2 + \beta_{j,k} \alpha_j \right)   \right) .
\end{align*}
Plugging in \eref{eq:model-right-ev-y} and \eref{eq:model-right-ev-z} the right-hand side vanishes using
\eqref{eq:constraints-interfacial-pressures}.\\
Finally, we investigate \eqref{eq:RiemannInvariants-material-other-d}. Introducing the phasic specific enthalpies
\begin{align*}
\oh_j(\rho_j) := \oee_j(\rho_j) + \frac{p_j(\rho_j)}{\rho_j}
\end{align*}
the gradient is determined by
\begin{align*}
  &\pdiff{\Psi}{\alpha_l} =  -\left(v_{j,n}-V_{I,n}\right) \pdiff{V_{I,n}}{\alpha_l},\\
  &\nabla_{\bv_l} \Psi =  \left(v_{j,n}-V_{I,n}\right) \delta_{j,l}\bn -
   \left(v_{j,n}-V_{I,n}\right)\nabla_{\bv_l}  V_{I,n},\\
  &\pdiff{\Psi}{p_l} = h_j'(\rho_j) \rho_j'(p_j)  \delta_{j,l} -\left(v_{j,n}-V_{I,n}\right) \pdiff{V_{I,n}}{p_l}.
\end{align*}
Then it holds
\begin{align*}
  \br_{I,k}\cdot \nabla_{\bw} \Psi =
  \kappa_0 \left(  \sum_{l=1}^K\left( \left(v_{j,n}-V_{I,n}\right) \delta_{j,l} \oby_{l,k}\cdot \bn +
                    h_j'(\rho_j) \rho_j'(p_j)\delta_{j,l}  \oz_{l,k}\right)
            -  \left(v_{j,n}-V_{I,n}\right)   \br_{I,k}\cdot \nabla_{\bw} V_{I,n} 
                                                                           \right)    .
\end{align*}
Again, the last term on the right-hand side vanishes because $\Psi_I=V_{I,n}$ is a Riemann invariant.
Using  \eref{eq:model-right-ev-y}, \eref{eq:model-right-ev-z} and \eref{eq:drift} the remaining terms on the right-hand side become
\begin{align*}
  \br_{I,k}\cdot \nabla_{\bw} \Psi =
     \frac{\kappa_0}{\alpha_j\sigma_j}
                \left(
           \left(\delta^n_j\right)^2 \gamma_{j,k} \left( 1 - \rho_j h_j'(\rho_j) \rho_j'(p_j) \right)  +
           \alpha_j \beta_{j,k} \left(\cbar_j^2 \rho_j h_j'(\rho_j) \rho_j'(p_j) - \left(\delta^n_j\right)^2 \right)
                \right) .
\end{align*}
From \eqref{eq:eos-k} and \eqref{eq:eos-c} we deduce $\rho_j'(p_j) =  1/\cbar_j^2$ implying $ \oh'_j(\rho_j) = \cbar_j^2/\rho_j$.
Then we obtain with \eref{eq:drift}
\begin{align*}
  \br_{I,k}\cdot \nabla_{\bw} \Psi =
     \frac{\kappa_0}{\sigma_j}  \beta_{j,k} \left(\cbar_j^2  - \left(\delta^n_j\right)^2 \right) =
  - \kappa_0 \beta_{j,k}   .
\end{align*}
Due to the assumptions on the interfacial velocity and the interfacial pressures we obtain by
definition \eref{eq:beta_kl} of the coefficients $\beta_{j,k}$ and
the interfacial velocities \eref{thermodynamics-properties-interfacial-2} with coefficients \eref{thermodynamics-properties-interfacial-2a}
\begin{align*}
  \beta_{j,k} = \frac{\ogamma_j}{\hat \rho} (p_j-p_K)
\end{align*}
for $j,k\in\{1,\ldots, K-1\}$. From the assumption $\veloI = \bv_K$ we conclude $\ogamma_i=0$, $i=1,\ldots,K-1$, and $\ogamma_K=1$. Thus, the right-hand side vanishes. This verifies the assertion.

\proofendmy

\begin{remark}
Finally, we would like to comment on existing results in the literature.
In \cite{saleh-seguin:2020} the eigenstructure of the 1D homogeneous 
barotropic
Baer-Nunziato model  is investigated for the particular choice of the interfacial velocity $V_I=v_1$ and the interfacial pressures
\begin{align}
   &k=1:\ P_{1,l} = p_l,\quad l=2,\ldots,K \nonumber\\
   &k>1:\ P_{1,l} = p_l,\quad l=1,\ldots,K,\ l\ne k ,\nonumber
\end{align}
specifying the eigenvalues and the corresponding right eigenvectors for non-resonance states. For this particular setting the authors investigate the corresponding characteristic fields. In particular, they determine the Riemann invariants for the linearly degenerated fields corresponding to the eigenvalues $\lambda_{I,k}$, $k=1,\ldots, K-1$.
Note that the Riemann invariants \eqref{eq:RiemannInvariants-material-other} coincide with those given in \cite{saleh-seguin:2020}.
In contrast to this,  we give the full eigenstructure and the corresponding characteristic fields for the quasi-1D model using a much larger class of interfacial velocity and interfacial pressures. \\
For a three-component model, i.e., $K=3$,  the eigenvalues  are also given in \cite{boukili-herard:2018} for the quasi-1D model but without the corresponding eigenvectors. There, in addition, the Riemann invariants are derived for the 1D model.

We would like to point out that in \cite{boukili-herard:2018,saleh-seguin:2020}
the authors make explicit use of the ``energies'' $\oee_k$ introduced in Remark \ref{rem:energy-barotropic-fluid} which coincide with $e_k$ only for isentropic fluids.

\end{remark}

\subsection{Symmetrization}
\label{subsec:bn-neq-Symmetrization}


From Kato's theorem \cite{Kato:1975} it follows that there exists a local-in-time smooth solution to the Cauchy problem of the projected system \eref{eq:sa-model-primitive-coupled-normal}, if the problem is symmetrizable, i.e., there exists a symmetric positive definite  matrix $\bP_n=\bP_n(\bw)$ such that the matrix $\bP_n \bB_n$ is symmetric.
It can be verified that the quasi-1D system \eref{eq:sa-model-primitive-coupled-normal} can be symmetrized.

\begin{theorem}(Symmetrization)
\label{theorem:symmetrization}
If the non-resonance condition \eref{eq:non-resonance} holds, then the quasi-1D system \eref{eq:sa-model-primitive-coupled-normal} is symmetrizable.
\end{theorem}

\proof
To construct a symmetrizer for $\bB_n$  we first observe that the matrix
\[
  \bP_{k,n} =
  \left(
  \begin{matrix}
  \bT_n \bD_{k,n} \bT_n^t    & \bzero   \\[2mm]
  \bzero^t               & 1/(2 \cbar_k^4)
  \end{matrix}
  \right)
\]
with $\bT_n:=(\bn,\bt_1,\ldots,\bt_{d-1})$ an orthonormal matrix and $\bD_{k,n}:=\diag(\lambda_{n,k},\lambda_1,\ldots,\lambda_{d-1})$ a diagonal matrix with entries $\lambda_{k,n}:=\rho_k^2/(2 \cbar_k^2)$ and $\lambda_i:=1$, $i=1,\ldots, d-1$
is a symmetrizer of the phasic problem
\[
  \pdifft{\bw_k} + \bB_{k,n}(\bw_k) \, \pdiff{\bw_k}{\xi} = \bzero ,
\]
i.e., $\bP_{k,n}$ is a symmetric positive definite matrix and  the matrix $\bP_{k,n} \bB_{k,n}$ is symmetric.
Then we make the following ansatz for a symmetrizer of \eref{eq:sa-model-primitive-coupled-normal}:
\[
  \bP_n=
  \left(
  \begin{matrix}
     K\,P_{\alpha,n} \bI_{K-1}   & \bP_{1,\alpha,n}^t & \cdots & \bP_{K,\alpha,n}^t \\[1mm]
     \bP_{1,\alpha,n}         & \bP_{1,n}            &        & \\
     \vdots                   &                    & \ddots & \\[1mm]
     \bP_{K,\alpha,n}         &                    &        & \bP_{K,n}
  \end{matrix}
  \right) ,
\]
where
\[
  \bP_{k,\alpha,n} = \bL_{k,n}^t \left( \bLambda_{k,n}- V_{I,n}\bI_{d+1}\right)^{-1} \bR_{k,n}^t \bP_{k}  \bA_{k,n}
\]
with $\bA_{k,n}$, $\bR_{k,n}$, $\bL_{k,n}$ and $\bLambda_{k,n}$ defined by \eref{eq:akn}, \eref{eq:rkn}, \eref{eq:lkn} and \eref{eq:lambdakn}.
Note that $\bP_{k,\alpha,n}$ is well-defined if the non-resonance condition \eref{eq:non-resonance} holds true. Obviously, $\bP_n$ is symmetric.
It turns out that $\bP_n \bB_n$ is symmetric whenever $\bP_{k,\alpha,n}^t\bA_{k,n}$ is symmetric.  The latter holds true provided that $\bP_{k,n}\bR_{k,n}=\bL_{k,n}^t$.
It remains to verify that $\bP_n$ is positive definite.
For this purpose we have to verify for any $\ba=(\ba_\alpha^t,\ba_1^t,\ldots,\ba_K^t)^t\ne\bzero$ with $\ba_\alpha\in\R^{K-1}$, $\ba_k\in\R^{d+1}$, $k=1,\ldots,K$ that $\ba^t \bP_n \ba$ is positive. A straightforward calculus yields
\[
  \ba^t \bP_n \ba =
   P_{\alpha,n}\sum_{k=1}^{K-1} \sum_{i=1}^{K-1} \left( a_{\alpha,i} + (\bP_{k,\alpha,n}^t\ba_k)_i /P_{\alpha,n}\right)^2 +
                 \sum_{k=1}^{K} P_{\alpha,n}^{-1} \ba_k^t \bQ \ba_k
\]
with $\bQ:= P_{\alpha,n} \bP_{k} -  \bP_{k,\alpha,n} \bP_{k,\alpha,n}^t$.
Since $\bP_k$ is symmetric positive definite,  the Cholesky decomposition $\bP_k=\bC_k\bC_k^t$ exists. Furthermore, the matrix
$\bE_k: = \bC_k^{-1} \bP_{k,\alpha,n}^t \bP_{k,\alpha,n}\bC_k^t$ is symmetric and, thus, there exists an orthogonal matrix $\bT_k$ such that
$\bT_k \bE_k \bT_k^t = \bD_k$ where $\bD_k$ is a diagonal matrix with the eigenvalues $\mu^k_i$ of $\bE_k$ as entries. Then we obtain
\[
  \ba_k^t \bQ \ba_k =  \bbb_k^t (P_{\alpha,n} \bI_{d+1} - \bD_k) \bbb_k =
  \sum_{i=1}^{d+1} b_{k,i}^2  \left(P_{\alpha,n}-\mu_i^k\right)
\]
with $\bbb_k:= \bT_k^t\bC_k^t\ba_k$. Choosing $P_{\alpha,n}> \max_{i,k}\left\{ |\mu^k_i| \right\} > 0$ the term  $\ba_k^t \bQ \ba_k$ is non-negative and is positive for $\ba_k\ne\bzero$.

\proofendmy

The idea of the proof is very similar to the proof  in case of the quasi-1D full Baer-Nunziato model, see \cite{Mueller-Hantke-Richter:16}. However, there the proof only works for the genuine 1D case, i.e., $d=1$, because the symmetrizer for the corresponding phasic matrices $B_{k,n}$ was chosen too specific.
It is worthwhile mentioning that modifying the symmetrizer $\bP_k$ for the phasic model in \cite{Mueller-Hantke-Richter:16}, Section 4.3,  replacing $0.5(\rho_k/\cbar_k)^2\,\bI_{d}$ by $\bT_n \bD_{k,n} \bT_n^t $, the proof also works for the quasi-1D case.
Furthermore, we would like to mention that  Theorem \ref{theorem:symmetrization} generalizes the result for the genuinely 1D case in \cite{saleh-seguin:2020}.

\section{Mathematical Entropy}
\label{subsec:bn-neq-thermo}

From a mathematical point of view, the concept of entropy-entropy flux pairs, cf.~\cite{Godlewski-Raviart:1991}, has been introduced to
characterize a unique weak solution of an initial (boundary) value problem of (inhomogeneous) conservation laws
that in quasi-conservative form reads
\beq
  \label{sec:eep-1}
  \pdifft{\bu} + \sum_{i=1}^{d} \bA_i(\bu) \pdiffx{\bu}{i} = \bS(\bu) - \sum_{i=1}^{d} \bB_i(\bu) \pdiffx{\bu}{i},\quad \bA_i(\bu) := \pdiff{\bff_i}{\bu}(\bu)
\eeq
where $\bu:\R_+\times \Omega \to \calDD\subset \R^m$ with $\Omega\subset \R^d$,
$\bff_i:\calDD \to \R^m $, $i=1,\ldots,d$ and $\bS:\calDD\to \R^m$, denote the vector of $m$ conserved quantities, the fluxes in the $i$th coordinate direction, $i=1,\ldots,d$, and the source function, respectively.
Here the last term on the right-hand side contains the non-conservative products in the original system determined by
\eqref{eq:sa-model-alphak}, \eqref{eq:sa-model}, \eqref{eq:sa-model-full-Ek-local}.
Motivated by thermodynamics, the entropy inequality
\beq
  \label{sec:eep-2}
  \pdifft{U(\bu)} + \sum_{i=1}^{d} \pdiffx{\bF_i(\bu)}{i} \le 0
\eeq
has to hold in a weak sense for any convex function $U:\calDD\to\R$ and functions $F_i:\calDD\to\R$, $i=1,\ldots, d$, referred to as entropy and entropy flux, that satisfy the compatibility conditions
\beq
  \label{sec:eep-3}
  \nabla_{\vects{\bu}} U(\bu)^T \, \bA_i(\bu)  = \nabla_{\vects{\bu}} F_i(\bu)^T,\ i=1,\ldots, d.
\eeq
Due to these conditions we infer for smooth solutions of \eref{sec:eep-1} the entropy equation
\beq
  \label{sec:eep-4}
  \pdifft{U(\bu)} + \sum_{i=1}^{d} \pdiffx{F_i(\bu)}{i} = \nabla_{\vects{\bu}} U(\bu)^T \,\left(\bS(\bu) - \sum_{i=1}^{d} \bB_i(\bu) \pdiffx{\bu}{i}\right),
\eeq
Obviously, the entropy inequality \eref{sec:eep-2} holds if and only if the entropy production is non-positive, i.e.,
\beq
   \nabla_{\vects{\bu}} U(\bu)^T \,\left(\bS(\bu) - \sum_{i=1}^{d} \bB_i(\bu) \pdiffx{\bu}{i}\right) \le 0.
\eeq
A sufficient condition is given by
\beq
  \label{sec:eep-4n}
   \nabla_{\vects{\bu}} U(\bu)^T \,\bS(\bu) \le0 \quad \mbox{and} \quad \nabla_{\vects{\bu}} U(\bu)^T \sum_{i=1}^{d} \bB_i(\bu) \pdiffx{\bu}{i} = 0.
\eeq

\subsection{Isothermal case}
\label{subsec:bn-neq-thermo-isothermal}

For an isothermal fluid we have to account for heat exchange in the energy equation.
Similar to the case of isothermal Euler equations, cf.~\cite{Thein:2018},  we therefore have to start from the full Baer-Nunziato system  to derive an appropriate entropy-entropy flux pair. This can be found in
\cite{Mueller-Hantke-Richter:16}. Neglecting viscosity but accounting for heat exchange the full model reads
\begin{subequations}
\label{eq:sa-model-full}
\begin{align}
\label{eq:sa-model-full-alphak}
 &\pdifft{\alpha_k} + \vprod{\veloI}{\pgrad{\alpha_k}} = \sourcemy{\alpha}{k} ,\\
\label{eq:sa-model-full-rhok}
&\pdifft{(\alpha_k\,\rho_k)} + \pdiv{(\alpha_k\,\rho_k\,\velo{k})} =
\sourcemy{\alpha\rho}{k} ,\\[2mm]
\label{eq:sa-model-full-mk}
&\pdifft{(\alpha_k\,\rho_k\,\velo{k})} + \pdiv{(\alpha_k\,\rho_k\,\dyade{\velo{k}} + \alpha_k\,p_k\,\identity)} + \sum_{l=1,\ne k}^{K} P_{k,l}\, \pgrad{\alpha_l}=
 \Source{\alpha\rho\vects{\velo{}}}{k} ,\\[2mm]
\label{eq:sa-model-full-Ek}
&\pdifft{(\alpha_k\,\rho_k\,E_k)}
+ \pdiv{(\alpha_k\,\rho_k\,\velo{k}\,(E_k+p_k/\rho_k))}
+\sum_{l=1,\ne k}^{K} P_{k,l}\, \vprod{\veloI} \pgrad{\alpha_l} =
- \pdiv{(\alpha_k\, \bq_k)}
+ \sourcemy{\alpha\rho E}{k}
\end{align}
\end{subequations}
with the heat flux $\bq_k$ and total $E_k = e_k + \velo{k}^2/2$ determined by
the internal energy  $e_k$ and kinetic energy $u_k=\velo{k}^2/2$. This model is complemented
for isothermal fluids by the EoS
\begin{equation}
\label{eq:eos-full-k}
 p_k = p_k(\rho_k) \quad\mbox{and}\quad e_k = e_k(\rho_k) .
\end{equation}
From \eref{eq:sa-model-full} we derive equations for the density and internal energy
\begin{align}
\label{eq:sa-model-full-rhok-single}
&\pdifft{\rho_k} + \frac{\rho_k}{\alpha_k}\,\vprod{(\velo{k}-\veloI)}\pgrad{\alpha_k}
+ \vprod{\velo{k}}\pgrad{\rho_k} + \rho_k\pdiv{\velo{k}} = \sourcemy{\rho}{k}
,\\
\label{eq:sa-model-full-ek}
&\pdifft{e_k} + \vprod{\velo{k}}(\pgrad{e_k}) - \frac{1}{\alpha_k \rho_k}
\sum_{l=1,\ne k}^{K} \vprod{P_{k,l} \left(\velo{k}- \veloI  \right) } \pgrad{\alpha_l}  +  \frac{p_k}{\rho_k}\pdiv{\velo{k}}=
- \frac{1}{\alpha_k \rho_k} \pdiv{ (\alpha_k\bq_k)}   + \sourcemy{e}{k}
\end{align}
with relaxation terms
\begin{align}
\label{eq:density-relax}
&\sourcemy{\rho}{k} := \frac{1}{\alpha_k} \left(S_{\alpha\rho,k} - \rho_k S_{\alpha,k} \right),\\
\label{eq:energy-relax}
&\sourcemy{e}{k} :=
\frac{1}{\alpha_k \rho_k} \left( \sourcemy{\alpha\rho E}{k} -
\vprod{\velo{k}}\Source{\alpha\rho\vects{\velo{}}}{k} + \sourcemy{\alpha\rho}{k} (u_k - e_k) \right) .
\end{align}
In order to derive the evolution equation for the entropy
$s_k=s_k(\rho_k)$
we rewrite
\eref{thermo-entropy} as
\beq
  \label{thermo-gibbs-relation}
  \Tref d s_k = d e_k - \frac{p_k}{\rho_k^2} d\rho_k = \left(e_k' - \frac{p_k}{\rho_k^2}\right) d\rho_k,
\eeq
where due to \eref{eq:isothermalfluid}
\begin{align*}
  T_k=\Tref=const,\quad \forall\, k=1,\ldots,K.
\end{align*}
By means of the evolution equations \eref{eq:sa-model-full-rhok-single} and
\eref{eq:sa-model-full-ek} for the density
and the internal energy we then deduce the entropy law
\begin{align}
\label{eq:sa-model-full-sk}
\pdifft{s_k} + \vprod{\velo{k}}\pgrad{s_k}
 =
\frac{1}{\alpha_k \rho_k } \left( \Pi_k-\frac{1}{\Tref} \pdiv{ (\alpha_k\bq_k)} \right)  + \sourcemy{s}{k}
\end{align}
with relaxation term
\begin{align}
\label{eq:entropy-relax}
\sourcemy{s}{k} :=
\frac{1}{\alpha_k \rho_k \Tref}
\left( \alpha_k \rho_k \sourcemy{e}{k} -
\frac{p_k}{\rho_k} \sourcemy{\alpha\rho}{k}
 + p_k \sourcemy{\alpha}{k} \right)
\end{align}
and production term
\begin{align}
\label{eq:entropy-prod-k}
\Pi_k:=
\frac{1}{\Tref}
\left(
\sum_{l=1,\ne k}^{K} \vprod{P_{k,l} \left(\velo{k}- \veloI  \right)} \pgrad{\alpha_l} +
 p_k \vprod{(\velo{k}-\veloI)}{\pgrad{\alpha_k}}  \right) .
\end{align}
Finally,  we obtain together with \eref{eq:sa-model-full-rhok} for the volume specific entropy
\beq
\label{eq:sa-model-primitive-rhosk}
\pdifft{(\alpha_k\rho_k s_k)} + \pdiv{(\alpha_k\rho_k s_k\velo{k})} +
\frac{1}{\Tref} \pdiv{ (\alpha_k\bq_k)} =
\Pi_k
 \sourcemy{\alpha\rho s}{k}
\eeq
with relaxation term
\beq
\label{eq:rhoentropy-relax}
\sourcemy{\alpha\rho s}{k} :=
\alpha_k \rho_k \sourcemy{s}{k} +
 s_k \sourcemy{\alpha\rho}{k}  .
\eeq
Next we determine from the energy equation \eref{eq:sa-model-full-Ek} the heat flux
\begin{align}
\label{eq:sa-model-full-heatflux-k}
\pdiv{(\alpha_k\, \bq_k)} =
\sourcemy{\alpha\rho E}{k}
-\pdifft{(\alpha_k\,\rho_k\,E_k)}
- \pdiv{(\alpha_k\,\rho_k\,\velo{k}\,(E_k+p_k/\rho_k))}
-\sum_{l=1,\ne k}^{K} P_{k,l}\, \vprod{\veloI} \pgrad{\alpha_l} .
\end{align}
Plugging the heat flux into the entropy equation \eref{eq:sa-model-primitive-rhosk}
we obtain
\begin{align}
\label{eq:sa-model-full-free-energy-k}
&\pdifft{(\alpha_k\,\rho_k\,(\Tref s_k -E_k))} +
\pdiv{(\alpha_k\,\rho_k\,\velo{k}\,(\Tref s_k-E_k-p_k/\rho_k))} =
\\
&\hspace*{50mm}
\Tref \Pi_k + \sum_{l=1,\ne k}^{K} P_{k,l}\, \vprod{\veloI} \pgrad{\alpha_l}  + \Tref \sourcemy{\alpha\rho E}{k} .
\nonumber
\end{align}
Summing these evolution equations over all components and using the conservation constraint
\eref{eq:constraints-interfacial-pressures-gen} we obtain
\begin{align}
\label{eq:sa-model-F}
&\pdifft{\left(\sum_{k=1}^K \alpha_k\,\rho_k\,(E_k-\Tref s_k)\right)} +
\pdiv{\left(\sum_{k=1}^K  ( \alpha_k\,\rho_k\,\velo{k}\,(E_k-\Tref s_k+p_k/\rho_k) ) \right)} =
\nonumber\\
&\hspace*{50mm}
-\Tref\sum_{k=1}^K \source{\alpha \rho s}{k} - \Tref\sum_{k=1}^K  \Pi_k .
\end{align}
Comparing \eref{eq:sa-model-F} with \eref{sec:eep-4} we may choose for the entropy-entropy flux pair
\begin{subequations}
  \label{sec:eep-isothermal}
\begin{align}
  \label{sec:eep-isothermal-a}
  &U(\bu) :=
\sum_{k=1}^K \alpha_k\,\rho_k\,(E_k-\Tref\, s_k) =
\sum_{k=1}^K \alpha_k\,\rho_k\,(f_k+ 1/2 \velo{k}^2)  ,\\
  \label{sec:eep-isothermal-b}
  &F_i(\bu):=
\sum_{k=1}^K  ( \alpha_k\,\rho_k\,\velo{k}\,(E_k-\Tref s_k+p_k/\rho_k) ) =
\sum_{k=1}^K  ( \alpha_k\,\rho_k\,\velo{k}\,(g_k+ 1/2\velo{k}^2) )
\end{align}
\end{subequations}
with the specific free energy $f_k$ and Gibbs free energy $g_k$
\begin{align}
  f_k=e_k-\Tref s_k,\quad
  g_k=f_k+p_k/\rho_k .
\end{align}
The compatibility conditions \eref{sec:eep-3} hold by derivation of \eref{eq:sa-model-E}, see also
\cite{Mueller-Hantke-Richter:2014} in case of the full Baer-Nunziato model.


\begin{theorem}(Convexity of entropy function)
\label{theorem:Convexity}
Let $e_k$ be a convex function of $\tau_k$, $k=1,\ldots, K$.
Then the Hessian of $U$ is positive semi-definite, i.e.,  the entropy $U$ is a convex function of $\bu$ but not necessarily strictly convex.
\end{theorem}

\proof

To verify that
\beq
\label{def:u_k}
 U(\bu) :=
\sum_{k=1}^{K} \alpha_k \rho_k (e_k- \Tref s_k + \bv_k^2/2) =
\sum_{k=1}^{K} \alpha_k U_k(\bu_k)
\eeq
is a convex function of the quantities $\bu:= (\balpha^t, \alpha_1 \bu_1^t, \ldots,\alpha_K\bu_K^t)^t$ with
$\balpha:=(\alpha_1,\ldots,\alpha_{K-1})^t$ and $\bu_k:=(\rho_k,\rho_k \bv_k^t)^t$
we proceed in analogy to \cite{Mueller-Hantke-Richter:16} for the full multi-component Baer-Nunziato model but using an isothermal fluid.

For this purpose,
we need to prove that the Hessian is positive semi-definite. First of all, we note that by \eref{eq:saturation}
\[
  \frac{\partial \alpha_k}{\partial \alpha_l} = \delta_{k,l}-\delta_{k,K},\quad
  \frac{\partial \bu_k}{\partial \alpha_l} =
   -\frac{1}{\alpha_k} \bu_k (\delta_{k,l}-\delta_{k,K}),\
  \frac{\partial \bu_k}{\partial \alpha_l \bu_l} = \frac{1}{\alpha_k} \delta_{k,l} \bI_{d+1}
\]
holds for $k=1,\ldots,K,\ l=1,\ldots,K-1$. Then it follows for the gradient of $U$
\beq
  \label{eq:convexity-Hessian-0}
  \frac{\partial U}{\partial \alpha_l}(\bu)  = U_l(\bu_l) - U_K(\bu_K) - \vprod{\frac{\partial U_l}{\partial \bu_l}(\bu_l)}{\bu_l}
                                               + \vprod{\frac{\partial U_K}{\partial \bu_K}(\bu_K)}{\bu_K},\
   \frac{\partial U}{\partial \alpha_l \bu_l}(\bu) = \frac{\partial U_l}{\partial \bu_l}(\bu_l).
\eeq
The Hessian of $U$ is determined by the second order derivatives
\begin{eqnarray*}
  \frac{\partial^2 U}{\partial \alpha_k \partial \alpha_l}(\bu) &=&
  \delta_{k,l} \frac{1}{\alpha_l} \bu_l^t \frac{\partial^2 U_l}{\partial^2 \bu_l}(\bu_l) \bu_l +
               \frac{1}{\alpha_K} \bu_K^t \frac{\partial^2 U_K}{\partial^2 \bu_K}(\bu_K) \bu_K,\ k,l=1,\ldots,K, \\
  \frac{\partial^2 U}{\partial \alpha_k \partial \alpha_l \bu_l}(\bu) &=&
  - \frac{1}{\alpha_l} (\delta_{k,l}-\delta_{K,l})\frac{\partial^2 U_l}{\partial^2 \bu_l}(\bu_l) \bu_l, l=1,\ldots,K,\ k=1,\ldots,K-1,\\
  \frac{\partial^2 U}{\partial \alpha_k\bu_k \partial \alpha_l\bu_l}(\bu) &=&
  \frac{1}{\alpha_l} \delta_{k,l} \frac{\partial^2 U_l}{\partial^2 \bu_l}(\bu_l),\ k,l=1,\ldots,K.
\end{eqnarray*}
For a compact representation of the Hessian we introduce the notation
\begin{eqnarray*}
  \bU_{\vects{\balpha}, \vects{\balpha}} &:=&
  \left(\frac{\partial^2 U}{\partial \alpha_k \partial \alpha_l}(\bu)\right)_{l,k=1,\ldots,K-1} \in \R^{(K-1)\times (K-1)},\\
  \bU_{\alpha_k\vects{\bu_k}, \alpha_k\vects{\bu_k}} &:=&
  \frac{\partial^2 U}{\partial \alpha_k\bu_k \partial \alpha_k\bu_k}(\bu)\in \R^{(d+1)\times (d+1)},\\
  \bU_{\alpha_k\vects{\bu_k},\vects{\balpha}} &:=&
  \left( \frac{\partial^2 U}{\partial \alpha_k \bu_k \partial \alpha_1 }(\bu),\ldots,
         \frac{\partial^2 U}{\partial \alpha_k \bu_k \partial \alpha_{K-1} }(\bu)  \right) \in \R^{(d+1)\times (K-1)}  .
\end{eqnarray*}
According to the above second order derivatives these are determined by
\begin{eqnarray}
  \label{eq:convexity-Hessian-1}
  \bU_{\vects{\balpha}, \vects{\balpha}} &=&
  \frac{1}{\alpha_K} \bu_K^t \bU_K'' \bu_K \bone_{K-1} +
  \diag \left( \left(\frac{1}{\alpha_k} \bu_k^t \bU_k'' \bu_k \right)_{k=1,\ldots,K-1} \right),\\
  \label{eq:convexity-Hessian-2}
  \!\!\!\!\!\!\!\!\!\bU_{\alpha_k\vects{\bu_k}, \alpha_k\vects{\bu_k}} &=&
   \frac{1}{\alpha_k} \bU_k'',\\
  \label{eq:convexity-Hessian-3}
  \bU_{\alpha_k\vects{\bu_k},\vects{\balpha}} &=&
  \left( -\frac{1}{\alpha_k} (\delta_{k,l}-\delta_{K,k}) \bU_k'' \bu_k\right)_{l=1,\ldots,K-1} = \bU_{\balpha,\alpha_k\bu_k}^t,
\end{eqnarray}
where $\bU_k''$ denotes the Hessian of the entropy $U_k=U_k(\bu_k)$ of component $k$.
Then the Hessian can be represented as block-matrix
\beq
  \label{eq:convexity-Hessian}
  \bU''(\bu) =
  \begin{pmatrix}
    \bU_{\vects{\balpha},\vects{\balpha}}       & \bU_{\vects{\balpha},\alpha_1\vects{\bu_1}}       & \ldots & \bU_{\vects{\balpha},\alpha_1\vects{\bu_K}}       \\
    \bU_{\alpha_1\vects{\bu_1},\vects{\balpha}} & \bU_{\alpha_1\vects{\bu_1},\alpha_1\vects{\bu_1}} &        &                                 \\
    \vdots                      &                                   & \ddots &                                 \\
    \bU_{\alpha_K\vects{\bu_K},\vects{\balpha}} &                                   &        & \bU_{\alpha_1\vects{\bu_K},\alpha_K\vects{\bu_K}} \\
  \end{pmatrix} .
\eeq
To verify positive semi-definiteness of the Hessian we introduce the vector $\bx = (\ba^t,\bbb_1^t,\ldots,\bbb_K^t )^t$
with $\ba\in\R^{K-1}$ and $\bbb_k\in\R^{d+1}$, $k=1,\ldots,K$. Then we obtain by the block-structure
\eref{eq:convexity-Hessian} of the Hessian
\beq
  \label{eq:convexity-positivity}
  \bx^t \bU'' \bx =
  \ba^t \bU_{\vects{\balpha},\vects{\balpha}} \ba +
  \sum_{k=1}^{K} \ba^t \bU_{\vects{\balpha},\alpha_k\vects{\bu_k}} \bbb_k +
  \sum_{k=1}^{K} \bbb_k^t \left( \bU_{\alpha_k\vects{\bu_k},\vects{\balpha}} \ba + \bU_{\alpha_k\vects{\bu_k},\alpha_k\vects{\bu_k}} \bbb_k \right) .
\eeq
By means of \eref{eq:convexity-Hessian-1}, \eref{eq:convexity-Hessian-2} and \eref{eq:convexity-Hessian-3} we determine
\begin{eqnarray*}
  &&\ba^t \bU_{\vects{\balpha},\vects{\balpha}} \ba =
    \frac{1}{\alpha_K} (a\bu_K)^t \bU_K'' (a\bu_K) +
   \sum_{k=1}^{K-1} \frac{1}{\alpha_k} (a_k\bu_k)^t \bU_k''\, (a_k\bu_k),\  a: = \sum_{l=1}^{K-1} a_l , \\
  &&\ba^t \bU_{\vects{\balpha},\alpha_k\vects{\bu_k}} \bbb_k =
  - \sum_{l=1}^{K-1} \frac{1}{\alpha_k} (\delta_{k,l}-\delta_{k,K}) \bbb_k^t \bU_k''\, (a_l\bu_k),\\
 && \bbb_k^t \bU_{\vects{\balpha},\alpha_k\vects{\bu_k}} \ba =
  \frac{1}{\alpha_K} \delta_{k,K}  \bbb_K^t \bU_K''\, (a \bu_K) -  \frac{1}{\alpha_k} (1-\delta_{k,K})  \bbb_k^t \bU_k'' \, (a_k\bu_k) ,  \\
 &&\bbb_k^t \bU_{\alpha_k\vects{\bu_k},\alpha_k\vects{\bu_k}} \bbb_k =
  \frac{1}{\alpha_k} \bbb_k^t \bU_k''\, \bbb_k.
\end{eqnarray*}
Incorporating this into \eref{eq:convexity-positivity} we finally conclude after some calculus with
\begin{align*}
  \bx^t \bU'' \bx =
  \sum_{k=1}^{K-1} \frac{1}{\alpha_k} (\bbb_k- a_k \bu_k)^t \bU_k'' \, (\bbb_k- a_k \bu_k)
    +
  \frac{1}{\alpha_K} (\bbb_K- a \bu_K)^t \bU_K'' \, (\bbb_K- a \bu_K) .
\end{align*}

Finally, we check that the Hessians $\bU_k''$ are positive semi-definite where $\bU_k$ is introduced in
\eqref{def:u_k}.
For this purpose we consider the energy and the entropy to be given in terms of volume and temperature, i.e.\
$\te_k(\tau_k,T)=e_k(\tau_k,s_k(\tau_k,T))$. The first derivatives of the energy and the pressure with respect to the volume are given by
\begin{align}
	\frac{\partial \te_k}{\partial\tau_k} = \frac{\partial e_k}{\partial\tau_k} + \frac{\partial e_k}{\partial s_k}\frac{\partial s_k}{\partial\tau_k}
	\stackrel{\eqref{thermo-pres-temp}}{=} -p_k + \Tref\frac{\partial s_k}{\partial\tau_k}\quad\text{and}\quad
	\frac{\partial p_k}{\partial\tau_k} &\stackrel{\eqref{eq:sound-fdt-T-tau}}{=} -\frac{\cisoT_k^2}{\tau_k^2}.
\end{align}
From this we conclude
\begin{align}
	\frac{\partial}{\partial \rho_k} (\rho_k(e_k- \Tref s_k)) &= e_k- \Tref s_k + \frac{p_k}{\rho_k} ,\\
  \frac{\partial^2}{\partial \rho_k^2} (\rho_k(e_k- \Tref s_k)) &=
  \frac{\cisoT_k^2}{\rho_k} > 0.
\end{align}
For the Hessian we then determine
\begin{align}
  \label{eq:Hessian-u-k}
  \bU_k'' =
  \frac{1}{\rho_k}
  \left(
  \begin{matrix}
      \cisoT_k^2 + \bv_k^2 & & -\bv_k^t \\
      -\bv_k & & \bI_{d}
  \end{matrix}
  \right)
\end{align}
that can be easily verified to be positive definite. This implies $\bx^t \bU'' \bx \ge 0$.
Note that for $\bx \ne \bzero$ we cannot ensure
$\bx^t \bU'' \bx$ to be positive even if $\bU_k''$ is strictly convex because all the terms $\bbb_k-a_k\bu_k$, $k=1,\ldots,K-1$, and $\bbb_K-a\bu_K$ may vanish at the same time.

\proofendmy

Furthermore, to ensure that the entropy inequality \eref{sec:eep-2} holds we have to verify that the right-hand side of \eref{eq:sa-model-F} is not positive, see also \cite{Mueller-Hantke-Richter:2014} in case of the full Baer-Nunziato model.

\subsubsection{Entropy production due to interfacial velocity and pressure}
\label{thermodynamics-properties-interfacial-isothermal}


Since we cannot control the sign of $\Pi_k$, we determine
the interfacial pressures $P_{k,l}$ and the interfacial velocity $\veloI$ such
that the sum $\Pi:=\sum_{k=1}^{K} \Pi_k$ vanishes and the conservation constraints \eref{eq:constraints-interfacial-pressures-gen} or, equivalently, \eref{eq:constraints-interfacial-pressures} hold.
Assuming that the interfacial velocity can be written as a linear combination of the velocities of the components according to \eref{thermodynamics-properties-interfacial-2} there exists a unique closure for the interfacial pressures.

\begin{theorem}(Entropy production due to interfacial states)
\label{theorem:entropy-production-interfacial-states-isothermal}
For any convex combination \eref{thermodynamics-properties-interfacial-2} for the interfacial velocity $\veloI$
there uniquely exist  interfacial pressures
\begin{equation}
\label{theorem:entropy-production-interfacial-states-Pkl-isothermal}
P_{k,l} =   \beta_k p_l  + p_k (1- \beta_k)
\end{equation}
such that the production term $\Pi= \Tref  \sum_{k=1}^{K} \Pi_k$ vanishes and
the conservation constraints \eref{eq:constraints-interfacial-pressures-gen} or, equivalently,
\eref{eq:constraints-interfacial-pressures} hold.
In particular, the interfacial pressures are all positive and
\begin{equation}
\label{theorem:entropy-production-interfacial-states-PI-isothermal}
  P_I =   \sum_{k=1}^K  p_k (1- \beta_k) .
\end{equation}
\end{theorem}

\proof

The proof is in complete analogy to the proof of Theorem 6 in \cite{Mueller-Hantke-Richter:16}
for the full multi-component Baer-Nunziato model
assuming local thermal equilibrium where we have to replace the temperatures $T_k$ by $\Tref$. This is due to the fact that the terms $\Pi_k$ as well as the conservation constraints coincide.
Thus, the resulting interfacial pressures \eref{theorem:entropy-production-interfacial-states-Pkl-isothermal} are a special case of  those presented in \cite{Mueller-Hantke-Richter:16}. We emphasize that the resulting interfacial states are the same for both  isentropic and  isothermal fluids, see Theorem
\ref{theorem:entropy-production-interfacial-states-isentropic}.

\proofendmy

We would like to mention that the above result is motivated by Gallou\"et et al.~\cite{Gallouet-Herard-Seguin:04} and
H{\'e}rard \cite{Herard:07} for a
two-phase and a three-phase model, respectively,
in case of
the full Baer-Nunziato model.
In \cite{Mueller-Hantke-Richter:16} it was proven for a multi-component fluid with an arbitrary number of components.

Using for the convex combination \eref{thermodynamics-properties-interfacial-2} the coefficients
\eref{thermodynamics-properties-interfacial-2a},
then the interfacial pressure and the interfacial velocity are given by
\beq
\label{theroem:entropy-production-interfacial-states-special-3-isothermal}
P_I= \sum_{k=1}^{K} p_k (1-c_k\alpha_k\rho_k /{\hat\rho} ),\quad
\veloI = \left.\sum_{i=1}^Kc_i\alpha_i \rho_i \bv_i\right/\sum_{k=1}^K c_k\alpha_k\rho_k.
\eeq
This closure of the interfacial states ensures consistency with an entropy inequality
and the fields corresponding to the material waves are linearly degenerated.

For special choices of $\bc\in[0,1]^K$ the interfacial values coincide with those in the literature in case of two and three components. For instance, choosing $\bc=\be_i$ for some $i\in\{1,\ldots,K\}$ we obtain
\beq
\label{theroem:entropy-production-interfacial-states-special-1-isothermal}
P_I= \sum_{k=1,\ne i}^{K} p_k,\quad \veloI = \velo{i}.
\eeq
For $i=1$ these coincide with those given in \cite{Gallouet-Herard-Seguin:04} and
\cite{Herard:07} for $K=2$ and $K=3$, respectively.
In case of uniform coefficients $c_k=1/K$, $k=1,\ldots,K$, the interfacial pressure and velocity are given by
\beq
\label{theroem:entropy-production-interfacial-states-special-2-isothermal}
P_I= \sum_{k=1}^{K} p_k (1-\alpha_k\rho_k /\rho)
= \sum_{k=1}^{K} p_k \sum_{j=1,\ne k}^{K}\alpha_j\rho_j /\rho
,\quad
\veloI = \left.\sum_{i=1}^K\alpha_i \rho_i \bv_i\right/\sum_{k=1}^K \alpha_k\rho_k = \bv.
\eeq
where $\rho$ and $\bv$ are the density and the velocity of the mixture, see \eref{eq:mixture-cons}, respectively.

Obviously,
in case of two components, i.e., $K=2$, the  Abgrall-Saurel closure \cite{Saurel-Abgrall:1999}
\beq
\label{eq:abgrall-saurel-closure}
P_I= p_1 \frac{\alpha_2 \rho_2}{\rho} +  p_2 \frac{\alpha_1 \rho_1}{\rho}
,\quad
\veloI = v_1 \frac{\alpha_1 \rho_1}{\rho} +  v_2 \frac{\alpha_2 \rho_2}{\rho}
\eeq
may violate the 2nd law of thermodynamics because the sum of the entropy production terms due to the interfacial states does not vanish.

\begin{remark}
Finally, we would like to point out that for barotropic fluids another entropy-entropy flux pair (EEP) has been derived in \cite{boukili-herard:2018,saleh-seguin:2020} based on the ``energies'' $\oee_k$, see Remark \ref{rem:energy-barotropic-fluid}.
As these ``energies'' coincide with the energies $e_k$  for isentropic fluids, we present this entropy law in Sect.~\ref{subsec:bn-neq-thermo-isentrop}.
Since isothermal fluids are a special case of a barotropic fluid, this  EEP is also an EEP in the isothermal case. It is worthwhile mentioning that the other EEP results in the same interfacial pressures as given in Thm.~\ref{theorem:entropy-production-interfacial-states-isothermal}.
\end{remark}

\subsubsection{Entropy production due to relaxation}
\label{sec:relax-terms-isothermal}

To be admissible with the entropy inequality \eref{sec:eep-4n}  the  production term
\beq
\label{thermodynamics-properties-entropy-relaxation-total-isothermal}
  S_{U} :=  - \Tref \sum_{k=1}^{K}  \sourcemy{\alpha\rho s}{k}
\eeq
has to be
non-positive.
Here we use \eref{eq:rhoentropy-relax}, \eref{eq:entropy-relax} and \eref{eq:energy-relax} to determine the
phasic
terms
\beq
\label{thermodynamics-properties-entropy-relaxation-isothermal}
  \sourcemy{\alpha\rho s}{k} =
 \frac{1}{\Tref}
\left(  \sourcemy{\alpha\rho E}{k} + \left( \frac{1}{2} \velo{k}^2-g_k\right)  \sourcemy{\alpha\rho}{k} + p_k \sourcemy{\alpha}{k} - \velo{k}\cdot \Source{\alpha\rho\vects{\velo{}}}{k} \right) .
\eeq
Note that in abuse of physical notation we call this term entropy production rather than energy production with respect to the mathematical concept of entropy-entropy flux pairs because energy is known to be conserved.
In the following we consider one-by-one the contributions of mechanical relaxation and chemical relaxation.
The analysis is similar to \cite{Mueller-Hantke-Richter:16} in case of the physical entropy production.
Note that  all the relaxation processes are in agreement with the constraints due to conservation \eref{eq:constraints}.

\begin{theorem}(Entropy production due to mechanical relaxation)
\label{theorem:entropy-production-mechanical-relaxation-isothermal}
Let the relaxation parameters be non-negative, i.e., $\theta_p$, $\theta_v\ge 0$.
Then the entropy production due to mechanical relaxation is
non-positive,
i.e.,
\begin{equation}
\label{eq:relaxation-entropy-constraints-mechanical-relaxation-isothermal}
   S_{U}^{p,v} =    \Tref \sum_{k=1}^{K} \sourcemy{\alpha\rho s}{k}^{p,v}
   \le
   0.
\end{equation}
\end{theorem}

\proof
According to  \eref{thermodynamics-properties-entropy-relaxation-isothermal}
and \eref{pres-relax-comp}, \eref{velo-relax-comp}
the phasic entropy production terms for velocity and pressure relaxation are determined by
\beq
\label{eq:velo-pres-relax-comp-thermo-isothermal}
  S^v_{\alpha\rho s, k} =  - \theta_v \frac{\alpha_k\rho_k}{\Tref} (\velo{}-\bv_k)^2 ,\quad
  S^p_{\alpha\rho s, k} = - \theta_p \frac{\alpha_k}{\Tref} (p-p_k)^2 .
\eeq
This immediately implies the assertion.

\proofendmy

For the relaxation of chemical potentials we discuss different scenarios for two-phase and three-phase models.

In case of the two-component mixture, the entropy production term due to relaxation of the chemical potentials is determined by  \eref{thermodynamics-properties-entropy-relaxation-isothermal}
and \eref{mass-relax-two-comp} as
\beq
\label{eq:mass-relax-comp-thermo-K2-isothermal}
  S^\mu_{\alpha\rho s, k} =
  (-1)^{k}
  \relaxG\, \dot m\, \frac{1}{\Tref}
  \left(
  \epsilon -  g_k + \frac{p_k}{\varrho}   - \frac{1}{2} (\velo{k}- \hbv)^2
  \right), \quad k=1,2.
\eeq
However choosing the parameters  $\epsilon$, $\varrho$ and $\hbv$ appropriately
the sum of the entropy production terms is non-negative.

\begin{theorem}(Entropy production due to relaxation of Gibbs free energies)
\label{theorem:entropy-production-Gibbs-relaxation-isothermal}
Let the relaxation parameter $\theta_\mu$ be positive.
The mass flux is chosen as the kinetic relation
\begin{align}
\label{eq:kinetic-isothermal}
	\dot m=a (g_2-g_1),\ a\ge 0.
\end{align}
Choosing in \eref{mass-relax-two-comp} the parameters
\begin{align}
\label{eq:interfacial-velocity-convexity-factors-K2-isothermal}
\beta_1^{v}  = \beta_2^{v} = \frac{1}{2} ,\quad
\varrho = \frac{1}{c}\frac{p_2-p_1}{g_2-g_1},\ c \le 1,
\end{align}
then the entropy production is
non-positive,
i.e.,
\beq
\label{eq:mass-relax-comp-thermo-mixture-isothermal}
  S^\mu_{U} = - \Tref \sum_{k=1}^2  S^\mu_{\alpha\rho s, k}
  \le
  0 .
\eeq
\end{theorem}

\proof
First of all, we note that by  \eref{eq:interfacial-velocity-convexity-factors-K2-isothermal} the velocity $\hbv$ is the average of the phasic velocities and, thus, the entropy production does not depend on the velocity because
\beq
\label{eq:no-velo-int-isothermal}
   \frac{1}{2} \left( (\hbv - \velo{1})^2 -  (\hbv - \velo{2})^2\right) =
   \frac{1}{2}  (\velo{2} - \velo{1}) \cdot (2 \hbv - \velo{1} - \velo{2}) =
  \bzero.
\eeq
Then the entropy production reduces to
\[
  S^\mu_{U} =
  - \relaxG\, \dot m\, \left( g_2 - g_1 + \frac{p_1-p_2}{\varrho}  \right) =
  -  \relaxG\, a\,(g_2-g_1)^2 ( 1- c)
  \le
  0.
\]
Thus, the energy production $S_{U}^\mu$ is
non-positive.

\proofendmy

Note that the kinetic relation \eref{eq:kinetic-isothermal} is in agreement with the kinetic relation in
\cite{Dreyer-Duderstadt-Hantke-Warnecke:2012}. Furthermore, choosing $\hbv$ as the average of the phasic velocities is reasonable because the entropy production \eref{eq:mass-relax-comp-thermo-mixture-isothermal} should be a product of the relaxed mass flux $\relaxG\, \dot m$ and an interfacial entropy should not depend on any velocity.

Note that \eref{theroem:entropy-production-interfacial-states-special-3-isothermal}
and the mixture velocities \eref{chem-relax-velo} with coefficients \eref{eq:interfacial-velocity-convexity-factors-K2-isothermal} differ. In \cite{Herard-Hurisse:2012} it was recommended to use different interfacial velocities for the convective system and the relaxation terms.  This is admissible  because the conservation constraint \eref{eq:constraints}  is satisfied for any convex combination \eref{thermodynamics-properties-interfacial-2}.
Moreover, the source term cannot be derived from the ensemble averaging procedure, see \cite{Drew-Passman:1999}, Chapter 11, but the averaged model has to be closed
by modeling these terms appropriately. Therefore, we are free to choose another velocity in the chemical relaxation model.
Note that in the Drew-Passman model different interfacial velocities have been introduced in the evolution equations for volume fraction, momentum and energy, see \cite{Drew-Passman:1999}, 
Eqns.~(11.8), (11.39) and (11.41).

In case of the three-component mixture, the entropy production terms due to the relaxation of chemical potentials is determined by \eref{thermodynamics-properties-entropy-relaxation}
and \eref{mass-relax-three-comp} as
\begin{subequations}
\label{eq:mass-relax-comp-thermo-isothermal}
\begin{align}
\label{eq:mass-relax-comp-thermo-1-isothermal}
  &S^\mu_{\alpha\rho s, 1} =
  \relaxG\, \dot m\, \frac{1}{\Tref}
  \left(
  \epsilon_1 - g_1 + \frac{p_1}{\varrho_1}  +  \frac{1}{2} (\hbv-\velo{1})^2
  \right), \\
\label{eq:mass-relax-comp-thermo-2-isothermal}
  &S^\mu_{\alpha\rho s, 2} =
  -\relaxG\, \dot m\, \frac{1}{\Tref}
  \left(
  \epsilon_2 - g_2 - \frac{p_2}{\varrho_2}  +  \frac{1}{2} (\hbv-\velo{2})^2
  \right), \\
\label{eq:mass-relax-comp-thermo-3-isothermal}
  &S^\mu_{\alpha\rho s, 3} =
  \relaxG\, \dot m\,\frac{1}{\Tref}
  \left(
    \epsilon_2 - \epsilon_1 - p_3(\frac{1}{\varrho_1}+\frac{1}{\varrho_2})
  \right) .
\end{align}
\end{subequations}
The phasic relaxation terms may become negative also when $\relaxG$ and $\dot m$ are positive.
However, choosing the parameters $\varrho_1$, $\varrho_2$ and $\hbv$ appropriately the total production term is non-negative.

In particular, we are interested in the three-component model, i.e., $K=3$, with water vapor ($k=1$), liquid water ($k=2$) and inert gas ($k=3$).
The source terms describing the mass transfer depend on the chemical potentials of the vapor and the liquid. In the model under consideration the gas is a mixture of water vapor and some other constituent, where both components are modeled as an ideal gas. Accordingly the chemical potential of the vapor is given by
\begin{equation}
\label{mu_gas}
 \mu_1=g_1+(\gamma_1-1) c_{v,1} \Tref
 \ln\left(\frac{\alpha_1}{\alpha_1 +  \alpha_3}\right).
\end{equation}
In the special case of vanishing third component, i.e., $\alpha_3 = 0$,
the chemical potential of the vapor reduces to the vapor Gibbs free energy.
The chemical potential $\mu_2$ of the liquid equals its Gibbs free energy $g_2$, i.e.,
\[
 \mu_2=g_2\,.
\]
In chemical equilibrium the chemical potentials of the vapor and the liquid
equal each other.
For details see the book of M\"uller and M\"uller \cite{MuellerMueller}, Section 8.2.4.

In the following we consider a homogeneous mixture where velocity, pressure and temperature are in equilibrium.
In the total entropy  we therefore additionally have to account for the mixture entropy given by
\beq
\label{eq:mixture-entropy}
  S_M = -\sum_{k=1,3} \alpha_k \rho_k  (\gamma_k-1)c_{v,k}  \ln\left(\frac{\alpha_k}{\alpha_1+\alpha_3}\right),
\eeq
see \cite{IMueller}, p.~54, 298, 320.
Again, the chemical potential of the vapor phase is given  by \eref{mu_gas} and
the chemical potential  of the liquid phase equals its Gibbs free energy, i.e., $ \mu_2=g_2$.
In chemical equilibrium the chemical potentials of the vapor and the liquid phase equal each other.
Accordingly, the mass flux is now a function of  $\mu_2-\mu_1$, i.e.,
\beq
\label{eq:kinetic-K3}
	\dot m=a (\mu_2-\mu_1)
\eeq
with $a\geq0$
that again is an agreement with the kinetic relation in \cite{Dreyer-Duderstadt-Hantke-Warnecke:2012}.

\begin{theorem}(Entropy production due to relaxation of chemical potentials)
\label{theorem:entropy-production-chemicval-potential-relaxation}
For a homogeneous mixture, i.e., $\bv_k=\bv=const$, $p_k=p=const$ and $T_k=T=const$,  the total entropy production
is
non-positive,
i.e.,
\begin{align}
\label{eq:2ndlawthermodynamics-K3}
 S^\mu_{U} - \Tref S^\mu_{S_M} \le 0,
\end{align}
if the relaxation parameter $\theta_\mu$
as well as the equilibrium temperature $T$ are positive.
\end{theorem}

\proof
For a homogeneous mixture the entropy production
\eref{thermodynamics-properties-entropy-relaxation-total-isothermal} reduces to
\[
  S^\mu_{U} =
  - \dot m \,\relaxG\,  (g_2-g_1) =
  - \dot m \,\relaxG\,
  \left(\mu_2-\mu_1+(\gamma_1-1)c_{v,1}\Tref\ln\left(\frac{\alpha_1}{\alpha_1 + \alpha_3}\right)\right)   .
\]
In analogy to the derivation of the entropy production, we determine the entropy production of the mixture entropy $S_M$
\beq
  S^\mu_{S_M} =
 -\dot m \,\relaxG\, (\gamma_1-1)c_{v,1} \ln\left(\frac{\alpha_1}{\alpha_1 + \alpha_3}\right)   .
\eeq
Then the total entropy production is given by
\beq
   S^\mu_{U} - \Tref S^\mu_{S_M} =   -\dot m \,\relaxG\,  \left(\mu_2-\mu_1\right)    .
\eeq
Thus, by \eref{eq:kinetic-K3} we conclude \eref{eq:2ndlawthermodynamics-K3}.
\proofendmy

\subsection{Isentropic case}
\label{subsec:bn-neq-thermo-isentrop}

In contrast to the isothermal case, we do not have to account for heat exchange in the energy equation.
Since for isentropic fluids the ``energies'' $\oee_k$ coincide with the energies $e_k$, see Remark \ref{rem:energy-barotropic-fluid},
we follow H{\'e}rard's original work  \cite{herard:2016}
for barotropic fluids
and choose
for the mathematical entropy function the total specific energy $E_k=e_k(\rho_k) + 0.5 \bv_k^2$ with internal energy $e_k$ determined by
\begin{align}
\label{eq:intener-k}
  e_k'(\rho_k) = \frac{p_k(\rho_k)}{\rho_k^2}
\quad  \mbox{or, equiv., }\quad
e_k'(\tau_k) = -p_k(\tau_k)
\end{align}
motivated by \eqref{thermo-energy}.
For smooth solutions of the model
\eref{eq:sa-model-alphak}, \eref{eq:sa-model}
we obtain the evolution equations for the fractional total energies
\begin{align}
\label{eq:sa-model-Ek}
\pdifft{(\alpha_k\,\rho_k\,E_k)} + \pdiv{(\alpha_k\,\rho_k\,\velo{k}\,(E_k+ p_k/\rho_k))} =
\osource{\alpha \rho E}{k} + \oPi_k ,
\end{align}
with relaxation term $\osource{\alpha \rho E}{k}$ and production term $\oPi_k$
\begin{align}
\label{thermodynamics-properties-entropy-relaxation}
\osource{\alpha \rho E}{k} &=
(e_k+p_k/\rho_k- 0.5 \bv_k^2)\source{\alpha \rho}{k} - p_k \source{\alpha}{k} +
\bv_k\cdot \Source{\alpha\rho\vects{\velo{}}}{k} , \\
\label{eq:sa-model-production-interfacial}
\oPi_k&=
p_k (\veloI-\bv_k) \cdot \nabla \alpha_k    - \sum_{l=1,\ne k}^{K} P_{k,l} \bv_k\cdot \nabla \alpha_l
\nonumber\\
& =
- \sum_{l=1,\ne k}^{K} \left( p_k (\veloI-\bv_k) + P_{k,l} \bv_k\ \right) \cdot \nabla \alpha_l .
\end{align}
We emphasize that the relaxation term $\osource{\alpha \rho E}{k}$ is different from the corresponding term determined by the relaxation models \eref{pres-relax-comp}, \eref{velo-relax-comp},
\eref{mass-relax-two-comp} and \eref{mass-relax-three-comp}.
It is interesting to note that in contrast to the single-phase case the energy equation is not redundant in the multi-component case due to the non-conservative products and the relaxation terms.
Similar to the isothermal case, see Eqn.~\eqref{eq:sa-model-full-heatflux-k}, we determine the heat flux as
\begin{align}
\pdiv{(\alpha_k\, \bq_k)} =
\sourcemy{\alpha\rho E}{k}-\osource{\alpha\rho E}{k}
-\sum_{l=1,\ne k}^{K} P_{k,l}\, \vprod{\veloI} \pgrad{\alpha_l}
-\oPi_{k}
\end{align}
comparing the difference of
\eqref{eq:sa-model-full-Ek-local} and \eqref{eq:sa-model-Ek}.

Summing these evolution equations over all components we obtain the evolution equation of the total energy
\begin{align}
\label{eq:sa-model-E}
\pdifft{\left(\sum_{k=1}^K \alpha_k\,\rho_k\,E_k\right)} +
\pdiv{\left(\sum_{k=1}^K  ( \alpha_k\,\rho_k\,\velo{k}\,(E_k+p_k/\rho_k) ) \right)} =
\sum_{k=1}^K\osource{\alpha \rho E}{k} + \sum_{k=1}^K\oPi_k ,
\end{align}
Comparing \eref{eq:sa-model-E} with \eref{sec:eep-4} we may choose for the  entropy-entropy flux pair
\begin{subequations}
  \label{sec:eep-isentropic}
\begin{align}
  \label{sec:eep-isentropic-a}
  &U(\bu) :=  \sum_{k=1}^{K} \alpha_k \rho_k E_k ,\\
  \label{sec:eep-isentropic-b}
  &F_i(\bu):=  \sum_{k=1}^K  ( \alpha_k\,\rho_k\,v_{k,i}\,(E_k+p_k/\rho_k) ),\ i=1,\ldots,d.
\end{align}
\end{subequations}
The compatibility conditions \eref{sec:eep-3} hold by derivation of \eref{eq:sa-model-E}.
In the isentropic case
non-strict convexity of $U$ has been proven in \cite{saleh-seguin:2020}.
Alternatively, the proof of Theorem \ref{theorem:Convexity} for the isothermal case
also applies in case of isentropic fluids where
\beq
 U(\bu) =  \sum_{k=1}^{K} \alpha_k \rho_k (e_k + \bv_k^2/2) = \sum_{k=1}^{K} \alpha_k U_k(\bu_k)
\eeq
with
$e_k=e_k(\tau_k,s_k)$ and $s_k=const$.
For the Hessian $\bU_k''$ we again obtain \eref{eq:Hessian-u-k}.
Then convexity of
$e_k$ with respect to $\tau_k$
according to \eref{thermo-energy-convex} ensures that the Hessian is positive semi-definite.

We emphasize that in case of an isothermal fluid the mixture total energy is \textit{not} an admissible entropy function because \eref{eq:sa-model-Ek} does not account for heat exchange.
To ensure the entropy inequality \eref{sec:eep-4n} we have to verify that the right-hand side of \eref{eq:sa-model-E} is not positive.
This will provide us with admissibility criteria for the interfacial pressures as well as the relaxation terms.

\subsubsection{Entropy production due to interfacial velocity and pressure}
\label{thermodynamics-properties-interfacial-isentropic}


Obviously, we cannot control the sign of $\Pi_k$. Therefore we determine
the interfacial pressures $P_{k,l}$ and the interfacial velocity $\veloI$ such
that the sum $\oPi:=\sum_{k=1}^{K} \oPi_k$ vanishes and the conservation constraints \eref{eq:constraints-interfacial-pressures-gen} or, equivalently, \eref{eq:constraints-interfacial-pressures} hold.
Assuming that the interfacial velocity can be written as a linear combination of the velocities of the components according to \eref{thermodynamics-properties-interfacial-2} there exists a unique closure for the interfacial pressures.

\begin{theorem}(Entropy production due to interfacial states)
\label{theorem:entropy-production-interfacial-states-isentropic}
For any convex combination \eref{thermodynamics-properties-interfacial-2} for the interfacial velocity $\veloI$
there uniquely exist  interfacial pressures
\begin{equation}
\label{theorem:entropy-production-interfacial-states-Pkl-isentropic}
P_{k,l} =   \beta_k p_l  + p_k (1- \beta_k)
\end{equation}
such that the production term $\oPi=\sum_{k=1}^{K} \oPi_k$ vanishes and
the conservation constraints \eref{eq:constraints-interfacial-pressures-gen} or, equivalently, \eref{eq:constraints-interfacial-pressures} hold. In particular, the interfacial pressures are all positive and
\begin{equation}
\label{theorem:entropy-production-interfacial-states-PI-isentropic}
  P_I =   \sum_{k=1}^K  p_k (1- \beta_k) .
\end{equation}
\end{theorem}

\proof

First of all, we note that by the conservation constraint \eref{eq:constraints-interfacial-pressures-gen} and the definition \eref{eq:sa-model-production-interfacial}  of $\oPi_k$ it holds
\begin{align}
\oPi = \sum_{k=1}^K \oPi_k = \sum_{k=1}^K \sum_{l=1,\ne k}^K  (P_{k,l}-p_k) (\veloI-\bv_k) \cdot \pgrad{\alpha_l} .
\end{align}
Obviously, this term coincides with $\Tref \Pi= \sum_{k=1}^K \Pi_k$ with $\Pi_k$ determined by \eqref{eq:entropy-prod-k} in the isothermal case.
Since by the saturation constraint \eref{eq:saturation} we may substitute
$\pgrad{\alpha_K}$ by the linearly independent gradients
$\pgrad{\alpha_k}$, $k=1,\ldots,K-1$, we may rewrite $\oPi$ in terms of the linearly independent gradients.
Then $\oPi$ vanishes if and only if the cofactors vanish, i.e.,
\beq
  \label{thermodynamics-properties-interfacial-1}
    \sum_{k=1,k\ne l}^{K-1} \left( (P_{k,l}-p_k)  -(P_{k,K}-p_k) \right) (\velo{k} -\veloI) +
     (P_{K,l}-p_K) (\velo{K} -\veloI) = \bzero, \quad l=1,\ldots, K-1 .
\eeq
Next we use that
the interfacial velocity $\veloI$ is a convex combination of the single component velocities $\velo{k}$, i.e.,
\eref{thermodynamics-properties-interfacial-2} holds. Then we may rewrite the velocity differences $\velo{k} -\veloI$ in terms of the  differences $\velo{i}-\velo{i+1}$, $i=1,\ldots,K-1$.
Rearranging \eref{thermodynamics-properties-interfacial-1} in terms of these
yields $(K-1)^2$ equations for the $(K-1)K$ unknowns
$P_{k,l}$, $k,l=1,\ldots,K$, $k\ne l$. Additional $K-1$ equations are provided by the conservation
constraints \eref{eq:constraints-interfacial-pressures}. Altogether we obtain a linear system of equations for the interfacial pressures. 
This system is identical to the linear system derived in \cite{Mueller-Hantke-Richter:16}, Eqns.~(6.20) and (6.22), where the temperatures are formally replaced by 1. 
This system has a unique solution. The proof is identical to the one of the analogue theorem in \cite{Mueller-Hantke-Richter:16} in case of
the full Baer-Nunziato model.

\proofendmy

We would like to mention that in \cite{herard:2016} the interfacial pressures in case of
barotropic fluids
 are given for $K=2,3,4$. The proof is shortly sketched without investigating the regularity of the resulting linear system.
Furthermore, as in the isothermal case the interfacial pressures \eref{theorem:entropy-production-interfacial-states-Pkl-isentropic} coincide with those presented in
\cite{Mueller-Hantke-Richter:16} for
the full multi-component Baer-Nunziato model
at local thermal equilibrium.
Thus, using for the convex combination \eref{thermodynamics-properties-interfacial-2} the coefficients
\eref{thermodynamics-properties-interfacial-2a},
then the interfacial pressure and velocity again are given by
\eref{theroem:entropy-production-interfacial-states-special-3-isothermal}, \eref{theroem:entropy-production-interfacial-states-special-1-isothermal} and \eref{theroem:entropy-production-interfacial-states-special-2-isothermal}.

\subsubsection{Entropy production due to relaxation}
\label{sec:relax-terms-isentropic}

To be admissible with the
entropy inequality \eref{sec:eep-4n}  the  production $\sum_{k=1}^{K} \osource{\alpha\rho E}{k}$ has to be  non-positive.
Note that in abuse of physical notation we call this term entropy production rather than energy production with respect to the mathematical concept of entropy-entropy flux pairs because energy is known to be conserved.
In the following we consider one-by-one the contributions of 
pressure and velocity relaxation.
The analysis is similar to \cite{Mueller-Hantke-Richter:16} in case of the physical entropy production.
Note that  all the relaxation processes are in agreement with
the constraints due to conservation \eref{eq:constraints}.

\begin{theorem}(Entropy production due to pressure relaxation)
\label{theorem:entropy-production-pressure-relaxation-isentropic-herard}
Let the relaxation parameter be non-negative, i.e., $\theta_p\ge 0$. Furthermore, let
the coefficients $d_{k,l}$ be positive and symmetric, i.e.
\begin{align}
  \label{eq:pres-relaxation-entropy-constraint-isentropic}
  0 \le d_{k,l}=d_{l,k},\quad l,k=1,\ldots,K,\ l\ne k.
\end{align}
Then the entropy production due to
pressure relaxation \eref{pres-relax-comp-herard} is non-positive, i.e.,
\begin{equation}
\label{eq:relaxation-entropy-constraints-pressure-relaxation-isentropic}
   S_{U}^{p} =    \sum_{k=1}^{K} \sourcemy{U}{k}^{p} \le 0
\end{equation}
with  entropy production of component $k=1,\ldots,K$
\begin{align}
\label{eq:partial-energy-production-pressure-relaxation-isentropic}
  \sourcemy{U}{k}^{p}= \osource{\alpha\rho E}{k}^{p}  = -\theta_p p_k \sum_{l=1}^{K} d_{k,l} (p_k-p_l) .
\end{align}
In particular, this holds true for the  model \eref{pres-relax-comp}.
\end{theorem}

\proof
First of all, we note that by symmetry of arbitrary  coefficients $a_{k,l}=a_{l,k}$ it holds
\begin{align*}
\sum_{k=1}^{K-1} \sum_{l=k+1}^{K} a_{k,l} =
\sum_{l=2}^{K} \sum_{k=1}^{l-1} a_{k,l} =
\sum_{k=2}^{K} \sum_{l=1}^{k-1} a_{l,k}
\end{align*}
From this we conclude with $a_{k,l}=p_k^2 d_{k,l}$ and $a_{k,l}=p_k d_{k,l} p_l$
\begin{align*}
 & \sum_{l,k=1,l\ne k}^K p_k^2 d_{k,l} =
  \sum_{k=2}^{K} \sum_{l=1}^{k-1} p_k^2 d_{k,l} + \sum_{k=1}^{K-1} \sum_{l=k+1}^{K} p_k^2 d_{k,l} =
  \sum_{k=2}^{K} \sum_{l=1}^{k-1} (p_k^2 + p_l^2) d_{k,l}, \\
 & \sum_{l,k=1,l\ne k}^K p_k p_l d_{k,l} =
  \sum_{k=2}^{K} \sum_{l=1}^{k-1} p_k p_l d_{k,l} + \sum_{k=1}^{K-1} \sum_{l=k+1}^{K} p_k  p_l d_{k,l} =
  2 \sum_{k=2}^{K} \sum_{l=1}^{k-1} p_k p_l d_{k,l} .
\end{align*}
Combining these results we obtain
\begin{align*}
  \sum_{l,k=1,l\ne k}^K p_k d_{k,l} (p_k-p_l) =
  \sum_{k=2}^{K} \sum_{l=1}^{k-1} d_{k,l} (p_k^2 - 2 p_k p_l + p_l^2 ) =
  \sum_{k=2}^{K} \sum_{l=1}^{k-1} d_{k,l} (p_k -p_l )^2 .
\end{align*}
Obviously, this term is non-negative and vanishes only if all pressures coincide. Thus, by assumption
the energy production $\osource{\alpha\rho E}{k}^{p}$ is non-positive.
In particular, this holds true for the special model  \eref{pres-relax-comp} because $d_{k,l}=\alpha_k \alpha_l$ satisfies the constraint \eref{eq:pres-relaxation-entropy-constraint-isentropic}

\proofendmy

\noindent
Since the proof holds for an arbitrary number of components $K$  the theorem generalizes the result in \cite{herard:2016} for $K=4$. For $K=2$ the result is well-known.

For the velocity relaxation we obtain a similar result.

\begin{theorem}(Entropy production due to velocity relaxation)
\label{theorem:entropy-production-velocity-relaxation-isentropic-herard}
Let the relaxation parameter be non-negative, i.e., $\theta_v\ge 0$. Furthermore, let
 the coefficients $e_{k,l}$ be non-negative and symmetric, i.e.
\begin{align}
  \label{eq:velo-relaxation-entropy-constraint-isentropic}
  0 \le e_{k,l}=e_{l,k},\quad l,k=1,\ldots,K,\ l\ne k.
\end{align}
Then the entropy production due to
velocity relaxation \eref{velo-relax-comp-herard} is non-positive, i.e.,
\begin{equation}
\label{eq:relaxation-entropy-constraints-velocity-relaxation-isentropic}
   S_{U}^{v} =    \sum_{k=1}^{K} \sourcemy{U}{k}^{v} \le 0
\end{equation}
with  entropy production of component $k=1,\ldots,K$
\begin{align}
\label{eq:partial-energy-production-velocity-relaxation-isentropic}
  \sourcemy{U}{k}^{v} = \osource{\alpha\rho E}{k}^{v}  = -\theta_v  \sum_{l=1}^{K} e_{k,l} \, \velo{k}\cdot(\velo{k}-\velo{l})  .
\end{align}
In particular, this holds true for the model \eref{velo-relax-comp}.
\end{theorem}

\proof
The proof is in full analogy to the pressure relaxation where we replace $d_{k,l}$, $p_k$ and $p_l$ by
$e_{k,l}$, $v_{k,i}$ and $v_{l,i}$, respectively. Then we obtain
\begin{align*}
  \sum_{l,k=1,l\ne k}^K  e_{k,l} \velo{k}\cdot (\velo{k}-\velo{l}) =
  \sum_{i=1}^d \sum_{k=2}^{K} \sum_{l=1}^{k-1} e_{k,l} (v_{k,i} -v_{l,i} )^2 =
   \sum_{k=2}^{K} \sum_{l=1}^{k-1} e_{k,l} (\velo{k}-\velo{l} )^2 .
\end{align*}
Obviously, this term is non-negative and vanishes only if all velocities coincide. Thus, by assumption
the energy production $\osource{\alpha\rho E}{k}^{v}$ is non-positive.
In particular, this holds true for the special model  \eref{velo-relax-comp} because $e_{k,l} = \alpha_k \rho_k \alpha_l \rho_l/\sum_{j=1}^K \alpha_j \rho_j$ satisfies the constraint \eref{eq:velo-relaxation-entropy-constraint-isentropic}.

\proofendmy

Since there is no temperature relaxation and no relaxation of chemical potentials, there is no corresponding energy production.

\section{Relaxation to equilibrium}
\label{subsec:relax}

The main objective of our work is to develop new, efficient and robust relaxation procedures. For this purpose
we split the relaxation process from the fluid motion by formally performing an operator splitting.
Since the relaxation times differ for the different relaxation types,
we solve the initial value problem
\beq
\label{eq:sa-model-vec-relax-op-type}
\odifft{\bw(t)} =
 \bS^{\xi}(\bw(t)),\quad t\in [t_n,t_{n+1}],
\qquad \bw(t_n) = \bw^{0}
\eeq
with
$\bS^{\xi} = ((S^{\xi}_{\alpha_k})_{k=1,\ldots,K-1}, (S^{\xi}_{\rho,k},\bS^{\xi}_{\rho\bv,k})_{k=1,\ldots,K})^T$
separately for each relaxation type $\xi$.
To avoid the explicit computation of the relaxation times $\relaxall$ in
\eref{pres-relax-comp}, \eref{velo-relax-comp}, \eref{mass-relax-two-comp} and \eref{mass-relax-three-comp},
we may perform the change of variables
$\ot:= (t_n-t)\theta_{\xi}$ and $\obw(\ot) := \bw(t)$ in
\eref{eq:sa-model-vec-relax-op-type}, i.e.,
\beq
\label{eq:sa-model-vec-relax-op-type-mod}
\odiff{\obw(\ot)}{\ot} =
 \frac{1}{\theta^{\xi}} \bS^{\xi}(\obw(\ot)),\quad \ot\in [0,\Delta t\theta_{\xi} ],\qquad
\obw(0) =
\obw^0.
\eeq
Then by definition of the relaxation terms, the relaxation parameter cancels on the right-hand side
of \eref{eq:sa-model-vec-relax-op-type-mod}. Since for all relaxation processes the
conservation constraints \eref{eq:constraints} are satisfied
we conclude from \eref{eq:sa-model-vec-relax-op-type} that
\beq
\label{eq:sa-model-vec-relax-bulk}
\sum_{k=1}^{K}\odiff{\oalpha_k(\ot)}{\ot} = 0,\
\odiff{\orho(\ot)}{\ot} = 0,\
\odiff{(\orho\,\obv)(\ot)}{\ot} = \bzero.
\eeq
Hence, the bulk quantities for mixture density, momentum and energy
as well as the saturation condition \eref{eq:saturation}
remain constant during the relaxation process and it holds
\beq
\label{eq:sa-model-vec-relax-bulk-equilibrium}
\sum_{k=1}^{K}\alpha_k^{\infty} = \sum_{k=1}^{K}\alpha_k^{0} = 1,\
\rho^{\infty} = \rho^{0},\
(\rho\,\bv)^{\infty} = (\rho\,\bv)^{0}.
\eeq
Furthermore, the conservation of bulk mass and momentum imply that the bulk velocity remains constant
\begin{equation}
  \label{eq:ode-interfacial-velo}
  \odiff{\obv }{\ot} =\bzero
\end{equation}
and, thus, it holds
\beq
\label{eq:sa-model-vec-relax-interfacial-velocity-equilibrium}
 \bv^{\infty} = \bv^{0} .
\eeq
We are only interested in the equilibrium state rather the transient relaxation behavior
as considered, for instance, in \cite{Herard--Jomee:23,Herard--Jomee:23-preprint}. 
Therefore
we assume that the relaxation process is infinitely fast, i.e.,
$\theta_{\xi}\to\infty$ and we do not need to determine the relaxation parameters.
To compute the equilibrium state, where the source terms \eref{pres-relax-comp}, \eref{velo-relax-comp},
\eref{mass-relax-two-comp} and \eref{mass-relax-three-comp} vanish, we perform integration of
the ODE system to infinity. This results in a system of algebraic equations for the equilibrium state that will be
derived and solved in the subsequent sections. For ease of notation we will use $\bw$ instead of
$\obw$ in \eref{eq:sa-model-vec-relax-op-type-mod}.
In the following we consider one by one the different relaxation
processes to equilibrium. Note that the equilibrium state is independent of the order of the relaxation procedures.

\textit{Velocity relaxation}.
First of all, we investigate velocity relaxation. This relaxation process is the same for isentropic and isothermal fluids. The procedure to determine the velocity equilibrium we  proceed similar to Saurel and Le M\'etayer in
\cite{Saurel-Le} for
the full Baer-Nunziato model.
The equilibrium state of the velocity relaxation process is determined by solving the  system of ODEs
\begin{subequations}
\begin{align}
  \label{eq:velo-ode-1}
  &
  \odiff{\alpha_k}{\ot}
   = 0,\\
  \label{eq:velo-ode-2}
  &
  \odiff{\alpha_k\,\rho_k}{\ot}
   = 0,\\
  \label{eq:velo-ode-3}
  &
  \odiff{\alpha_k\,\rho_k\, \bv_k}{\ot}
   = \alpha_k\,\rho_k(\velo{}-\bv_k)  
\end{align}
\end{subequations}
resulting from
\eref{eq:sa-model-vec-relax-op-type-mod}
with the source terms \eref{velo-relax-comp}. Since we assume that the relaxation process is infinitely fast,
the solution of the system of ODEs converges towards the equilibrium state
where the right-hand side  vanishes. This holds true for
\begin{equation}
  \label{eq:velo-eq}
 \bv_k = \velo{}^{\infty},\quad
 \ot\to\infty.
\end{equation}
Integration of \eref{eq:velo-ode-1}, \eref{eq:velo-ode-2} over $[0,\infty]$
then results in the algebraic equations
\begin{subequations}
\label{eq:velo-algebraic}
\begin{align}
  \label{eq:velo-algebraic-1}
  &\alpha_k^\infty = \alpha_k^0 ,\\
  \label{eq:velo-algebraic-2}
  &\alpha_k^\infty\,\rho_k^\infty = \alpha_k^0\,\rho_k^0
  \quad\mbox{equiv.}\quad  \rho_k^\infty = \rho_k^0.
\end{align}
\end{subequations}
From the equilibrium condition \eref{eq:velo-eq} and
\eref{eq:sa-model-vec-relax-interfacial-velocity-equilibrium} we conclude that
\begin{equation}
  \label{eq:velo-algebraic-4}
  \bv_k^\infty = \velo{}^\infty = \velo{}^0  .
\end{equation}
Finally, we end up with the algebraic system
\eref{eq:sa-model-vec-relax-interfacial-velocity-equilibrium},
\eref{eq:velo-algebraic-1}, \eref{eq:velo-algebraic-2} and
\eref{eq:velo-algebraic-4}
by which we determine the velocity equilibrium state.
The procedure is summarized in the following algorithm.
\begin{algorithm}[Velocity relaxation procedure]~
\label{alg:velo-relax}
 \begin{enumerate}
   \item  Compute the equilibrium mixture velocity according to \eref{eq:mixture-cons} as
          \beq
            \bv^\infty = \left.\sum_{k=1}^{K} \alpha_k^0 \rho_k^0 \bv_k^0 \right/ \sum_{k=1}^{K} \alpha_k^0 \rho_k^0 ;
          \eeq
   \item Update the velocities of the components  using \eref{eq:velo-algebraic-4};
 \end{enumerate}
\end{algorithm}

For pressure relaxation, thermal relaxation and chemical relaxation we have to distinguish between isentropic fluids and isothermal fluids.

\subsection{Isothermal case}
\label{subsubsec:relax-isothermal}

In
this section
  we always assume for each component an isothermal stiffened gas, see Example \ref{ex:isotherm-stiff-gas-Eos}.

\subsubsection{Pressure relaxation}\label{trelax1}

For the pressure relaxation process we have to solve the system of ODEs
\begin{subequations}
\label{relax-pres-ODE}
\begin{align}
&\odiff{ \alpha_k}{\ot}=\alpha_k\,(p_k- p),\label{eq:relax:tem:al}\\
&\odiff{ \alpha_k\,\rho_k}{\ot}=0,\label{eq:relax:tem:alrho}\\
&\odiff{ \alpha_k\,\rho_k\,\bv_k}{\ot}=0\label{eq:relax:tem:alrhou}
\end{align}
\end{subequations}
resulting from \eref{eq:sa-model-vec-relax-op-type-mod} with the source terms
\eref{pres-relax-comp}. Since we assume that the relaxation process is infinitely fast,
the solution of the system of ODEs converges towards the equilibrium state where the right-hand side
vanishes. This holds true for
\begin{equation}
  \label{eq:temp-eq}
  p_k^{\infty} = p^\infty,\qquad  \ot\to\infty.
\end{equation}
Integration of \eref{eq:relax:tem:alrho}, \eref{eq:relax:tem:alrhou}  over $[0,\infty]$
then results in the algebraic equations
\begin{subequations}
\label{eq:pres-algebraic}
\begin{align}
	\label{masscons}
	&\alpha_k^\infty\rho_k^\infty=\alpha_k^0\rho_k^0,\\
	&\mathbf{v}_k^\infty=\mathbf{v}_k^0,\, k=1,\dots,K.
\end{align}
\end{subequations}
In the equilibrium state it must hold
\begin{align}\label{eqcondT}
	p_k^\infty=p^\infty,\, k=1,\dots,K.
\end{align}
In addition, for an isothermal fluid all temperatures are in equilibrium, i.e., $T_k=\Tref$, $k=1,\dots,K$.
Assume all components to be stiffened gases than it holds
\begin{align}\label{eosT}
	p_k=C_k\rho_k-\pi_k,
\end{align}
where $C_k=\cisoT_k^2=T c_{v,k}(\gamma_k-1)>0$. From the saturation condition
\begin{align}\label{satcond}
	\sum_{k=1}^K\alpha_k=1
\end{align}
and mass conservation \eqref{masscons} we obtain
\begin{align}\label{polrhoT}
	\prod_{k=1}^K\rho_k^\infty=\sum_{k=1}^K(\alpha_k^0\rho_k^0)\prod_{j=1,j\ne k}^K\rho_j^\infty.
\end{align}
Let us fix an arbitrary component $k_0$. Then we obtain from \eqref{eosT}
\begin{align}\label{denT}
  \rho_k^\infty=
  \frac{C_{k_0}}{C_k}\rho_{k_0}^\infty
   +\frac{\pi_k-\pi_{k_0}}{C_k},\,
  k=1,\dots,K,\, k\ne k_0.
\end{align}
Putting \eqref{denT} into \eqref{polrhoT} we get a polynomial for $\rho_{k_0}^\infty$
\begin{align}
\label{eq:polynomial-rho-T}
  \prod_{k=1}^K\left( \frac{C_{k_0}}{C_k}\rho_{k_0}^\infty+\frac{\pi_k-\pi_{k_0}}{C_k} \right) =
  \sum_{k=1}^K(\alpha_k^0\rho_k^0) \prod_{j=1,j\ne k}^K\left(\frac{C_{k_0}}{C_j}\rho_{k_0}^\infty+\frac{\pi_j-\pi_{k_0}}{C_j} \right).
\end{align}
In order to obtain a polynomial of smallest possible degree we introduce the following sets.
\begin{align}
\label{eq:setN}
{\cal{E}}_{k_0}:=\{k\,:\,\pi_k=\pi_{k_0},\,k\in\{1,\dots,K\}\} ,\quad
{\cal{N}}_{k_0}=\{k\,:\,\pi_k\ne \pi_{k_0},\,k\in\{1,\dots,K\}\} .
\end{align}
Depending on the choice of $k_0$ the polynomial to determine the equilibrium density $\rho_{k_0}^\infty$ is of degree $\#{\cal{N}}_{k_0}+1$. Accordingly we recommend $k_0$ to be chosen such that the cardinality of ${\cal{N}}_{k_0}$ is smallest.

\textbf{Special case 1.}
Let us assume that $\pi_1=\pi_2\dots=\pi_K=:\pi$.
Then there exists only one root of the polynomial
\eref{eq:polynomial-rho-T}. Applying \eref{denT} the equilibrium phasic densities $\rho_j^{\infty}$ are determined by
\begin{align}
	\label{eqdens1}
	\rho_j^\infty=
        \frac{1}{C_j}\sum_{k=1}^K\alpha_k^0\rho_k^0C_k=
        \frac{\sum_{k=1}^K\alpha_k^0 (p_k^0+\pi)}{C_j} \quad \mbox{for }j=1,\dots,K.
\end{align}
The equilibrium volume fractions can be calculated employing mass conservation \eref{masscons}.

\textbf{Special case 2.}
Let us assume that $K=2$ with $\pi_1=0$ and $\pi_2\ne0$.
Then it holds
\begin{subequations}
\label{eqdens2}
\begin{align}
\label{eqdens2-expr1}
&X_1
=\frac{-\pi_2+\sum_{k=1}^2\alpha_k^0\rho_k^0C_k}{C_1},\\
\label{eqdens2-expr2}
&X_2
=-\frac{\pi_2}{C_1}(\alpha_1^0\rho_1^0),\\
\label{eqdens2-roots}
&	\rho_{1,\pm}^\infty=\frac{X_1}{2}\pm \sqrt{\frac{X_1^2}{4}-X_2}.
\end{align}
\end{subequations}
Because $-\pi_2<0$ one has to choose the upper sign, i.e., $\rho_1^\infty=\rho_{1,+}^\infty$. Using the saturation
condition \eqref{satcond} and mass conservation \eqref{masscons}
one can find the equilibrium volume fractions and finally $\rho_2^\infty$.

\textbf{Special case 3.}
Let us assume that $K=3$ with $\pi_1=\pi_3=0$ and $\pi_2\ne0$.
Then it holds
\begin{subequations}
\label{eqdens3}
\begin{align}
\label{eqdens3-expr1}
&X_1
=\frac{-\pi_2+\sum_{k=1}^3\alpha_k^0\rho_k^0C_k}{C_1},\\
\label{eqdens3-expr2}
&X_2
=-\frac{\pi_2}{C_1^2}(\alpha_1^0\rho_1^0C_1+\alpha_3^0\rho_3^0C_3),\\
\label{eqdens3-roots}
&	\rho_{1,\pm}^\infty=\frac{X_1}{2}\pm \sqrt{\frac{X_1^2}{4}-X_2}.
\end{align}
\end{subequations}
Because $-\pi_2<0$
one has to choose the upper sign. Analogously one can find $\rho_3^*$. Again, using the saturation condition
\eqref{satcond} and mass conservation \eqref{masscons}
one can find the equilibrium volume fractions and finally $\rho_2^\infty$.

Finally, after having determined the equilibrium phasic masses the equilibrium pressure can be computed by
\begin{align}
	p^\infty=\sum_{k=1}^K\alpha_k^\infty p_k^\infty =
       \sum_{k=1}^K\alpha_k^\infty \left( C_k\rho_k^\infty-\pi_k \right) =
       \sum_{k=1}^K \left( C_k\alpha_k^0\rho_k^0-\alpha_k^\infty \pi_k \right) .
	\label{eqpres1}
\end{align}

For convenience of the reader we summarize the relaxation procedure  in the following algorithm.
\begin{algorithm}[Pressure relaxation procedure in case of isothermal stiffened gases]~
\label{alg:pres-relax-isothermal-stiffgas}
 \begin{enumerate}
 \item Determine the index $k_0$  such that
    the cardinality of the set ${\cal N}_{k_0}$  determined by \eref{eq:setN} is smallest.
 \item Determine the roots $\rho_{k_0}^{\infty,i}$, $i=1,\ldots,\#{\cal N}_{k_0}+1$,  of the polynomial
	\eref{eq:polynomial-rho-T};
 \item Single out the unique physically admissible solution by verifying the conditions
	 $\alpha_k^{\infty,i}\in(0,1)$ and
   $\rho_k^{\infty,i}>0$ for all $k=1,\ldots, K$ where the volume fractions and phasic densities are determined by
	 \eref{denT} and \eref{masscons};
 \item By means of this solution  update the phasic densities $\rho_k^{\infty}$ and volume fractions
   $\alpha_k^{\infty}$ using \eref{denT} and \eref{masscons}, respectively.
 \item Compute the equilibrium pressure using \eref{eqpres1};
 \end{enumerate}
\end{algorithm}

For the special cases discussed above we can prove the existence and uniqueness of admissible equilibrium states,
see Step 4 in the above algorithm.
\begin{theorem}[Properties of equilibrium state]~\\
For all of the three cases
(1) $K$ components with $\pi_1= \ldots = \pi_K$
(2) two components with $\pi_2\ne \pi_1 =0$,
(3) three components with $\pi_1=\pi_2\ne\pi_3$
there exist  unique  equilibrium pressure, equilibrium volume fractions and equilibrium phasic densities with
$p^\infty>-\pi$, $\alpha_k^\infty\in[0,1]$  and $\rho_k^\infty>0$ with
minimal pressure $\pi:=\max\{\pi_k\,;\, k=1,\ldots,K\}$
provided that $p_k^0 > -\pi_k$, $k=1,\ldots, K$.
\end{theorem}
\noindent
\textbf{Proof}: \\
{\bf Case 1}: In this case there only exists one root $\rho_{k_0}^\infty$  of \eref{eq:polynomial-rho-T} for arbitrary but fixed $k_0\in\{1,\ldots,L\}$. By means of \eref{denT} the remaining equilibrium phasic densities are determined by \eref{eqdens1}. All of them are positive. Furthermore, the equilibrium volume fractions are determined by the mass conservation \eref{masscons} as
\begin{align*}
\alpha_j^\infty =
\frac{\alpha_j^0 \rho_j^0}{\rho_j^\infty} =
\frac{\alpha_j^0 \rho_j^0}{\frac{1}{C_j}\sum_{k=1}^K\alpha_k^0\rho_k^0C_k} \in [0,1].
\end{align*}

{\bf Case 2}: In this case  $\pi_2\ne\pi_1=0$. Since the discriminant in \eref{eqdens2-roots} is positive,
there exist two roots of the polynomial \eref{eq:polynomial-rho-T} from which we choose $\rho_1^\infty=\rho_{1,+}^\infty$. By \eref{eosT} we can rewrite the expression \eref{eqdens2-expr1} as
\begin{align*}
X_1
=\frac{\alpha_1^0\rho_1^0 C_1 + \alpha_2^0 \,(p_2^0+\pi_2)}{C_1} >0
\end{align*}
that is positive if $p_2$ is larger than the minimal pressure $-\pi_2$. This implies
$\rho_1^\infty \ge X_1 >0$ a
and, thus, because of \eref{denT}
\begin{align*}
  \rho_2^\infty =
  \frac{C_1}{C_2} \rho_1^\infty + \frac{\pi_2}{C_2}   >0.
\end{align*}
For the equilibrium volume fractions we conclude using mass conservation \eref{masscons} and saturation \eqref{satcond}
\begin{align*}
  &0 \le  \alpha_1^\infty = \frac{\alpha_1^0 \rho_1^0}{\rho_1^\infty} <
  \frac{\alpha_1^0 \rho_1^0 C_1}{\alpha_1^0 \rho_1^0 C_1 + \alpha_2^0\,(p_2^0+\pi_2)} \le 1,\\
  &0 \le \alpha_2^\infty = 1- \alpha_1^\infty\le 1.
\end{align*}

{\bf Case 3}: In this case  we assume that $\pi_1=\pi_2\ne\pi_3$. Since the discriminant in \eref{eqdens3-roots} is positive,
there exist two roots of the polynomial \eref{eq:polynomial-rho-T} from which we choose $\rho_1^\infty=\rho_{1,+}^\infty$. By \eref{eosT} we can rewrite the expression \eref{eqdens3-expr1} as
\begin{align*}
X_1
=\frac{\alpha_1^0\rho_1^0 C_1 + \alpha_3^0\rho_3^0 C_3 +\alpha_2^0 \,(p_2^0+\pi_2)}{C_1} >0
\end{align*}
that is positive if $p_2$ is larger than the minimal pressure $-\pi_2$. This implies
$\rho_1^\infty \ge X_1  >0$
and, thus, because of \eref{denT}
\begin{align*}
  \rho_2^\infty = \frac{C_1}{C_2} \rho_1^\infty + \frac{\pi_2}{C_2}   >0,\quad
  \rho_3^\infty = \frac{C_1}{C_3} \rho_1^\infty > 0.
\end{align*}
For the equilibrium volume fractions we conclude using mass conservation \eref{masscons} and saturation \eqref{satcond}
\begin{align*}
  &0 \le  \alpha_1^\infty = \frac{\alpha_1^0 \rho_1^0}{\rho_1^\infty} <
  \frac{\alpha_1^0 \rho_1^0 C_1}{\alpha_1^0 \rho_1^0 C_1 + \alpha_3^0 \rho_3^0 C_3 + \alpha_2^0\,(p_2^0+\pi_2)} \le 1,\\
  &0 \le  \alpha_3^\infty = \frac{\alpha_3^0 \rho_3^0}{\rho_3^\infty}
    =  \frac{\alpha_3^0 \rho_3^0 C_3}{C_1 \rho_1^\infty} <
  \frac{\alpha_3^0 \rho_3^0 C_3}{\alpha_1^0 \rho_1^0 C_1 + \alpha_3^0 \rho_3^0 C_3 + \alpha_2^0\,(p_2^0+\pi_2)} \le 1,\\
  &0 \le \alpha_2^\infty = 1- \alpha_1^\infty  - \alpha_3^\infty \le 1.
\end{align*}

Finally, for all cases we conclude
for the equilibrium pressure \eref{eqpres1}  with \eref{eosT} and mass conservation \eref{masscons}
\begin{align*}
p^\infty =
\sum_{k=1}^K \left( \alpha_k^0 p_k^0 - (\alpha_k^\infty -\alpha_k^0) \pi_k \right).
\end{align*}
Assuming that the phasic pressures are larger than the minimal pressures, i.e.,
$p_k^0 > -\pi_k$,
then we can estimate  the equilibrium pressure by $-\pi$ using the saturation \eqref{satcond}.


\subsubsection{Relaxation of chemical potentials }
\label{sec:chem-relax}

For two components the equilibrium state of the chemical potential relaxation process is
determined by solving the system of ODEs
\begin{eqnarray}
	\label{eq:chem-ode-2-1}
	\hspace*{-16mm}&&\odiff{\alpha_k}{\ot} = (-1)^{k+1} \frac{\dot m}{\varrho},\ k=1,2,
	\\
	\label{eq:chem-ode-2-2}
	\hspace*{-16mm}&&\odiff{\alpha_k\,\rho_k}{\ot} = (-1)^{k+1}\dot m,\ k=1,2,
	\\
	\label{eq:chem-ode-2-3}
	\hspace*{-16mm}&&\odiff{\alpha_k\,\rho_k\, \bv_k}{\ot} = (-1)^{k+1}\dot m\,\bV_I,\ k=1,2.
\end{eqnarray}
resulting from \eref{eq:sa-model-vec-relax-op-type-mod} with the source terms \eref{mass-relax-two-comp}.
Again we assume that the relaxation process is infinitely fast
such that the solution of the system of ODEs
converges towards the equilibrium state,
where the right-hand side vanishes.
According to \cite{Zein:2010,Zein-Hantke-Warnecke:2013} this holds true for
\begin{eqnarray}
	\label{peqmu2}
	p_1^{\infty}=p_2^{\infty}&=&p^{\infty},\quad\ot\to\infty,\\
	\label{mueq2}
	g_{1}^{\infty}&=&g_2^{\infty},\quad \ot\to\infty.
\end{eqnarray}
This implies $\dot{m}=\dot{m}(g_2-g_1)=0$ and $p^{\infty}=p(\Tref)$
as well as $\rho_k^{\infty}=\rho_{k,ref}$.
Here $p(\Tref)$ denotes the pressure at saturation temperature and $\rho_{k,ref}$ the corresponding densities. Since we consider isothermal fluids, the temperature uniquely defines the equilibrium state on the saturation line where both phases coexist.
Note that an equilibrium state does not necessarily exist.
This can easily be checked as follows.

The saturation condition and conservation of mass imply
\begin{equation}
	\label{eq:mass-con-2}
	M:=\alpha_1^{0}\rho_1^{0}+(1-\alpha_1^{0})\rho_2^{0}=\alpha_1^{\infty}\rho_1^{\infty}+(1-\alpha_1^{\infty})\rho_2^{\infty}\,.
\end{equation}
Then one of the three cases applies:
(i) if \eqref{eq:mass-con-2} has a solution $\alpha_1^{\infty}\in(0,1)$, then an equilibrium state exists;
(ii)  for non-positive solutions the vapor phase
completely vanishes;
(iii) otherwise the liquid phase will be extinct.
In the latter two cases we set the volume fraction of the vanishing phase to a small positive value $\varepsilon$ and determine the densities using pressure equilibrium $p_1^\infty=p_2^\infty$.

The full relaxation procedure is given in the following algorithm.
\begin{algorithm}[Chemical Relaxation Procedure for $K=2$]~
	\label{alg:chem2}
	\begin{enumerate}
		\item
		Solve equation \eref{eq:mass-con-2} to identify (i) equilibrium, (ii) total condensation or (iii) total evaporation case
		\item
		Determine the missing quantities
		\begin{itemize}
			\item
			if (i): $\alpha_1^{\infty}$ is the solution of \eref{eq:mass-con-2}, $p^{\infty}=p(\Tref)$, $\rho_k^{\infty}=\rho_{k,ref}$, $k=1,2$
	        \item
	        if (ii): $\alpha_1^{\infty}:=\varepsilon$, $\alpha_2^{\infty}=1-\alpha_1^{\infty}$,
$\rho_1^\infty=\frac{\cisoT_2^2M-\alpha_2^\infty(\pi_2-\pi_1)}{\alpha_2^\infty \cisoT_1^2+\alpha_1^\infty \cisoT_2^2}$,
$\rho_2^\infty=\frac{M-\alpha_1^\infty\rho_1^\infty}{\alpha_2^\infty}$
	        \item
	        if (iii): $\alpha_2^{\infty}:=\varepsilon$, $\alpha_1^{\infty}=1-\alpha_2^{\infty}$,
$\rho_1^\infty=\frac{\cisoT_2^2M-\alpha_2^\infty(\pi_2-\pi_1)}{\alpha_2^\infty \cisoT_1^2+\alpha_1^\infty \cisoT_2^2}$,
$\rho_2^\infty=\frac{M-\alpha_1^\infty\rho_1^\infty}{\alpha_2^\infty}$
		\end{itemize}
	\end{enumerate}
\end{algorithm}
Note that in case (ii) and (iii) we determine $\rho_2^\infty$ by the mass conservation \eqref{eq:mass-con-2}. From this we deduce together with \eqref{eq:eos-iso-p} for an isothermal stiffened gas the equilibrium density $\rho_1^\infty$.
Since here $k=1$ represents water vapor modeled by an ideal gas, it holds $\pi_1=0$.

A direct consequence of the algorithm is the following theorem.
\begin{theorem}
The chemical relaxation procedure given by Alg.~\ref{alg:chem2} provides a unique solution. In particular,
the resulting equilibrium pressure is positive.
\end{theorem}

The equilibrium state of the chemical potential relaxation process for three components is
determined by solving the system of ODEs
\begin{eqnarray}
  \label{eq:chem-ode-1}
   \hspace*{-16mm}&&\odiff{\alpha_k}{\ot} = \frac{\dot m}{\varrho_k},\ k=1,2, \quad
    \odiff{\alpha_3}{\ot} = -\dot m\,\left(\frac{1}{\varrho_1}+\frac{1}{\varrho_2}\right),
     \\
  \label{eq:chem-ode-2}
  \hspace*{-16mm}&&\odiff{\alpha_k\,\rho_k}{\ot} = (-1)^{k+1}\dot m,\ k=1,2, \quad
      \odiff{\alpha_3\,\rho_3}{\ot} = 0,
  \\
  \label{eq:chem-ode-3}
   \hspace*{-16mm}&&\odiff{\alpha_k\,\rho_k\, \bv_k}{\ot} = (-1)^{k+1}\dot m\,\bV_I,\ k=1,2, \quad
    \odiff{\alpha_3\,\rho_3\, \bv_3}{\ot} = \bzero.
\end{eqnarray}
resulting from
\eref{eq:sa-model-vec-relax-op-type-mod}
with the source terms \eref{mass-relax-three-comp}.
According to \cite{Zein:2010,Zein-Hantke-Warnecke:2013} at equilibrium it holds
\begin{eqnarray}
\label{peqmu}
p_1^{\infty}=p_2^{\infty}=p_3^{\infty}&=&p^{\infty},\quad\ot\to\infty,\\
\label{mueq}
\mu_{1}^{\infty}&=&g_2^{\infty},\quad \ot\to\infty.
\end{eqnarray}
Note that due to these equilibrium conditions the mass flux $\dot m$
vanishes at equilibrium and, thus, the right-hand sides in
\eref{eq:chem-ode-1},
\eref{eq:chem-ode-2} and \eref{eq:chem-ode-3}
become zero.
Obviously, the equations \eref{eq:chem-ode-1} are in agreement
with the saturation condition \eref{eq:saturation}, i.e.,
\beq
\label{satmu}
 \alpha_1^{\infty}+\alpha_2^{\infty}+\alpha_3^{\infty}=1.
\eeq
In the following we start with relaxed values for velocity and pressure obtained by the previous mechanical relaxation procedures and, thus, it holds
\beq
p_1^0=p_2^0=p_3^0, \quad
\bv_1^0=\bv_2^0=\bv_3^0.
\eeq
During the chemical relaxation procedure we now assume that
mechanical equilibrium is maintained but the corresponding equilibrium values may change.
Due to conservation of total mass and total momentum it is obvious that the equilibrium velocity  does not change, i.e.,
\beq
   \label{eq:chem-velo-eq}
  \velo{}^{\infty} = \bv_k^{\infty} = \bv_k^0 = \velo{}^0,\quad k=1,2,3.
\eeq
The mass equations \eref{eq:chem-ode-2} induce
\begin{eqnarray}
 \label{mass3}
\alpha_3^{\infty}\rho_3^{\infty}&=&\alpha_3^0\rho_3^0,\\
\label{massD}
\alpha_1^{\infty}\rho_1^{\infty}+\alpha_2^{\infty}\rho_2^{\infty}&=&
\alpha_1^0\rho_1^0+\alpha_2^0\rho_2^0 =:M\,.
\end{eqnarray}
Motivated by the equilibrium condition
\eref{mueq} the aim is now to
derive a function
\begin{equation}\label{eqmu}
 f_\mu(\alpha_1\rho_1):=\mu_{1}^\infty(\alpha_1\rho_1)-g_2^{\infty}(\alpha_1\rho_1)
\end{equation}
depending only on the product $\alpha_1\rho_1$
such that the root $\alpha_1^{\infty}\rho_1^\infty$ is the solution for the
relaxed mass density for the first component.
For this purpose we have to express $\alpha_1^{\infty}$, $\alpha_2^\infty$ and
$p^{\infty}$
in terms of
$\alpha_1^{\infty}\rho_1^\infty$, see below. Then the chemical potential  for water vapor $\mu_1$ and
the Gibbs free energies $g_k$, $k=1,2$, can be written as
 \begin{eqnarray}
  \label{eq:chem-pot}
  \mu_{1}^{\infty}(\alpha_1\rho_1)&=&g_1^{\infty}(\alpha_1\rho_1)+
  c_1^2 \ln\left(\frac{\alpha_1(\alpha_1\rho_1)}{1-\alpha_2(\alpha_1\rho_1)}\right),\\[2mm]
  \label{eq:chem-gibbs}
  g_k^{\infty}(\alpha_1\rho_1)&=&c_k^2\ln\left(\frac{p^{\infty}(\alpha_1\rho_1)+\pi_k}{p_{k,ref}+\pi_k}\right) .
 \end{eqnarray}
Here $p_{k,ref}$ denotes the saturation pressure to a given reference temperature. For the particular configuration at hand it holds $p_{1,ref}=p_{2,ref}=p(\Tref)$.
If an inert component is present a root of \eref{eqmu} always exists.
 Nevertheless, the equilibrium state can be very close to total condensation or total evaporation. To avoid numerical instabilities a small amount of the particular phase should be kept.
In a first step one has to figure out, which
of the following four  cases occurs:
\begin{itemize}
\item[(i-a)] condensation process with equilibrium solution,
\item[(i-b)] nearly total condensation,
\item[(ii-a)] evaporation process with equilibrium solution,
\item[(ii-b)] nearly total evaporation.
\end{itemize}
In case (i-b) and (ii-b) the result can be directly obtained.
In the cases (i-a) and (ii-a) a bisection method is provided to find the equilibrium state. Due to the fact that the pressure relaxation method is simple and always gives a unique, physical solution,
we base the bisection method to find the equilibrium state on the pressure relaxation procedure.

According to thermodynamics a condensation process (i) and an evaporation process (ii) are characterized by
a positive or negative sign of $\mu_1-g_2$, respectively.
Thus, using the data from the pressure relaxation procedure we may
identify condensation and evaporation processes.

{\bf Condensation.}
If a condensation process is identified by
$f_\mu(\alpha_1^0\rho_1^0) = \mu_1(\alpha_1^0\rho_1^0)-g_2(\alpha_1^0\rho_1^0)>0$
the expression $\alpha_1^0\rho_1^0$ is too large
and $\alpha_1\rho_1$ has to decrease. The smallest admissible
value for this expression is $\alpha_1^*\rho_1^*=tol>0.$
Using the pressure relaxation procedure with $\alpha_1^*\rho_1^*$ instead of $\alpha_1^0\rho_1^0$
we determine the corresponding values for all
variables of the phases. Using these data one has to check the sign of the difference of the chemical potentials. If still
$\mu_1-g_2\ge 0$ holds, then
nearly total condensation will occur. We keep $\alpha_1^*\rho_1^*=tol>0$ and the corresponding data.
Otherwise the interval $[\alpha_1^*\rho_1^*=tol, \alpha_1^0\rho_1^0]$
is admissible for the bisection method.

{\bf Evaporation.}
If an evaporation process is identified by
$f_\mu(\alpha_1^0\rho_1^0) = \mu_1(\alpha_1^0\rho_1^0)-g_2(\alpha_1^0\rho_1^0)<0$
the expression $\alpha_1^0\rho_1^0$ is too small
and $\alpha_1\rho_1$ has to increase.
The largest admissible
value for this expression is $\alpha_1^*\rho_1^*=M-tol.$
Analogously to the previous case we determine the corresponding values for all
variables of the phases and check the sign of the difference of the chemical potentials. Either we identify nearly total evaporation and keep the data or we find the admissible interval $[\alpha_1^0\rho_1^0,M-tol]$
for the bisection method.\\

We summarize the relaxation procedure in the following algorithm.

\begin{algorithm}[Chemical Relaxation Procedure for $K=3$]~
\label{alg:chem3}
\begin{enumerate}
\item
Identify condensation or evaporation processes (i) or (ii) by the sign of
$f_\mu(\alpha_1^0\rho_1^0) = \mu_1(\alpha_1^0\rho_1^0)-g_2(\alpha_1^0\rho_1^0)$
and determine the initial interval for the bisection method
\item
\begin{itemize}
	\item if (i-a): Apply the bisection method to find the root $\overline {\alpha_1 \rho_1}$ of \eref{eqmu}
	                 in the interval $[\alpha_1^*\rho_1^*=tol, \alpha_1^0\rho_1^0]$
	                 and perform pressure relaxation with $\overline {\alpha_1 \rho_1}$;
	\item if (i-b): Apply the pressure relaxation procedure to $\alpha_1^*\rho_1^*=tol$;
	\item if (ii-a): Apply the bisection method to find the root $\overline {\alpha_1 \rho_1}$ of \eref{eqmu}
	                  in the interval $[\alpha_1^0\rho_1^0,\alpha_1^*\rho_1^*]$ with
	                  $\alpha_1^*\rho_1^*=M-tol$
	                  and perform pressure relaxation with $\overline {\alpha_1 \rho_1}$;
	\item if (ii-b): Apply the pressure relaxation procedure to $\alpha_1^*\rho_1^*=M-tol$.
\end{itemize}
\end{enumerate}
\end{algorithm}

A direct consequence of the algorithm is the following theorem.

\begin{theorem}
The chemical relaxation procedure given by
Alg.~\ref{alg:chem3}
provides a unique solution. In particular,
the resulting equilibrium pressure is positive.
\end{theorem}

The above relaxation
procedures Algs.~\ref{alg:chem2} and \ref{alg:chem3}
can be considered as an essential improvement
of the original ones presented in
\cite{Zein:2010,Zein-Hantke-Warnecke:2013}
due to the following aspects: In the original procedure
the Gibbs free energies are relaxed, i.e.,
the influence of the mixture entropy,
which is described by the difference of the chemical potential and the Gibbs free energy of the vapor phase, is neglected.
This extra term cancels in pure phases. Moreover, in the original procedure a
nested iteration method was used. We now  reduce the numerical costs significantly by simplifying the equilibrium system to a \emph{scalar} equation depending only on $\alpha_1\rho_1$ that has to be solved by a \emph{single} iteration process. Moreover, for $K=2$ the solution can  be determined directly.
Thus, the computational costs are significantly reduced.

Finally we point out that  in complete analogy the above procedure can also be applied to mixtures with an arbitrary number of additional inert ideal gases.

\subsection{Isentropic case}
\label{subsubsec:relax-isentropic}


Similar to  isothermal fluids we have to solve the
system of ODEs \eref{relax-pres-ODE} for the pressure relaxation process implying conservation of mass and momentum \eref{masscons}.

For the full multi-component Baer-Nunziato model
pressure relaxation  has been investigated in
\cite{Lallemand-Saurel:2000}
for two- and multi-component models. Lallemand et al.~\cite{Lallemand-Saurel:2000} give eight different procedures
based on a  closure of the interfacial pressure that become physically non-admissible for more than two components, see \cite{Mueller-Hantke-Richter:16}.
These procedures predict different values for  the equilibrium pressure.

\textit{Pressure-temperature relaxation.}
Alternatively, uniqueness can be enforced by the additional condition of temperature equilibrium.
For this purpose it was suggested in \cite{Han-Hantke-Mueller:2017} to simultaneously relax
pressure and temperature and, thus, determine a unique equilibrium pressure  by avoiding iteration processes. This results in an efficient and stable procedure.
However, for isentropic fluids this approach does not work as can be verified for ideal gases:
in the equilibrium state it must hold
\begin{align}\label{eqcond}
	p_k^\infty=p^\infty,\,T_k^\infty=T^\infty,\, k=1,\dots,K.
\end{align}
Assume $K=2$ and both components to be ideal gases, see Example \ref{ex:isentrop-stiff-gas-Eos} with $\pi_k=0$. Then from the equations of state we have the relations
\begin{align}\label{eos2}
	p_k=A_k\rho_k^{\gamma_k},\, T_k=B_k\rho_k^{\gamma_k-1}
\end{align}
with constants
$A_k$, $B_k$ determined by \eref{eq:isentr-stiffgas-coeff}.
From \eqref{eqcond} and \eqref{eos2} we obtain
\begin{align}
	(\rho_1^\infty)^{\gamma_1-\gamma_2}=
	\left(\frac{A_2}{A_1}\right)^{\gamma_2-1}\left(\frac{B_1}{B_2}\right)^{\gamma_2}\,.
\end{align}
Obviously, for $\gamma_1\ne\gamma_2$ the parameters of the equations of state define a unique equilibrium density, that does not depend on the initial state. Analogously one can find $\rho_2^\infty$. Using
\eqref{masscons}
one can obtain the corresponding volume fractions $\alpha_1^\infty$ and $\alpha_2^\infty$, that usually will \textit{not} satisfy the saturation condition
\begin{align}
	\alpha_1^\infty+\alpha_2^\infty=1\,.
\end{align}

\textit{Pressure  relaxation.}
Therefore for isentropic fluids we relax the pressure without simultaneously relaxing the temperature, i.e.,
in the equilibrium state it must hold
\begin{align}\label{eqcond3}
	p_k^\infty=p^\infty,\, k=1,\dots,K.
\end{align}
Assume all $K$ components to be stiffened gases. Then from the equations of state we have the relations
\begin{align}\label{eosid}
	p_k+\pi_k=A_k\rho_k^{\gamma_k}
\end{align}
with constant
$A_k$ determined by \eref{eq:isentr-stiffgas-coeff-Ak}.
Then conservation of mass
\eqref{masscons}
and the equilibrium conditions \eqref{eqcond3} lead to
\begin{align}
\label{eqcond3-1}
	A_1\left(\frac{\alpha_1^0\rho_1^0}{\alpha_1^\infty}\right)^{\gamma_1} -\pi_1=
	A_k\left(\frac{\alpha_k^0\rho_k^0}{\alpha_k^\infty}\right)^{\gamma_k}-\pi_k,\, k=2,\dots,K\,.
\end{align}
Solving these equations for $\alpha_k^\infty$ gives
\begin{align}\label{relal}
	\alpha_k^\infty=\left[
	\frac{A_k(\alpha_k^0\rho_k^0)^{\gamma_k}(\alpha_1^\infty)^{\gamma_1}}{A_1(\alpha_1^0\rho_1^0)^{\gamma_1} +(\pi_k-\pi_1)(\alpha_1^\infty)^{\gamma_1}}
	\right]^{\frac{1}{\gamma_k}}
\end{align}
depending on the only unknown $\alpha_1^\infty$.
The numbering of the components should be chosen such that
\[
\min_k\{\pi_k,k=1,\dots,K\}=\pi_1.
\]
Using \eqref{relal} and the saturation condition we obtain
\begin{align}\label{Falpha1}
	F(\alpha_1^\infty)=\alpha_1^\infty+\sum_{k=2}^K \left[
	\frac{A_k(\alpha_k^0\rho_k^0)^{\gamma_k}(\alpha_1^\infty)^{\gamma_1}}{A_1(\alpha_1^0\rho_1^0)^{\gamma_1} +(\pi_k-\pi_1)(\alpha_1^\infty)^{\gamma_1}}
	\right]^{\frac{1}{\gamma_k}}-1=0.
\end{align}
It holds $F(0)<0$, $F(1)>0$ and the smooth function $F$ is strictly increasing for $\alpha_1^\infty\in[0,1]$. Accordingly equation \eqref{Falpha1} has a unique solution for $\alpha_1^\infty\in[0,1]$ which gives us a unique positive equilibrium pressure $p^\infty$ satisfying
\begin{align*}
\min_{k=1,\ldots,K} p_{0,k} \le p^\infty \le \max_{k=1,\ldots,K} p_{0,k} .
\end{align*}
This is not true in general for
the full Baer-Nunziato model
as discussed above.

The relaxation procedure is summarized in the following algorithm.
\begin{algorithm}[Pressure relaxation procedure in case of isentropic stiffened gases]~
\label{alg:pres-relax-isentropic-stiffgas}
 \begin{enumerate}
   \item Determine  $\alpha_1^\infty\in[0,1]$ as the unique root of the function \eref{Falpha1} in $[0,1]$;
   \item Determine  $\alpha_k^{\infty,i}$, $k=2,\ldots, K$,  from \eref{relal};
   \item Compute the equilibrium pressure $p^\infty$  using \eref{eqcond3-1} as
\begin{align}
\label{pres-relax-isentropic-stiffgas-pinfty}
	p^\infty = 	
        A_1\left(\frac{\alpha_1^0\rho_1^0}{\alpha_1^\infty}\right)^{\gamma_1} -\pi_1=
	A_k\left(\frac{\alpha_k^0\rho_k^0}{\alpha_k^\infty}\right)^{\gamma_k}-\pi_k,\, k=2,\dots,K\,.
\end{align}

   \item Update the phasic densities by $\rho_k^{\infty} = \rho_k(p^\infty)$.
 \end{enumerate}
\end{algorithm}

Since in general pressure and temperature cannot be in equilibrium at the same time
for isentropic fluids, we will not discuss temperature relaxation and chemical relaxation because pressures relax faster than temperatures and chemical potentials.

\section{Numerical results}
\label{sec:bn-neq-results}

\subsection{Discretization}
\label{discr}


For a condensed presentation of the discretization it is convenient to rewrite
the system \eref{eq:sa-model-rhok}, \eref{eq:sa-model-mk},
\eref{eq:sa-model-Ek} and \eref{eq:sa-model-alphak} in matrix-vector representation
\begin{align}
\label{eq:sa-model-complete}
\pdifft{\bu} + \pdiv{\bff^i(\bu)} + \sum_{i=1}^d \bG^i(\bu) \pdiff{\bu}{{x_i}} = \bs(\bu)
\end{align}
where we make use of the convention \eref{eq:convention}.
This forms a coupled system for the volume fractions
$\balpha:=(\alpha_1,\ldots,\alpha_{K-1})^T$  and the vectors
$\bu_k:=(\alpha_k\,\rho_k,\alpha_k\,\rho_k\,\bv_k)^T$ of the conserved quantities of phase $k=1,\ldots,K$
condensed in the vector $\bu:=(\balpha^T,\bu_1^T,\ldots,\bu_K^T)^T$. The flux $\bff^i$ in the $i$th coordinate direction and the source term $\bs$ are determined by
\begin{align}
\label{eq:sa-model-vector-fluxes}
  &\bff^i(\bu)^T :=
                       \left( \bzero_{K-1}^T,
                          \bff^i_1(\alpha_1,\bu_1)^T,
                          \ldots,
                          \bff^i_K(\alpha_K,\bu_K)^T
                      \right), \\[2mm]
\label{eq:sa-model-vector-sources}
  &\bs(\bu) :=
       \left( \bS_\alpha(\bu)^T, \bS_1(\bu)^T,\ldots,\bS_K(\bu)^T\right)
\end{align}
with phasic fluxes and sources
\begin{align}
  &\bff^i_k(\alpha_k,\bu_k) :=
 (\alpha_k \rho_k v_{k,i}, \alpha_k \rho_k \bv_1^T v_{k,i} + \alpha_k p_k \be_{d,i}^T,\\
  &\bS_k(\bu) := (\source{\alpha\rho}{k}(\bu), \Source{\alpha\rho\vects{\velo{}}}{k}^T(\bu))^T,\\
  &\bS_\alpha(\bu) := (\source{\alpha}{1}(\bu), \ldots,\source{\alpha}{K-1}(\bu))^T .
\end{align}
Here $\be_{d,i} \in\R^d$  denotes the
$ith$ unit vector in $\R^d$ and $\bzero_{K-1}$ the zero vector in $\R^{K-1}$.
The nonlinear product in \eref{eq:sa-model-complete} is characterized by the matrices
\begin{align}
\label{eq:sa-model-vector-nonconservative-products}
  \bG^i(\bu) =
   \left(
   \begin{matrix}
     V_{I,i} \bI_{K-1}  & \bzero_{(K-1)\times (K(d+1))}  \\
     \left(
     \begin{matrix}
       \bzero_{K-1}^T\\
       \be_{d,i}\otimes \bd_1
     \end{matrix}
     \right)
       & \bzero_{(d+1)\times (K(d+1))} \\
       \vdots & \vdots \\
     \left(
     \begin{matrix}
       \bzero_{K-1}^T\\
       \be_{d,i}\otimes \bd_K
     \end{matrix}
     \right)
       & \bzero_{(d+1)\times (K(d+1))}
   \end{matrix}
   \right) ,
\end{align}
where $\bI_{K-1}$ denotes the  identity matrix in  $\R^{(K-1)\times (K-1)}$ and the components of the vectors $\bd_k$, $k=1,\ldots, K$ are defined by $d_{k,l}=P_{k,l} - P_{k,K}$, $l=1,\ldots, K-1$. Note that $P_{k,k}$ by convention ist set to 0.

Since we are interested in flows near to equilibrium, we do not discretize the coupled system
\eref{eq:sa-model-complete}
but perform an operator splitting according to Godunov or Strang,
see \cite{Toro:1997},
resulting in the system of equations of fluid motion
\begin{align}
\label{eq:sa-model-fluid}
\pdifft{\bu} + \pdiv{\bff^i(\bu)} + \sum_{i=1}^d \bG^i(\bu) \pdiff{\bu}{{x_i}} = \bzero
\end{align}
and the system of relaxation
\begin{align}
\label{eq:sa-model-relax}
\odifft{\bu}= \bs(\bu) .
\end{align}
This procedure is in analogy to \cite{Saurel-Abgrall:1999}.
Note that the differential equation
\eref{eq:sa-model-fluid}
describes a variation in time and space, whereas the system
\eref{eq:sa-model-relax}
is an evolution in time only although the state $\bu$ depends on both time and space.
Therefore we use different symbols for the differentiation.

In each time step we thus perform alternately the evolution of the fluid and the relaxation process:

For the fluid equations \eref{eq:sa-model-fluid} we apply a path-conservative DG scheme. For this purpose, we extended our in-house Runge-Kutta DG solver \cite{HMS:2014,GerhardMueller:2016}
by a path-conservative discretization of the nonlinear products in \eref{eq:sa-model-fluid} following \cite{RhebergeBokhoveVanderVegt:2008}. For details on the derivation of the  spatial path-conservative DG scheme we refer to \cite{Blaut:2021}.

The efficiency of the scheme is improved by local grid adaption where we employ the
multi-resolution concept based on multiwavelets.
The key idea is to perform a multi-resolution analysis on a sequence of
nested grids providing a decomposition of the data on a coarse scale and a sequence
of details that encode the difference of approximations on
subsequent resolution levels. The detail coefficients become small when the underlying
data are locally smooth and, hence, can be discarded when dropping below a
threshold value $\varepsilon_{thresh}$.
By means of the thresholded sequence a new, locally refined grid
is determined. Details on this concept can be found in
\cite{HMS:2014,GerhardMueller:2016,GerhardMuellerSikstel:2021}.

Since our interest is in equilibrium processes we replace the transient relaxation procedure
\eref{eq:sa-model-relax} by integration to infinity, i.e., we determine the equilibrium states by algebraic equations according to Sect.~\ref{subsubsec:relax-isentropic} and Sect.~\ref{subsubsec:relax-isothermal} in case of an isentropic and isothermal fluid, respectively. The relaxation to equilibrium is performed one-by-one. First we determine  the equilibrium velocity by the velocity relaxation according to Algorithm \ref{alg:velo-relax}. Then we perform pressure relaxation applying Algorithm \ref{alg:pres-relax-isentropic-stiffgas} or \ref{alg:pres-relax-isothermal-stiffgas}. In addition, in case of an isothermal fluid  we determine the state at chemical equilibrium according to Algorithm \ref{alg:chem2} and \ref{alg:chem3} for a two-component and three-component mixture, respectively.
In case of a higher order DG scheme the solution on each element is given by a polynomial of higher degree. Since the equilibrium states are determined locally, for each element we have to perform a projection of the equilibrium states to the polynomial space.

For this purpose, we first determine for each element the equilibrium states in all quadrature points. From these we compute an interpolation polynomial that interpolates these equilibrium states. The polynomial degree of the interpolation polynomial is chosen smaller than the degree of the DG polynomial. Thus, projecting the interpolation polynomial onto the DG space will preserve the equilibrium states in the quadrature points. Note that in the DG scheme we only access the DG polynomials in the quadrature points.

We would like to remark that the relaxation process is applied whenever the states are out of equilibrium, i.e., after applying the limiter, performing grid adaptation or a Runge-Kutta stage in the time evolution.

\subsection{Validation}

For validation we consider two configurations. First, we investigate a
two-phase  configuration where each component is complemented by the isothermal  stiffened gas model, see Example \ref{ex:isotherm-stiff-gas-Eos}. The results are compared to a sharp-interface  isothermal Euler model, see \cite{Hantke2013,Hantke2019a}.
For another configuration we consider a four-component mixture of isothermal ideal gases. 
The results are compared to an isothermal phase-field model discussed in \cite{Hantke-Matern-Warnecke-Yaghi:2024}.

\subsubsection{How to choose material parameters}
In the literature one can find numerous numerical examples for two- and multi-phase flows with phase transition at similar states with
different sets of material parameters. Often these parameters are chosen arbitrarily. This in many cases leads to unphysical phase states, unphysical equilibrium states and unphysical processes. This can be avoided when appropriate tables are used, see for example \cite{STEAM}. For water vapor and liquid water we proceed as follows, where water vapor is modeled as an ideal gas and liquid water is modeled as a stiffened gas.
\begin{algorithm}[Procedure to determine material parameters]~
\label{alg:material-parameters-neu}
\begin{itemize}
    \item Vapor:
    \begin{itemize}
	\item
	Choose some
        reference temperature $\Tref$.
	\item
	Identify the corresponding saturation pressure $p_{ref}=p(\Tref)$ from the table.
	\item
	Identify the corresponding specific heat capacity at constant volume $c_{v,1}$, the specific volume $1/\rho_{ref,1}$ for the vapor phase and the specific entropy $s_{0,1}$ for the vapor phase from the table.

	\item
	$\pi_1=0$
	\item
	Determine $\gamma_1$ using \eref{eq:eos-iso-p}: $\gamma_1=1+\frac{p_{ref}}{\rho_{ref,1}c_{v,1}\Tref}$.

	
	\end{itemize}
	
        \item Liquid:
        
        \begin{itemize}

	\item
	Determine $q_1$ using \eref{eq:eos-iso-s}: $q_1=(s_{0,1}-c_{v,1})\Tref$.
	\item
	Identify the corresponding specific heat capacity at constant volume $c_{v,2}$, the specific volume $1/\rho_{ref,2}$, the specific entropy $s_{0,2}$ and the speed of sound $c_2$ for the liquid phase from the table.
	\item
	Determine $\gamma_2$ using \eref{eq:eos-stiff-p} and \eqref{eq:sound-fdt-T-rho-stiff}:
	$\gamma_2=\frac{1}{2}+\sqrt{\frac{1}{4}+\frac{c_2^2}{c_{v,2}\Tref}}$.
	\item
	Determine $\pi_2$ using \eref{eq:eos-iso-p}: $\pi_2=\rho_{ref,2}(\gamma_2-1)c_{v,2}\Tref-p_{ref}$.
		\item
	Determine $q_2$ using \eref{eq:eos-iso-s}: $q_2=(s_{0,2}-c_{v,2})\Tref$.
	\end{itemize}
\end{itemize}
\end{algorithm}

Choosing different values for $\Tref$ we find the following set of parameters
recorded in
Table  \ref{tab:material-parameters-2-neu}.
Note that the above procedure applies for
stiffened gas in general because we start from a reference state. Thus, we may apply it for both  the isothermal and the isentropic case, see Examples \ref{ex:isotherm-stiff-gas-Eos} and \ref{ex:isentrop-stiff-gas-Eos}.
Note that for water vapor 
 we do not need to specify the sound speed because for an ideal gas we set $\pi=0$.

\begin{table}[h]
	\begin{center}
		\begin{tabular}{|c|c|c|c|c|}
			\hline
			component & $1/\rho_{ref}$[m$^3$/kg]   & $c_v$ [J/kg/K] & $c$ [m/s]    & $s_0$ [J/kg]    \\
			\hline
			water vapor (1)  &  2.35915                  &  1530.4        & -            &  7478.1 \\
			liquid water (2)  &  0.00103594               &  3818.8        & 1552.7       &  1192.7 \\
			\hline
		\end{tabular}
		\caption{Material parameters taken from \cite{STEAM} for $\Tref=363.15$ K with $p_{ref}=70182.4$ Pa. }
		\label{tab:material-parameters-2-neu}
	\end{center}
\end{table}

\subsubsection{Numerical example: vapor--vapor compression problem}

To investigate nucleation phenomena we consider the following 1d-Riemann problem determined by  a compression problem  for an initially single component vapor. The initial data are listed in Table \ref{tab:initial-data-rp-HDW3-neu} with material parameters given in Table \ref{tab:material-parameters-2-neu}.
Two sets of  computations  are performed where we either perform mechanical relaxation, see 
Fig.~1, 
or  mechanical and chemical relaxation, see 
Fig.~2, 
i.e., the computations are performed with and without mass transfer.
In the figures we present the volume of water vapor (top, left), the pressure of the mixture (top, right), the velocity of the mixture (bottom, left) and the density of the mixture (bottom, right).

The computational domain is $\Omega=(-2,2)$. The simulations are run to time $t=0.001$.
We perform  second order $(p=2)$ computations using 3 cells on the base level and 10 levels of refinement. The CFL number is 0.1 
resulting in about  61,000 time steps to reach the final time. The constant $c_{thresh}$ for the thresholding performed in the grid adaptation step is chosen as $10^{-5}$. This small value was needed to ensure that all discontinuities are resolved on the highest refinement level where limiting is performed. At the final time the adaptive grid is about 82\% of the uniformly refined grid.

In this configuration we consider  ``pure'' water vapor that is slightly polluted with liquid water because the model does not allow for pure states. Pressure and velocity are chosen in mechanical equilibrium.
The initial pressure is slightly below the saturation pressure. At the center $x=0$  the fluid is compressed due to the opposed velocities causing two compression/shock waves running to the left and right of the
center. This results in a significant pressure increase near the center where the pressure exceeds the saturation pressure. According to the pressure relaxation we observe a significant drop in the water vapor volume
that is caused by a rise in the water vapor density.

There is a significant difference between the computations without and with mass transfer.
When no chemical relaxation is performed  no mass is transferred from  water vapor to liquid water. This is reflected in the results for the density of the mixture. In case of no mass transfer the value is still close to
the initial value. However, when mass transfer occurs the density of the mixture significantly increases in the vicinity of the center. Note that this behavior is not fully resolved by the Baer-Nunziato model because it
requires a much finer grid to fully capture the condensation of the water vapor.
Note that the wave structure consists of two outer shock waves and two phase boundaries in between, which enclose the nucleated liquid.
However, due to the very small volume of the created liquid this cell is not fully resolved and we refer to \cite{May2023} for further insights on the numerical difficulties.

To validate the above results 
we compare the results with computations performed for an isothermal Euler model where  two phases are separated by a sharp interface. Phase transition is accounted for  by the jump condition at the interface. For details we refer to \cite{Hantke2013,Hantke2019a}.
Numerical computations of this model are performed independently applying a finite volume scheme in both of the two domains to either side of the interface. At the interface a two-phase Riemann problem is solved to compute the fluxes where the jump conditions account for mass transfer or neglect mass transfer.\\
We observe a perfect agreement of the two models in case of no mass transfer, see Fig.~\ref{fig:vapor-vapor-expansion-no-mass-transfer}.
When accounting for mass transfer then wave speeds of the two models perfectly match but there is an obvious severe  discrepancy in the values of the intermediate states, see the red lines in
Fig.~\ref{fig:vapor-vapor-expansion-mass-transfer}. By numerical experiments we observe that in this case a scaling of the kinetic relation in the sharp-interface model by a factor of $300$ results in a perfect
agreement of the states. 
Note that in the Baer-Nunziato model we relax to the equilibrium state, i.e., the relaxation velocity is infinite,  whereas in the isothermal Euler model the relaxation velocity is finite. Thus, increasing the relaxation velocity in the isothermal Euler model it comes closer to the equilibrium state.

\begin{table}[h!]
\begin{center}
\begin{tabular}{|l|l|l|l|l|l|}
\hline
          $\alpha_{1,L}$ &  $p_{L}$ [Pa] & $v_{L}$ [m/s] & $\alpha_{1,R}$  & $p_{R}$ [Pa] & $v_{R}$ [m/s]\\
\hline
$ 1-10^{-8} $               &  69985.63            & 2.7             &  $1-10^{-8}$          &   69985.63          &  -2.7 \\
\hline
\end{tabular}
\caption{Initial data for Riemann problem: vapor--vapor compression problem.}
\label{tab:initial-data-rp-HDW3-neu}
\end{center}
\end{table}

\setcounter{figure}{0}

\begin{figure}[h!]
\label{fig:vapor-vapor-expansion-no-mass-transfer}
	    \includegraphics[angle=0,scale=0.3]{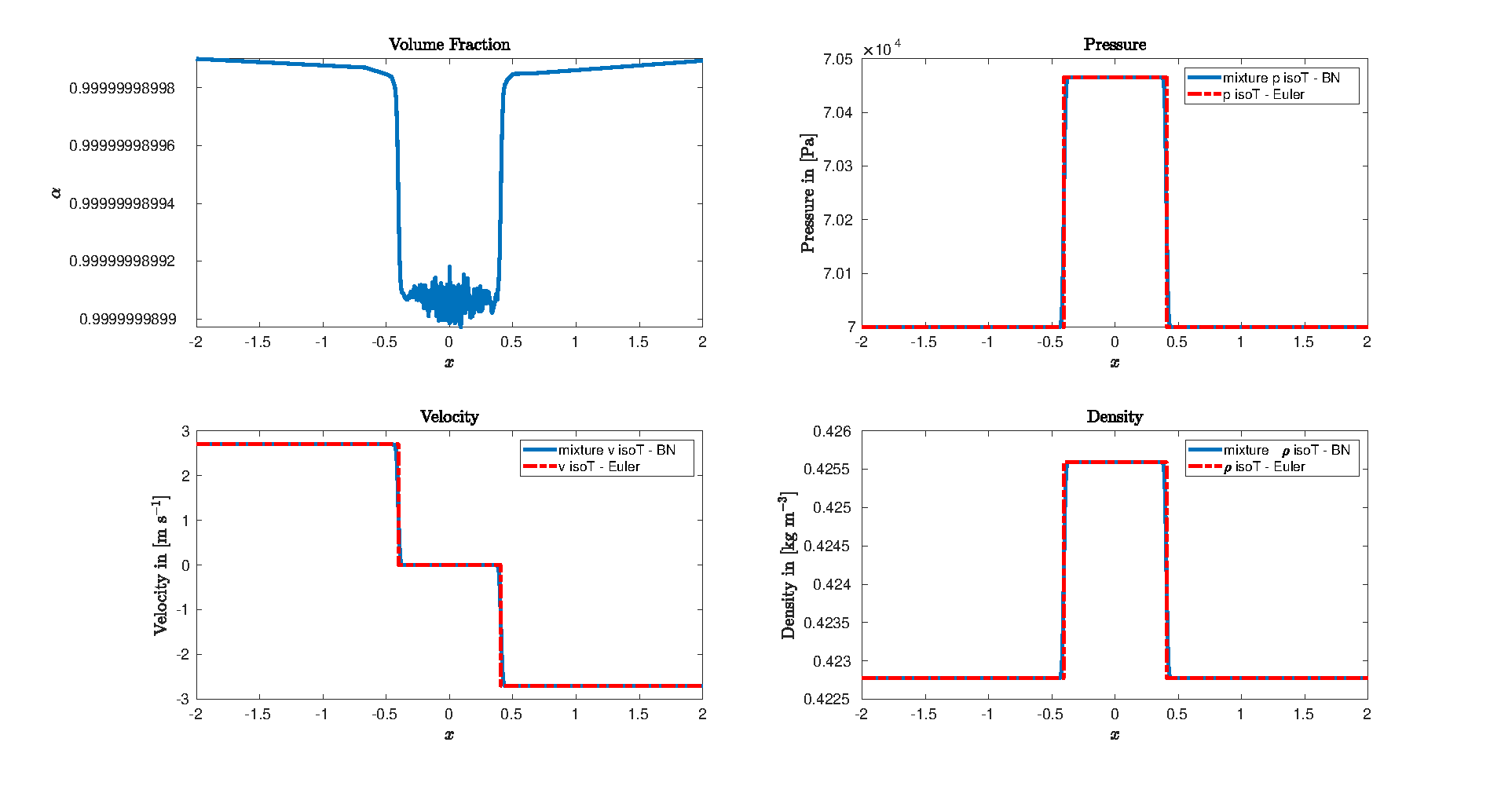}
\caption{Vapor-vapor compression: without phase transition.}
\end{figure}

	\begin{center}	
\begin{figure}[h!]
\label{fig:vapor-vapor-expansion-mass-transfer}
	    \includegraphics[angle=0,scale=0.3]{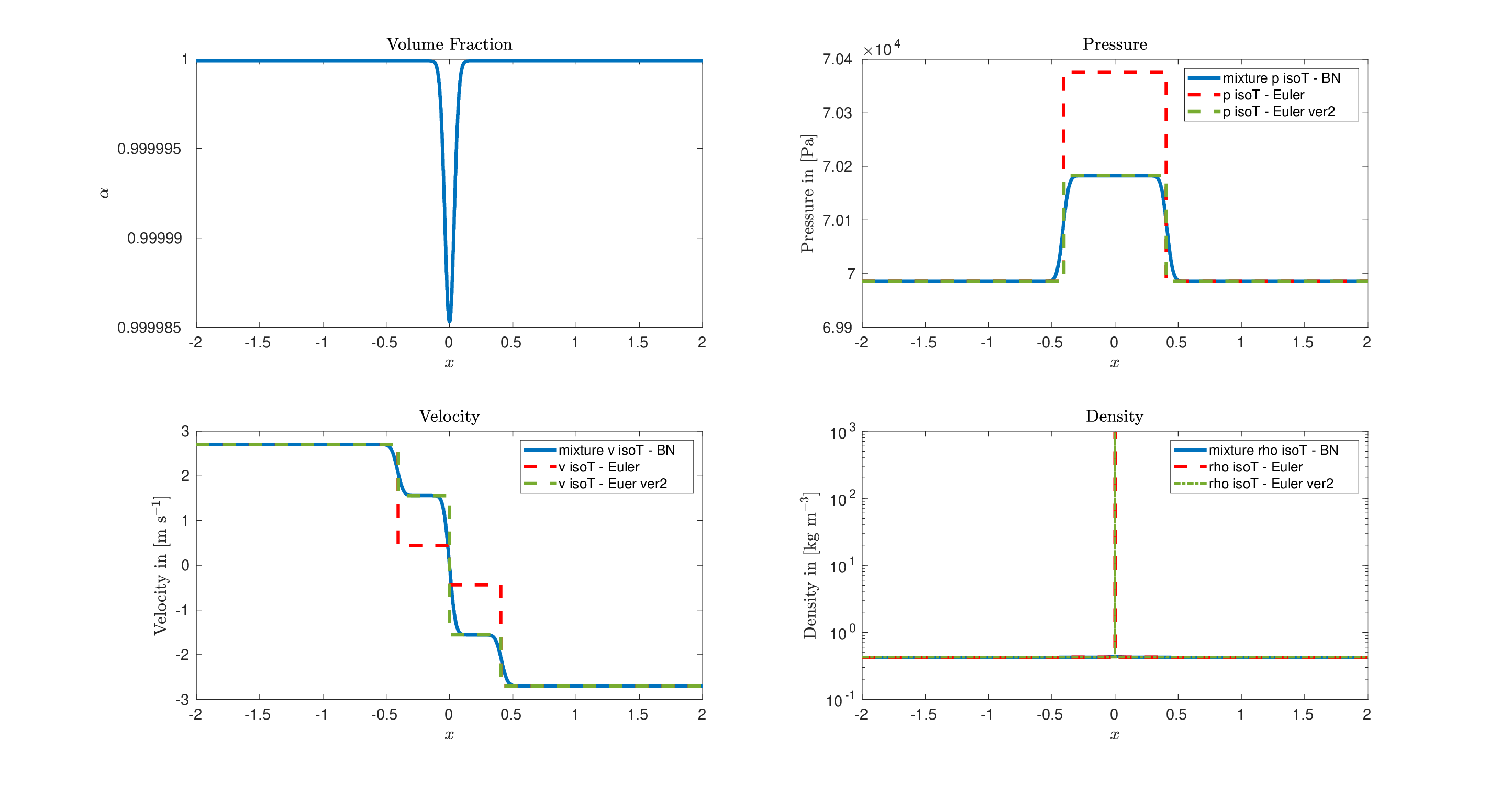}
\caption{Vapor-vapor compression: with phase transition.}
\end{figure}
	\end{center}

%
\subsubsection{Numerical example: two components --- two phases }

Another interesting test case for further validation of the model is given by the flow of 
two components with two phases, i.e.,
$k=1,2$ (component 1, liquid/gas) and $k=3,4$ (component 2, liquid/gas), 
in one dimension. 
The initial data are listed in Table \ref{tab:initial-data-rp-C5}.
All components are modeled by isothermal ideal gases, i.e.,
$\pi_1=\pi_2=\pi_3=\pi_4=0$ in \eqref{eq:eos-iso-p}. The isothermal speeds of sounds are chosen as
$c_1=600$ m/s, $c_2=1000$ m/s, $c_3=650$ m/s, $c_4=1100$ m/s. The heat capacities are
$c_{v,1}=1100$ J/kg\,K, $c_{v,2}=1200$ J/kg\,K, $c_{v,3}=2100$ J/kg\,K, $c_{v,4}=2400$ J/kg\,K.
The reference temperature is $T=293.15$ K.
From these values the parameters $\gamma_k$ can be computed by \eqref{eq:sound-fdt-T-rho-isotherm}.
Note that the liquid phases are also modeled by  ideal gases instead of stiffened gases.
This allows to rewrite the data of the Baer-Nunziato model into the data of the phase field model \cite{Hantke-Matern-Warnecke-Yaghi:2024} and the same material parameters can be used. For stiffened gases this is no longer obvious because of the minimal pressure $\pi$.

%

Computations for this configuration are only performed using velocity and pressure relaxation.
The computational domain is $\Omega=(-2,2)$. The simulations are run to time $t=0.001$.
We perform  second order $(p=2)$ computations using 3 cells on the base level and 10 levels of refinement. The CFL number is 0.1
resulting in about  30,000 time steps to reach the final time. The constant $c_{thresh}$ for the thresholding performed in the grid adaptation step is chosen as $10^{-5}$. This small value was needed to ensure that all discontinuities are resolved on the highest refinement level where limiting is performed. At the final time the adaptive grid is about 94\% of the uniformly refined grid.

In this configuration the pressure and velocity are chosen in mechanical equilibrium.
At the center $x=0$  the fluid is compressed due to the opposed velocities causing two compression/shock waves running to the left and right of the
center, see 
Fig.~3. 
These waves can be observed in the mixture velocity, the mixture pressure and the mixture  densities  of component 1 and component 2. The volume fractions only jump at the material interface near the center. We note that the mixture velocities and the mixture pressures are continuous across the material wave but the mixture densities suffer from a jump. This indicates that the material wave is a contact wave.


For validation of the Baer-Nunziato model we compare the results with computations using  the phase-field model \cite{Hantke-Matern-Warnecke-Yaghi:2024}.
Here each component is described by a single density while the phase-field parameter indicates the present phase. Accordingly the density 
$\rho_1$ of the phase-field model is related to $\alpha_1\rho_1+\alpha_2\rho_2$ of the Baer-Nunziato model and 
$\rho_2$ of the phase-field model is related to $\alpha_3\rho_3+\alpha_4\rho_4$ of the Baer-Nunziato model. 
Further we compare the mixture velocity and the mixture pressure of both models. The results of the two models  show a very good agreement except at the jumps were the phase-field model shows some smearing of the discontinuity.

\begin{table}[h!]
\begin{center}
\begin{tabular}{|l|l|l|l|l|l|l|l|l|l|}
\hline
          $\alpha_{1,L}$ &  $\alpha_{2,L}$ & $\alpha_{3,L}$ & $p_{L}$ [Pa] & $v_{L}$ [m/s] & $\alpha_{1,R}$ & $\alpha_{2,R}$ &  $\alpha_{3,R}$ & $p_{R}$ [Pa] & $v_{R}$ [m/s]\\
\hline
          $ 0.2$            &  $0.3$             & $0.1$             & $100000$        & $60$             & $0.5$              & $0.1$             & $0.2$              & $110000$         & $-50$\\
\hline
\end{tabular}
\caption{Initial data for Riemann problem: 
two components --- two phases.
}
\label{tab:initial-data-rp-C5}
\end{center}
\end{table}
\begin{figure}[h!]
\label{fig:two-components-two-phases}
	\begin{center}
		\includegraphics[scale=0.35]{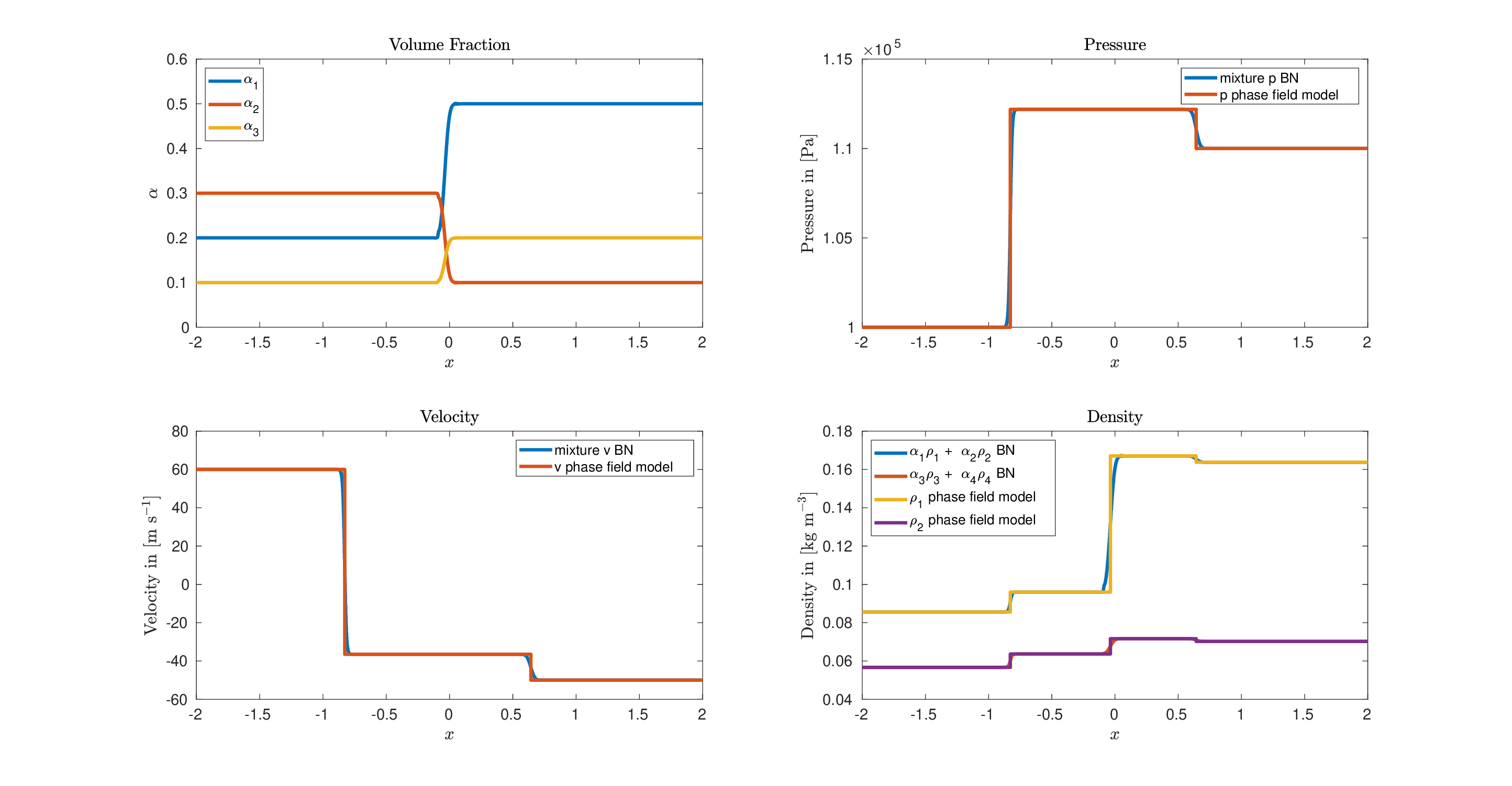}
	\end{center}
	\caption{two components --- two phases.}
\end{figure}

\section{Conclusion}
\label{sec:conclusion}

In this work we have investigated Baer-Nunziato-type  models for barotropic fluids with an arbitrary number of components. In the literature frequently barotropic fluids are confined to isentropic fluids, cf.~ \cite{herard:2016,boukili-herard:2018,saleh-seguin:2020} whereas here we also consider isothermal fluids.

First of all, we have determined eigenvalues and eigenvectors for the quasi-1D model in case of general barotropic fluids, i.e., the phasic pressures only depend on the phasic densities.  For non-resonance states a full set of linearly independent eigenvectors has been  derived. The different fields have been characterized as either genuinely nonlinear or linearly degenerated assuming that the phasic fundamental derivatives of thermodynamics are positive and choosing a particular convex combination for the interfacial velocity. For all fields we have given a full set of Riemann invariants whose gradients are linearly independent.
Furthermore we have verified that the quasi-1D equations is symmetrizable away from non-resonance states constructing a symmetrizer.

In case of isothermal fluids and isentropic fluids we have constructed appropriate entropy-entropy flux pairs and have verified that (i) the (mathematical) entropy functions are convex but not strictly convex even if the phasic entropy functions are strictly convex and (ii) the (mathematical) entropy inequality holds, i.e., the sum of the entropy production due to the non-conservative products and the relaxation processes
is non-increasing. It is worthwhile mentioning that for isothermal fluids and isentropic fluids we have used different entropy functions, namely, the mixture of the phasic entropies and the mixture of phasic total energies, respectively. In both cases we have to account for the heat flux in the balance laws for the phasic total energies to render these equations to be redundant in case of barotropic fluids. From a physical point of view this is mandatory. In the isothermal case heat production due to the work performed in the system has to be removed from the system to keep the temperatures constant. On the other hand, for isentropic fluids there is an energy production due to the relaxation processes.
For both cases corresponding heat fluxes have been identified.
Furthermore, since we cannot control the sign of the energy production due to the non-conservative products we have determined the interfacial velocity and the interfacial pressures such that this production term vanishes. Starting from an interfacial velocity determined by a convex combination of phasic velocities a unique set of interfacial pressures has been identified that coincides for both isothermal fluids and isentropic fluids.

The relaxation processes to the equilibrium state has been investigated for both the isothermal case and the isentropic case. Mechanical relaxation has been considered in both cases whereas relaxation of the chemical potentials is only meaningful for isothermal fluids. Algorithms for the computation of the corresponding equilibrium states are given.

Finally, we perform numerical computations in one space dimension by means of an adaptive path-conservative Runge-Kutta DG scheme. 
Computations for isothermal fluids have been performed for the configuration of a vapor-vapor expansion problem near pressure saturation. The results have been validated by means of a sharp-interface model separating two
pure phases modelled by the isothermal Euler equations. The results showed a perfect agreement when neglecting mass transfer. When accounting for phase change then the kinetic relation describing the mass transfer in
the sharp-interface model had to be scaled to obtain a perfect agreement. 
For another configuration we considered a mixture of two components with two phases. Computations were performed for mechanical equilibrium  but without relaxation of the chemical potentials. The results were compared to a phase field model. Again, the results showed a very good agreement between the models.  






\phantomsection
\addcontentsline{toc}{section}{\textbf{References}}
\renewcommand{\refname}{References}
\bibliographystyle{abbrv}
\bibliography{bn-bib}   

\begin{thebibliography}{10}

\bibitem{Baer-Nunziato:1986}
M.~Baer and J.~Nunziato.
\newblock A two-phase mixture theory for the deflagration-to-detonation
  transition ({DDT}) in reactive granular materials.
\newblock {\em International Journal of Multiphase Flow}, 12(6):861--889, 1986.

\bibitem{Blaut:2021}
B.~Blaut.
\newblock A {P}ath-conservative {DG} {S}cheme for {C}onservation {L}aws, 2021.
\newblock Master thesis (in German).

\bibitem{boukili-herard:2018}
{Boukili, Hamza} and {H\'erard, Jean-Marc}.
\newblock Relaxation and simulation of a barotropic three-phase flow model.
\newblock {\em ESAIM: M2AN}, 53(3):1031--1059, 2019.

\bibitem{Coquel_Herard_Saleh-Seguin:2014}
F.~Coquel, J.-M. H\'erard, K.~Saleh, and N.~Seguin.
\newblock Two properties of two-velocity two-pressure models for two-phase
  flows.
\newblock {\em Int.~J.~Multiphase Flows}, 12(3):593--600, 2014.

\bibitem{Drew:1983}
D.~Drew.
\newblock Mathematical modeling of two-phase flow.
\newblock {\em Ann. Rev. Fluid Mech.}, 15:261--291, 1983.

\bibitem{Drew-Passman:1999}
D.~Drew and S.~Passman.
\newblock {\em Theory of Multicomponent Fluids}, volume 135 of {\em Applied
  Mathematical Sciences}.
\newblock Springer, 1999.

\bibitem{Dreyer-Duderstadt-Hantke-Warnecke:2012}
W.~Dreyer, F.~Duderstadt, M.~Hantke, and G.~Warnecke.
\newblock Bubbles in liquids with phase transition. part 1: On phase change of
  a single vapor bubble in liquid water.
\newblock {\em Contin.~Mech.~Thermodyn.}, 24:461--483, 2012.

\bibitem{Flatten-Lund:2011}
T.~Fl\r{a}tten and H.~Lund.
\newblock Relaxation two-phase flow models and the subcharacteristic condition.
\newblock {\em Mathematical Models and Methods in Applied Sciences},
  21(12):2379--2407, 2011.

\bibitem{Gallouet-Herard-Seguin:04}
T.~Gallou{\"e}t, J.-M. H{\'e}rard, and N.~Seguin.
\newblock Numerical modelling of two-phase flows using the two-fluid
  two-pressure approach.
\newblock {\em Mathematical Models and Methods in Applied Sciences},
  14(5):663--700, 2004.

\bibitem{GerhardMueller:2016}
N.~Gerhard and S.~M{\"u}ller.
\newblock Adaptive multiresolution discontinuous {G}alerkin schemes for
  conservation laws: multi-dimensional case.
\newblock {\em Computational and Applied Mathematics}, 35(2):321--349, 2016.

\bibitem{GerhardMuellerSikstel:2021}
N.~Gerhard, S.~M{\"u}ller, and A.~Sikstel.
\newblock A wavelet-free approach for multiresolution-based grid adaptation for
  conservation laws.
\newblock {\em Communications on Applied Mathematics and Computation},
  4:108--142, 2021.

\bibitem{Goatin-LeFloch:2004}
P.~Goatin and P.~LeFloch.
\newblock {The Riemann problem for a class of resonant hyperbolic systems of
  balance laws}.
\newblock {\em Annales Inst.~Henri Poincare}, 21(6):881--902, 2004.

\bibitem{Godlewski-Raviart:1991}
E.~Godlewski and P.-A. Raviart.
\newblock {\em Hyperbolic {S}ystems of {C}onservation {L}aws}, volume 3-4 of
  {\em Math{\'e}matiques \& Applications (Paris)}.
\newblock Springer, 1991.

\bibitem{Godunov1996}
S.~K. {Godunov}, T.~Y. Mikha{\^{i}}lova, and E.~I. Romenski{\^{i}}.
\newblock {Systems of thermodynamically coordinated laws of conservation
  invariant under rotations}.
\newblock {\em Siberian Mathematical Journal}, 37(4):690--705, jul 1996.

\bibitem{Godrom2003}
S.~K. {Godunov} and E.~Romenski.
\newblock {\em {Elements of Continuum Mechanics and Conservation Laws}}.
\newblock Springer US, 2003.

\bibitem{Godunov:1995a}
S.~K. {Godunov} and E.~I. Romenski.
\newblock {Thermodynamics, conservation laws, and symmetric forms of
  differential equations in mechanics of continuous media}.
\newblock In {\em {Computational Fluid Dynamics Review 95}}, pages 19--31.
  {John Wiley, NY}, 1995.

\bibitem{Han-Hantke-Mueller:2017}
E.~Han, M.~Hantke, and S.~M{\"u}ller.
\newblock Efficient and robust relaxation procedures for multi-component
  mixtures including phase transition.
\newblock {\em Journal of Computational Physics}, 338:217--239, 2017.

\bibitem{Hantke2013}
M.~Hantke, W.~Dreyer, and G.~Warnecke.
\newblock Exact solutions to the {R}iemann problem for compressible isothermal
  {E}uler equations for two phase flows with and without phase transition.
\newblock {\em Quarterly of Applied Mathematics}, 71(3):509 -- 540, 2013.

\bibitem{Hantke-Matern-Warnecke-Yaghi:2024}
M.~Hantke, C.~Matern, G.~Warnecke, and H.~Yaghi.
\newblock The {R}iemann problem for a two-phase mixture hyperbolic system with
  phase function and multi-component equation of state.
\newblock {\em Quarterly of Applied Mathematics}, 82:451–466, June 2024.

\bibitem{Hantke-Mueller-Grabowsky:2020}
M.~Hantke, S.~M{\"u}ller, and L.~Grabowsky.
\newblock News on {B}aer-{N}unziato-type model at pressure equilibrium.
\newblock {\em Continuum Mechanics and Thermodynamics}, 2020.

\bibitem{Hantke2019a}
M.~Hantke and F.~Thein.
\newblock A general existence result for isothermal two-phase flows with phase
  transition.
\newblock {\em Journal of Hyperbolic Differential Equations}, 16(04):595--637,
  2019.

\bibitem{Hantke-Mueller:18}
{Hantke, M.} and {M\"uller, S.}
\newblock Analysis and simulation of a new multi-component two-phase flow model
  with phase transitions and chemical reactions.
\newblock {\em Quart. Appl. Math.}, 76(2):253--287, 2018.

\bibitem{Hantke-Mueller:19}
{Hantke, M.} and {M\"uller, S.}
\newblock Closure conditions for a one temperature non-equilibrium
  multi-component model of {B}aer-{N}unziato type.
\newblock {\em ESAIM: ProcS}, 66:42--60, 2019.

\bibitem{Herard:07}
J.-M. H{\'e}rard.
\newblock A three-phase flow model.
\newblock {\em Mathematical and Computer Modelling}, 45(5):732--755, 2007.

\bibitem{herard:2016}
J.-M. H{\'e}rard.
\newblock A class of compressible multiphase flow models.
\newblock {\em Comptes Rendus Mathematique}, 354(9):954--959, 2016.

\bibitem{Herard-Hurisse:2012}
J.-M. H{\'e}rard and O.~Hurisse.
\newblock A fractional step method to compute gas-liquid flows.
\newblock {\em Computers and Fluids}, 55:57--69, 2012.

\bibitem{Herard--Jomee:23}
{H\'erard, J.-M.} and {Jom\'ee, G.}
\newblock Pressure relaxation in some multiphase flow models.
\newblock {\em ESAIM: ProcS}, 72:19--40, 2023.

\bibitem{Herard--Jomee:23-preprint}
{H\'erard, J.-M.} and {Jom\'ee, G.}
\newblock {Relaxation process in a hybrid two-phase flow model}.
\newblock 2023.

\bibitem{Herard-Mathise:2021}
{H\'erard, J.-M.} and {Mathis, H.}
\newblock A three-phase flow model with two miscible phases.
\newblock {\em ESAIM: M2AN}, 53(4):1373--1389, 2019.

\bibitem{Herard-Hurisse:2021}
{H\'erard, Jean-Marc}, {Hurisse, Olivier}, and {Quibel, Lucie}.
\newblock A four-field three-phase flow model with both miscible and immiscible
  components.
\newblock {\em ESAIM: M2AN}, 55:S251--S278, 2021.

\bibitem{HMS:2014}
N.~Hovhannisyan, S.~M\"uller, and R.~Sch\"afer.
\newblock Adaptive multiresolution discontinuous galerkin schemes for
  conservation laws.
\newblock {\em Mathematics of Computation}, 83:113--151, 01 2014.

\bibitem{IMueller}
{I.~M\"uller}.
\newblock {\em Thermodynamics}.
\newblock Pitman, London, 1985.

\bibitem{Kapila-Menikoff-Bdzil-Son-Stewart:2001}
A.~Kapila, R.~Menikoff, J.~Bdzil, S.~Son, and D.~Stewart.
\newblock Two-phase modelling of { DDT} in granular materials: Reduced
  equations.
\newblock {\em Phys. Fluid}, 13:3002--3024, 2001.

\bibitem{Kato:1975}
T.~Kato.
\newblock The cauchy problem for quasi-linear symmetric hyperbolic systems.
\newblock {\em Arch. Rational Mech. Anal.}, 58(3):181--205, 1975.

\bibitem{Lallemand-Saurel:2000}
M.-H. Lallemand, A.~Chinnayya, and O.~{Le Metayer}.
\newblock Pressure relaxation procedures for multiphase compressible flows.
\newblock {\em International Journal for Numerical Methods in Fluids},
  49(1):1--56, 2005.

\bibitem{Linga-Flatten:2019}
{Linga, G.} and {Fl\r{a}tten, T.}
\newblock A hierarchy of non-equilibrium two-phase flow models.
\newblock {\em ESAIM: ProcS}, 66:109--143, 2019.

\bibitem{May2023}
S.~May and F.~Thein.
\newblock Explicit implicit domain splitting for two phase flows with phase
  transition.
\newblock {\em Physics of Fluids}, 35(1):016108, 2023.

\bibitem{MuellerMueller}
I.~M\"uller and W.~M\"uller.
\newblock {\em Fundamentals of Thermodynamics and Applications}.
\newblock Springer-Verlag, Berlin, 2009.

\bibitem{Mueller:2009}
S.~M\"uller.
\newblock Multiresolution schemes for conservation laws.
\newblock In R.~DeVore and A.~Kunoth, editors, {\em Multiscale, nonlinear and
  adaptive approximation}, pages 379--408. Berlin: Springer, 2009.
\newblock Dedicated to Wolfgang Dahmen on the occasion of his 60th birthday.

\bibitem{Mueller-Hantke-Richter:2014}
S.~M{\"u}ller, M.~Hantke, and P.~Richter.
\newblock Closure conditions for non-equilibrium multi-component models.
\newblock IGPM Preprint 414, RWTH Aachen University, 2014.

\bibitem{Mueller-Hantke-Richter:16}
S.~M{\"u}ller, M.~Hantke, and P.~Richter.
\newblock Closure conditions for non-equilibrium multi-component models.
\newblock {\em Continuum Mechanics and Thermodynamics}, 28(4):1157--1189, Jul
  2016.

\bibitem{Peshkov2018}
I.~Peshkov, M.~Pavelka, E.~Romenski, and M.~Grmela.
\newblock Continuum mechanics and thermodynamics in the hamilton and the
  {Godunov}-type formulations.
\newblock {\em Continuum Mechanics and Thermodynamics}, 30(6):1343--1378, Nov
  2018.

\bibitem{Petitpas-Saurel-Franquet-Chinnayya:2009}
F.~Petitpas, R.~Saurel, E.~Franquet, and A.~Chinnayya.
\newblock Modeling detonation waves in condensed materials: multiphase cj
  conditions and multidimensional computations.
\newblock {\em Shock Waves}, 19:377--401, 2009.

\bibitem{ReAbgrall:2020}
B.~Re and R.~Abgrall.
\newblock Non-equilibrium model for weakly compressible multi-component flows:
  The hyperbolic operator.
\newblock In {\em Lecture Notes in Mechanical Engineering}, pages 33--45.
  Springer International Publishing, 2020.

\bibitem{RhebergeBokhoveVanderVegt:2008}
S.~Rhebergen, O.~Bokhove, and J.~{van der Vegt}.
\newblock Discontinuous galerkin finite element methods for hyperbolic
  nonconservative partial differential equations.
\newblock {\em Journal of Computational Physics}, 227(3):1887--1922, 2008.

\bibitem{RioMartin2023}
L.~R{\'i}o-Mart{\'i}n and M.~Dumbser.
\newblock High-order ader discontinuous galerkin schemes for a symmetric
  hyperbolic model of compressible barotropic two-fluid flows.
\newblock {\em Communications on Applied Mathematics and Computation}, Nov
  2023.

\bibitem{Rodio-Abgrall:2015}
M.~G. Rodio and R.~Abgrall.
\newblock An innovative phase transition modeling for reproducing cavitation
  through a five-equation model and theoretical generalization to six and
  seven-equation models.
\newblock {\em International Journal of Heat and Mass Transfer}, pages~--, 05
  2015.

\bibitem{Romenski1998}
E.~Romenski.
\newblock Hyperbolic systems of thermodynamically compatible conservation laws
  in continuum mechanics.
\newblock {\em Math. Comput. Modell.}, 28(10):115--130, 1998.

\bibitem{Romenski2001}
E.~Romenski.
\newblock Thermodynamics and hyperbolic systems of balance laws in continuum
  mechanics.
\newblock {\em In: Toro E.F. (eds) {Godunov} Methods. Springer, Boston, MA},
  pages 745--761, 2001.

\bibitem{Romenski2009}
E.~Romenski, D.~Drikakis, and E.~Toro.
\newblock Conservative models and numerical methods for compressible two-phase
  flow.
\newblock {\em Journal of Scientific Computing}, 42(1):68, Jul 2009.

\bibitem{Romenski2007}
E.~Romenski, A.~D. Resnyansky, and E.~F. Toro.
\newblock Conservative hyperbolic formulation for compressible two-phase flow
  with different phase pressures and temperatures.
\newblock {\em Quart. Appl. Math.}, 65(2):259--279, 2007.

\bibitem{Ruggeri2021}
T.~Ruggeri and M.~Sugiyama.
\newblock {\em Classical and {R}elativistic {R}ational {E}xtended
  {T}hermodynamics of {G}ases}.
\newblock Springer, Cham, 2021.

\bibitem{Saleh:2012}
K.~Saleh.
\newblock {\em Analyse et Simulation Num{\'e}rique par Relaxation
  d'{\'E}coulements Diphasiques Compressibles}.
\newblock PhD thesis, Universit{\'e} Pierre et Marie Curie, Paris, 2012.

\bibitem{saleh-seguin:2020}
K.~Saleh and N.~Seguin.
\newblock Some mathematical properties of a barotropic multiphase flow model.
\newblock {\em ESAIM: ProcS}, 69:70--78, 2020.

\bibitem{Saurel-Abgrall:1999}
R.~Saurel and R.~Abgrall.
\newblock A multiphase {G}odunov method for compressible multifluid and
  multiphase flows.
\newblock {\em Journal of Computational Physics}, 150(2):425--467, 1999.

\bibitem{Saurel-Le}
R.~Saurel and O.~LeMetayer.
\newblock A multiphase flow model for compressible flows with interfaces,
  shocks, detonation waves and cavitation.
\newblock {\em J.~Fluid Mech.}, 43:239--271, 2001.

\bibitem{Saurel-Petitpas-Abgrall:2008}
R.~Saurel, F.~Petitpas, and R.~Abgrall.
\newblock Modelling phase transition in metastable liquids: Application to
  cavitating and flashing flows.
\newblock {\em Journal of Fluid Mechanics}, 607:313--350, 07 2008.

\bibitem{Thein:2018}
F.~Thein.
\newblock {\em Results for {T}wo {P}hase {F}lows with {P}hase {T}ransition}.
\newblock PhD thesis, Otto-von-Guericke Universit{\"a}t, Magdeburg, 2018.

\bibitem{Thein2022}
F.~Thein, E.~Romenski, and M.~Dumbser.
\newblock Exact and numerical solutions of the {R}iemann problem for a
  conservative model of compressible two-phase flows.
\newblock {\em Journal of Scientific Computing}, 93(3):83, Nov 2022.

\bibitem{Toro:1997}
E.~F. Toro.
\newblock {\em Riemann {S}olvers and {N}umerical {M}ethods for {F}luid
  {D}ynamics: A {P}ractical {I}ntroduction}.
\newblock Berlin: Springer, 1997.

\bibitem{STEAM}
W.~Wagner and H.-J. Kretzschmar.
\newblock {\em International Steam Tables}.
\newblock Berlin - Heidelberg: Springer, 2008.

\bibitem{Zein:2010}
A.~Zein.
\newblock {\em Numerical methods for multiphase mixture conservation laws with
  phase transition}.
\newblock PhD thesis, Otto-von-Guericke University, Magdeburg, 2010.

\bibitem{Zein-Hantke-Warnecke:2010}
A.~Zein, M.~Hantke, and G.~Warnecke.
\newblock Modeling phase transition for compressible two-phase flows applied to
  metastable liquids.
\newblock {\em J.~Comput.~Phys.}, 229(8):2964--2998, 2008.

\bibitem{Zein-Hantke-Warnecke:2013}
A.~Zein, M.~Hantke, and G.~Warnecke.
\newblock On the modelling and simulation of a laser-induced cavitation bubble.
\newblock {\em Int.~J.~Num.~Meth.~Fluids.}, 73(2):172--203, 2013.

\end{thebibliography}

\end{document}